%% file: TOIS2018.tex
\documentclass[format=acmsmall, review=False, screen=true]{acmart}

\usepackage{booktabs} 
\usepackage{lipsum} 
\usepackage{balance}
\usepackage{paralist}
\usepackage{filecontents}
\usepackage{etoolbox}
\usepackage{mathrsfs}
\usepackage{bm}
\usepackage{setspace}
\usepackage{amssymb}
\usepackage{graphicx}
\usepackage{caption}
\usepackage{subcaption}
\usepackage[ruled]{algorithm2e}
\usepackage{algpseudocode}

\newcommand{\revised}[1]{\textcolor{black}{{\bf}{#1}{\bf}}}

\acmJournal{TOIS}
\acmVolume{9}
\acmNumber{4}
\acmArticle{25}
\acmYear{2019}
\acmMonth{12}
\copyrightyear{2019}

\setcopyright{acmlicensed}

\acmDOI{0000001.0000001}

\begin{document}
\title{Explainable Product Search with a Dynamic Relation Embedding Model}


\author{Qingyao Ai}
\authornote{This work is done during his Ph.D. at University of Massachusetts Amherst.}
\affiliation{%
	\institution{School of Computing, University of Utah}
	\city{Salt Lake City} 
	\state{UT} 
	\country{USA}
	\postcode{84112-9205}
}
\email{aiqy@cs.utah.edu}

\author{Yongfeng Zhang}
\affiliation{%
	\institution{Department of Computer Sciences, Rutgers University}
	\city{Piscataway} 
	\state{NJ} 
	\country{USA}
	\postcode{08854-8019}
}
\email{yongfeng.zhang@rutgers.edu}

\author{Keping Bi}
\affiliation{%
	\institution{College of Information and Computer Sciences, University of Massachusetts Amherst}
	\city{Amherst} 
	\state{MA} 
	\country{USA}
	\postcode{01003-9264}
}
\email{kbi@cs.umass.edu}

\author{W. Bruce Croft}
\affiliation{%
	\institution{College of Information and Computer Sciences, University of Massachusetts Amherst}
	\city{Amherst} 
	\state{MA} 
	\country{USA}
	\postcode{01003-9264}
}
\email{croft@cs.umass.edu}

\begin{abstract}
Product search is one of the most popular methods for customers to discover products online.
Most existing studies on product search focus on developing effective retrieval models that rank items by their likelihood to be purchased.
They, however, ignore the problem that there is a gap between how systems and customers perceive the relevance of items.
Without explanations, users may not understand why product search engines retrieve certain items for them, which consequentially leads to imperfect user experience and suboptimal system performance in practice.
In this work, we tackle this problem by constructing explainable retrieval models for product search.
Specifically, we propose to model the ``search and purchase'' behavior as a dynamic relation between users and items, and create a dynamic knowledge graph based on both the multi-relational product data and the context of the search session.
Ranking is conducted based on the relationship between users and items in the latent space, and explanations are generated with logic inferences and entity soft matching on the knowledge graph.
Empirical experiments show that our model, which we refer to as the Dynamic Relation Embedding Model (DREM), significantly outperforms the state-of-the-art baselines and has the ability to produce reasonable explanations for search results.




\end{abstract}

%
%



\maketitle

\renewcommand{\shortauthors}{Q. Ai et al.}


\input{introduction}

\input{related}

\input{model}

\input{explanation}

\input{experiment_setup}

\input{experiment_results}


\input{conclusion}

\section{Acknowledgments}
This work was supported in part by the Center for Intelligent Information Retrieval and in part by an award from Amazon.com. Any opinions, findings and conclusions or recommendations expressed in this material are those of the authors and do not necessarily reflect those of the sponsor.

\bibliographystyle{ACM-Reference-Format-Journals}
\bibliography{sigproc} 

\newpage

\end{document}

%% file: introduction.tex
\section{Introduction}

Product search represents a special retrieval problem where users submit queries to a search engine in order to find products.
\revised{
As the e-commerce market keeps growing\footnote{\url{https://www.statista.com/statistics/534123/e-commerce-share-of-retail-sales-worldwide/}} and millions of new products are introduced every day, search has become one of the most effective and popular methods for people to discover products.
According to a recent marketing report\footnote{\url{https://www.mckinsey.com/industries/retail/our-insights/how-retailers-can-keep-up-with-consumers}}, the majority of sales on Amazon comes from its product search service.
Thus, the quality of product search affects both user experience with online shopping and the profits of e-commerce companies. 
}

Due to their applications in e-commerce, product search engines are usually optimized for user transactions.
In a typical product search scenario, users submit a query to a search engine, and the system returns a list of products for users to explore.
Items on the search engine result pages are often sorted by their likelihood to be purchased so that the number of user transactions can be maximized in each search session~\cite{duan2013supporting,van2016learning,ai2017learning}.

Showing relevant items on the top, however, is not enough to guarantee the effectiveness of product search.
Existing studies based on this paradigm often simplify the problem of product search by assuming that users will purchase an item as long as it is observed and relevant~\cite{van2016learning,ai2017learning}.
They ignore the fact that there is a significant gap between the item relevance perceived by search engines and e-shopping users~\cite{zhang2019sigir}.
Because purchasing is expensive and highly personal~\cite{ai2017learning}, users often need a good reason to justify their purchases.
As modern product search systems become increasingly sophisticated, it is difficult for normal users to understand why search engines retrieve certain items for them.
A direct consequence is that users may not perceive a retrieved item as relevant even when it satisfies their search intents.

Therefore, to actually optimize user purchases, a good product search engine needs to retrieve relevant products as well as providing good explanations of why retrieved items should be interesting to users.
Previous studies on product recommendation have shown that providing appropriate explanations significantly improves user acceptance for recommended items~\cite{herlocker2000explaining,tintarev2007survey}.
It benefits recommendation systems in multiple ways including user satisfaction, system transparency, debugging complexity, etc.~\cite{bilgic2005explaining,cramer2008effects,tintarev2011designing,zhang2014explicit}. 
It is reasonable to assume that providing explanations for retrieval results will be equally beneficial for product search. 



Despite its potential, the explainability of retrieval systems has not been well studied in product search.
There are two problems that limit the development of explainable product search systems.
First, purchasing is a complicated behavior as it depends on multiple factors such as user preference, product presentation, and search context.
To provide high-quality explanations, we need to consider the relationship between users and products from multiple angles (e.g., brands, categories, etc.).
As far as we know, no existing retrieval model can directly incorporate different product knowledge for product search.       
Second, producing a readable explanation requires the system to have logical reasoning.
For example, the system should be able to infer that ``\textit{Bob} likes \textit{Apple} products" after seeing him search and purchase multiple products from \textit{Apple}.
An explanation is reasonable and effective only when it is formulated based on well-grounded logic, while how to construct such a product retrieval model with logical reasoning ability is still an open question for the IR community. 

In this paper, we present our initial attempt to tackle the problem of explainable product search.
Inspired by the studies of relation prediction in knowledge base~\cite{bordes2011learning,bordes2013translating}, we propose to create a unified knowledge graph on multiple types of product data, and conduct retrieval with it. 
Our motivation is to integrate multi-relational product information for search, and generate explanations with logic inference on the knowledge graph.
Previous studies on knowledge graphs and embeddings mainly focus on modeling static data relationships that do not change.
In product search, however, the relationships between users and items are not deterministic among different search sessions.
For example, a camera lens is relevant to a photographer when she searches for ``camera'', but not when she searches for ``toothbrush''. 
To solve this problem, we propose a Dynamic Relation Embedding Model (DREM) that dynamically models the relationship between users and items based on the search context.
Although inferring the relationship of an arbitrary user-item pair with observed data is often infeasible due to data sparsity~\cite{zhang2016collaborative}, we show that reasonable explanation paths can be extracted with our proposed method through entity soft matching.
Empirical experiments and analysis with Amazon benchmark datasets show that incorporating different product knowledge with DREM has significant potential for explainable product search.

Our main contributions can be summarized as follows:
\begin{itemize}
\item We propose a Dynamic Relation Embedding Model to construct a session-dependent knowledge graph for product retrieval.
\item We propose a Soft Matching Algorithm to efficiently extract explainable paths with knowledge embeddings for search explanations. 
\item We conducted both retrieval experiments and case studies to verify the effectiveness of the proposed approach in product retrieval and explainable search.
\end{itemize}

The rest of this paper is organized as follows.
In Section~\ref{sec:related_work}, we discuss the related work.
Then we introduce our approach and how to extract explanations for product search in Section~\ref{sec:our_approach} and \ref{sec:explanation}.
We describe our experimental setup and results in Section~\ref{sec:setup} and \ref{sec:results}.
Finally, we conclude our work and discuss future studies in Section~\ref{sec:conclusion}.
  


%% file: related.tex
\section{Related Work}\label{sec:related_work}

There are four lines of studies that are related to our work: product search, explainable systems, knowledge embedding, and neural information retrieval.

\subsection{Product Search}
Product search refers to the problem of retrieving relevant products for customers to satisfy their purchase intents.
Previous studies on product search mainly focus on retrieving products based on their structured aspects such as brand, category etc.
For example, Lim et al.~\cite{lim2010multi} propose a document profile model to suggest semantic tags for each item based on their structural aspects, so that we can retrieve products by matching queries with multiple product aspects simultaneously.
Despite their success, searching with structured data cannot satisfy the need of e-shopping users today as their intents become more and more complicated.
As shown by Duan et al.~\cite{duan2013supporting}, writing structured search queries (e.g., SQL) is usually considered hard and inconvenient for search users.
In most cases, queries submitted to product search engines are free-form text that is difficult to structure.
However, there often exists a large vocabulary gap between the descriptions of products and user queries~\cite{van2016learning,ai2017learning}.
Nurmi et al.~\cite{nurmi2008product} find that the words customers used to write their shopping lists are often different from those sellers used to describe their products.
Because of these problems, IR researchers have proposed to construct semantic latent space for product search and conduct matching between queries and products with their latent representations.
For instance, Yu et al.~\cite{yu2014latent} construct a Latent Dirichlet Allocation model with product information and use it to diverse e-commerce search results.
Van Gysel et al.~\cite{van2016learning} introduce a Latent Semantic Embedding model that maps and matches n-grams from queries and product descriptions into a hidden space.
Ai et al.~\cite{ai2017learning,ai2019zero} constructed hierarchical and attentive embedding models that models the generative process of user purchases and product reviews so that items can be ranked by their likelihood to be purchased given the user and the query. 
Guo et al.~\cite{guo2018multi} propose a TranSearch model that can directly match text queries with product images. 
Bi et al.~\cite{bi2019leverage} use the embedding representations of clicked items as context information to refine the ranking of retrieved products.
Zhang et al.~\cite{zhang2018towards} and Bi et al.~\cite{bi2019conversational} also extract product aspects from review data and build embedding networks that encodes both items and their extracted aspects for conversational product search and recommendation.
Besides latent embedding techniques, there is a variety of studies~\cite{aryafar2017ensemble,karmaker2017application,Hu:2018:RLR:3219819.3219846} on extracting different text and product features and feeding them into learning-to-rank models for the optimization of different product retrieval objectives.
Wu et al.~\cite{wu2017ensemble} manually extract multiple statistic features from product search logs and construct an ensemble tree model to predict user clicks.
Wu et al.~\cite{wu2018turning} developed a special ranking loss function that optimizes product search engines by maximizing the revenue generated from online transactions.
While they are effective for product search on unstructured free text, it is difficult to extend these methods with structured metadata and their relationships.
As far as we know, our work is the first study that jointly models structured and unstructured multi-relational data in a single latent space for product search.


\subsection{Explainable System}
While there have been a variety of studies on the transparency of complicated systems, such as the interpretability of a machine learning model~\cite{hinton2015distilling,montavon2017explaining,krakovna2016increasing} and the importance of different ranking features~\cite{vidovic2016feature,hechtlinger2016interpretation}, the concept of explainability in this paper refers to a system's ability on generating human-readable explanations for its results, so that users are more likely to conduct certain behavior (e.g., clicks, purchases, etc.) after reading them.  
Explainability is an important criterion to measure the quality of a production system and has been extensively studied for social and product recommendation.
Strategies investigated for explainable systems include straightforward methods based on structured information (e.g. social tags, ``people also bought") and complicated methods based on the generated topic aspects from text data~\cite{zhang2017explainable}.
For example, Sharma and Cosley~\cite{sharma2013social} propose to incorporate social network information for music recommendation.
They generate recommendation explanations based on the social relationship between users and items.
Zhang et al.~\cite{zhang2014explicit} conducted product recommendation with extracted product features and user opinions from phrase-level sentiment analysis to provide explanations for the recommendation results accordingly.
He et al.~\cite{he2015trirank} applied an aspect extraction algorithm~\cite{zhang2014users} on product reviews and modeled the ternary relation between users, items and aspects to conduct explainable recommendation.
Tintarev and Masthoff~\cite{tintarev2011designing} studied different factors that affect the quality of an explanation, and tried to build an evaluation guideline for explainable recommendation. 
\revised{
Despite the popularity of explainable systems in recommendation, there are few studies on explaining search results for retrieval systems.
To the best of our knowledge, this paper is the first work that tries to tackle the problem of explainable search for product retrieval.
Also, different from previous studies that attempts to create explanations directly with the recommendation models, we propose a post-hoc explanation algorithm that leverages the knowledge representations built by the retrieval model to generate search result explanations. 
}

\subsection{Knowledge Embedding}
Knowledge embedding refers to a technique that models multi-relational data by constructing latent representations for entities and relationships.
The goal is to extract local or global connectivity patterns between entities and use these patterns to understand and generalize the observed relationships between a specific entity and others~\cite{bordes2013translating}. 
Inspired by the success of collaborative filtering in product recommendation~\cite{sarwar2001item}, most existing methods for multi-relational data are designed based on matrix factorization or related approaches.
For instance, Harshman and Lundy~\cite{harshman1994parafac} introduce a tensor factorization method for relation prediction.
Singh and Gordon~\cite{singh2008relational} propose a collective matrix factorization method improve relationship predictive accuracy by exploiting information from one relation while predicting another.
Nickle et al.~\cite{nickel2011three} propose a three-way model to conduct collective learning on multi-relational data.
\revised{
Liang et al.~\cite{liang2016factorization} propose to jointly decomposes the user-item interaction matrix and the item-item co-occurrence matrix with shared item embeddings.
There is also another line of studies trying to tackle the problem with non-parametric Bayesian frameworks.
Miller et al.~\cite{miller2009nonparametric} experiment with a non-parametric Bayesian model for link prediction on social networks.
Zhu et al.~\cite{zhu2016max} further improve the framework by introducing a max-margin technology.
More recently, the advance of deep learning techniques has led many researchers to explore the effectiveness of knowledge embedding with neural models.
Sochet et al.~\cite{socher2013reasoning} propose a neural tensor network to infer unobserved knowledge base entries based on the embedding of entities.
Bordes et al.~\cite{bordes2011learning} propose to embed symbolic representations from knowledge bases into a flexible continuous vector space with neural network architectures.
Furthermore, Bordes et al.~\cite{bordes2013translating} designed a translation-based embedding model (transE) to jointly model entities and relationships within a single latent space.
He et al.~\cite{he2018translation} propose a translation-based recommendation system to predict users' personalized sequential behaviors.
Yang et al.~\cite{yang2018towards} use the intermediate information embedding built by existing recommendation systems to analyze user-item relationships and create explanations.
Zhang et al.~\cite{zhang2018learning} and Ai et al.~\cite{ai2018learninghe} construct knowledge embedding graphs to generate post-hoc explanations for recommendation results.
}
In this work, we follow a similar paradigm with previous work but focus specifically on adapting the existing knowledge embedding techniques for explainable product search.



\subsection{Neural Information Retrieval}
Neural technology has attracted a lot of attention in the IR community in recent years. 
There have been a variety of neural models proposed for different information retrieval tasks such as ad-hoc retrieval~\cite{ai2016analysis}, question answering~\cite{yang2016anmm}, and recommendation~\cite{he2017neural}.
For example, Huang et al.~\cite{huang2013learning} propose to embed tri-grams with neural network and generate rank lists according to the cosine similarity between the aggregated tri-gram embeddings of queries and documents.
Vuli{\'c} and Moens~\cite{vulic2015monolingual} conduct multi-lingual retrieval by representing queries and documents with their averaged word embedding.
Guo et al.~\cite{guo2016deep} categorize existing neural retrieval models into two groups -- the representation-based models and the interaction-based models.
Based on this idea, they propose a deep relevance matching model that explicitly models the interaction patterns between queries and relevant documents.
There are also many studies on applying deep learning technology on other complicated retrieval problems such as user modeling~\cite{loyola2017modeling, borisov2016neural} and context-aware relevance modeling~\cite{ai2018learning,bello2018seq2slate}.
Similar to Van Gysel et al.~\cite{van2016learning} and Ai et al.~\cite{ai2017learning}, we use neural networks, especially the embedding-based neural retrieval framework, as the main technique for the development of product retrieval models. 
However, as shown in Section~\ref{sec:results}, we also conduct a systematic comparison of our proposed models and the state-of-the-art non-neural baselines.


	

%% file: model.tex

\begin{figure*}
	\centering
	\includegraphics[width=5in]{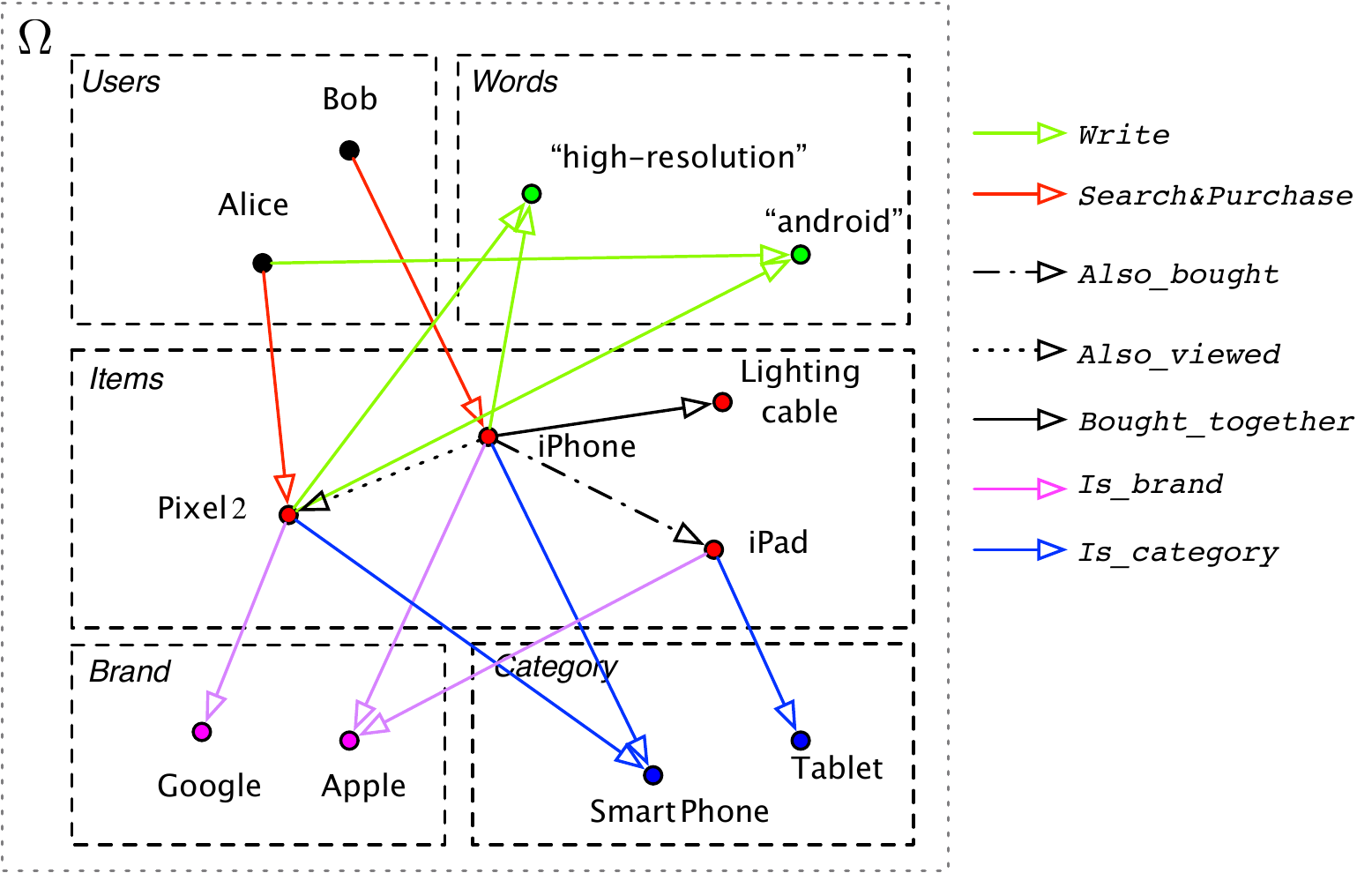}
	\caption{An example of a knowledge graph created by DREM for product search. }
	\label{fig:DREM}
\end{figure*}

\section{Model Description}\label{sec:our_approach}
We now provide detailed descriptions of the Dynamic Relation Embedding Model (DREM) for explainable product search.
DREM jointly models different user/product knowledge, and creates a knowledge graph with both static and dynamic relationships.
In this section, we first provide an overview of DREM and describe how to conduct product search with it.
Then we discuss the modeling of static and dynamic entity relationships in detail. 

\subsection{Overview}\label{sec:model_overview}

As discussed previously, explainable product search requires retrieval systems to be capable of modeling and conducting logical inference with different product information.  
The relationships between product-related entities usually are complicated and not injective.
For example, an item could belong to multiple categories and a category could include multiple items.
One of the most popular methods to model such multi-relational data is to construct a \textit{knowledge graph}. 
In a knowledge graph, each node represents an entity and each edge represents the existence of a certain relationship between two entities.
With this design, a knowledge graph satisfies the need of logical inference because any relationship between an arbitrary pair of entities can be inferred by the path between them.
In this paper, we design DREM to construct a knowledge graph for product data, and conduct explainable product search accordingly.

An example of a knowledge graph created by DREM is shown in Figure~\ref{fig:DREM}.
In DREM, each user, product and related entities are represented with vectors in a single latent space $\Omega$.
Two entities are linked with an edge if there is a relationship between them.
Each edge is labeled with a unique symbol to denote the type of the relationship.
In order to conduct product search with DREM, we create a special edge named as \textit{Search\&Purchase} to model the relationship between users and items.
In a search session, a user will be \textit{Search\&Purchase} with an item if he or she purchases the item.
Thus, the problem of product search is to find items that are likely to have the \textit{Search\&Purchase} relationship with the users.

Inspired by previous work on relationship prediction~\cite{bordes2013translating}, we assume that all relationships can be viewed as translations from one entity to another.
Suppose that there exists a relationship $r$ between two entities $x$ and $y$. 
Let $x$ be the head entity and $y$ be the tail entity, then we can directly get $y$ by translating $x$ with $r$ as
$$
\bm{y} = \bm{x} + \bm{r}
$$
Therefore, any relationship in $\Omega$ can be treated as a linear transformation between entities and represented with a latent vector that shares the same dimensionality with $\Omega$.
We refer to the latent vectors of entities and their relationships in $\Omega$ as \textit{entity embeddings} and \textit{relationship embeddings}, respectively.

The key of DREM is to effectively infer the embedding representations of entities and relationships. 
Although a variety of methods has been proposed for learning knowledge base embeddings~\cite{bordes2011learning,bordes2013translating}, none of them is applicable to DREM because the knowledge graph of DREM is not static.
Usually, user intent varies in different search sessions.
The relationship between users and items cannot be determined without the search context.
Therefore, \textit{Search\&Purchase} is a dynamic relationship that must be computed for product search on the fly.
In the next sections, we describe how to jointly model static and dynamic relationships in DREM.

\subsection{Static Relation Modeling}\label{sec:static_relation}
Given the formulation of entities and relationships in the latent space, a generic solution to estimate the embedding parameters is to use an EM-like algorithm~\cite{mclachlan2007algorithm}.
For example, we can iteratively minimize the empirical loss of $(x,r)$ by computing $\bm{r}$ as the mean of $\bm{y} - \bm{x}$ for all entity pairs with $r$, and computing $x$ as the mean of $\bm{y} - \bm{r}$ for all entity pairs with both $x$ and $r$. 
Such a method, however, is not appropriate for search problems because it does not explicitly differentiate entities with and without the relationship.
A trivial solution that gives similar representations to all $x$ and $y$ in $\Omega$ could still have a low empirical loss of $(x,r)$ in practice.
Based on this consideration, we propose to learn DREM by maximizing the posterior probability of observed relationships and minimizing the unobserved ones.

Let $r$ be a static relationship between head entity $x \in X_r$ and tail entity $y\in Y_r$.
$X_r$ and $Y_r$ are the sets of all possible entities that are of the same types with $x$ and $y$, respectively.
We refer to $r$ as static because it holds for $x$ and $y$ universally regardless of the search context. 
Inspired by the study of embedding-based generative framework~\cite{mikolov2013efficient, ai2016improving, ai2017learning}, we define the probability of observing tail entity $y$ given head entity $x$ and relationship $r$ as 
\begin{equation}
P(y|x,r) = \frac{\exp \big((\bm{x}+\bm{r})\cdot \bm{y}\big)}{\sum_{y' \in Y_r}\exp\big((\bm{x}+\bm{r})\cdot \bm{y'}\big)}
\label{equ:softmax}
\end{equation}
where $\bm{r} \in \mathbb{R}^{\alpha}$, $\bm{x} \in \mathbb{R}^{\alpha}$ and $\bm{y} \in \mathbb{R}^{\alpha}$ are the embedding representations of $r$, $x$ and $y$ with $\alpha$ dimensions. 
We directly optimize DREM through maximizing the log likelihood of observed $(x,r,y) \in S_{(x,r,y)}$ triples for all static relationships as
\begin{equation}
\mathcal{L}(S_{(x,r,y)}) = \!\!\!\!\!\!\!\!\sum_{(x,r,y) \in S_{(x,r,y)}} \!\!\!\!\!\!\!\! \log P(y|x,r)
\label{equ:static_loss}
\end{equation}
where $S_{(x,r,y)}$ is the set of all observed static $(x,r,y)$ triples in the training data. 
As shown in Equation~(\ref{equ:softmax}), $P(y|x,r)$ is a softmax function over $y$, which essentially assumes that $\sum_{y \in Y_r}P(y|x,r) = 1$. 
Therefore, the maximization of $\mathcal{L}(S_{(x,r,y)})$ will minimize the probability of unobserved $(x,r,y)$.

Optimizing $\mathcal{L}(S_{(x,r,y)})$ directly, however, is prohibitive in practice.
The computational complexity of $\mathcal{L}(S_{(x,r,y)})$ is $\mathcal{O}(\alpha|S_{(x,r,y)}||Y_r|)$, and the size of $S_{(x,r,y)}$ and $Y_r$ can be large (e.g., there are millions of items in Amazon product datasets).
To efficiently train DREM on large-scale data, we adopt a negative sampling strategy to approximate $P(y|x,r)$ in $\mathcal{L}(S_{(x,r,y)})$.
Negative sampling was first proposed by Mikolov et al.~\cite{mikolov2013distributed} and has been widely applied in machine learning and information retrieval tasks~\cite{mikolov2013distributed,le2014distributed,ai2016analysis,zhang2017joint,ai2017learning}. 
The idea of negative sampling is to approximate the denominator of softmax functions by randomly sampling some negative samples from the corpus.
Specifically, we sample negative instance $y'$ from $Y_r$ and compute $\log P(y|x,r)$ as
\begin{equation}
\log P(y|x,r) = \log \sigma\big((\bm{x}+\bm{r})\cdot \bm{y}\big) + k\cdot \mathbb{E}_{y'\sim P_r}[\log\sigma\big(-(\bm{x}+\bm{r})\cdot \bm{y'}\big)]
\label{equ:static_negative_sample}
\end{equation}
where $k$ is the number of negative instances, $\sigma(x)=\frac{1}{1+e^{-x}}$ is the sigmoid function and $P_r$ is a noise distribution for $y\in Y_r$.

In fact, as shown by previous studies~\cite{levy2014neural}, the optimization of $\mathcal{L}(S_{(x,r,y)})$ with the negative sampling strategy is theoretically principled as it essentially guides the model to factorizing the matrix of mutual information between relations and entities.
Let $\ell(x,r,y)$ be the expected loss on a specific relation triple $(x,r,y) \in S_{(x,r,y)}$ based on Equation~(\ref{equ:static_loss}) and Equation (\ref{equ:static_negative_sample}), then we have
\begin{equation}
\begin{split}
\ell(x, r, y) = \#(x, r, y)\cdot\log\sigma(\bm{y} \cdot (\bm{x} + \bm{r})) + k \cdot \#(x,r) \cdot P_r(y)\cdot \log\sigma(-\bm{y} \cdot (\bm{x} + \bm{r}))
\end{split}
\label{equ:static_local_object}
\end{equation}
where $\#(x, r, y)$ and $\#(x,r)$ are the numbers of observed relation triple $(x, r, y)$ and pair $(x ,r)$ in $S_{(x,r,y)}$.
If we derive the partial gradient of $\ell(x,r,y)$ with respect to $\bm{y} \cdot (\bm{x} + \bm{r})$ and let it be zero, we can easily get the following result:
\begin{equation}
\begin{split}
\bm{y} \cdot (\bm{x} + \bm{r}) & = \log(\frac{\#(x, r, y)}{\#(x,r)}\cdot \frac{1}{P_r(y)}) - \log k 
\end{split}
\label{equ:static_l_final_obj}
\end{equation}
In this work, we follow the common practice of defining $P_r(y)$ as the normalized frequency of $y$ in all observed $(x,r,y)$ for $r$, so the right hand side of Equation~(\ref{equ:static_l_final_obj}) is actually a shifted version of pointwise mutual information between $(x,r)$ and $y$, and optimizing $\mathcal{L}(S_{(x,r,y)})$ with negative sampling is similar to factorizing the mutual information matrix of $(x,r,y)\in S_{(x,r,y)}$. 


\subsection{Dynamic Relation Modeling}\label{sec:dynamic_relation}
In DREM, we create a relationship between users and products named as \textit{Search\&Purchase}.
Due to the nature of search tasks, this relationship is dynamic and cannot be determined without the search context.
For example, Canon cameras could be linked with users when the query is ``digital camera", but not when it is ``mobile phone".
Therefore, the embeddings of \textit{Search\&Purchase} are session-dependent and have to be computed on the fly.
For simplicity, we represent the context of a product search session with the query submitted by the user.
Note that other session information such as previous queries and clicks~\cite{shen2005context} can also be incorporated into the framework if needed.

Let $q$ be the query submitted by user $u$, $\{w_q\}$ be the words of the query, and $\bm{v}$ be the embedding representation of the relationship \textit{Search\&Purchase}.
Then we can compute $\bm{v}$ with a function of $q$ as
\begin{equation}
\bm{v} = f(q) = f(\{w_q|w_q\in q\})
\label{equ:f}
\end{equation}
Previous studies have proposed several methods to model search intents with queries in latent space~\cite{zamani2016estimating,vulic2015monolingual,van2016learning,ai2017learning}.
Ai et al.~\cite{ai2017learning} have explored and compared three options including averaged word vectors~\cite{vulic2015monolingual}, non-linear projections~\cite{van2016learning}, and recurrent neural networks (RNN)~\cite{palangi2016deep}.
In their experiments, the non-linear projection method usually produces the best and most robust results.
Suppose that the latent space of DREM has $\alpha$ dimensions, then the non-linear projection method defines $f(q)$ as
\begin{equation}
f(q) = \tanh(W \cdot \frac{\sum_{w_q \in q}\bm{w_q}}{|q|} + b)
\label{equ:fs}
\end{equation}
where $|q|$ is the length of $q$, $\bm{w_q}$ is the embedding of $w_q$, and $W\in\mathbb{R}^{\alpha \times \alpha}$, $b\in\mathbb{R}^{\alpha}$ are parameters to be learned in the training process.
In this work, we employ this non-linear projection function to compute $f(q)$.
We tried other query embedding functions~\cite{vulic2015monolingual,palangi2016deep} as well, but observed no significant performance improvement in our retrieval experiments. 

Similar to the modeling of static relationships, we use a softmax function to compute the conditional probability of item $i$ given user $u$ with the dynamic relationship $v$ as:
\begin{equation}
P(i|u,\bm{v}) = \frac{\exp \big((\bm{u}+\bm{v})\cdot \bm{i}\big)}{\sum_{i' \in I}\exp\big((\bm{u}+\bm{v})\cdot \bm{i'}\big)}
\label{equ:dynamic_softmax}
\end{equation}
where $I$ is the set of all items.
Again, we employ the negative sampling strategy to approximate the log likelihood of observed $(u,v,i)$ triples as
\begin{equation}
\log P(i|u,\bm{v}) = \log \sigma\big((\bm{u}+\bm{v})\cdot \bm{i}\big) + k\cdot \mathbb{E}_{i'\sim P_i}[\log\sigma\big(-(\bm{u}+\bm{v})\cdot \bm{i'}\big)]
\label{equ:dynamic_loss}
\end{equation}
where $P_i$ is an uniform distribution for $i \in I$.
Let $D_{(u,v,i)}$ be the set of observed $(u,v,i)$ triples in the training data, then the final optimization goal of DREM is to maximize the log likelihood of $D_{(u,v,i)}$ and $S_{(x,r,y)}$ as
\begin{equation}
\begin{split}
\mathcal{L} = \!\!\!\!\!\!\!\!\sum_{(u,v,i) \in D_{(u,v,i)}}& \!\!\!\!\!\!\!\!\!\! \log P(i|u,v) ~~~~+  \!\!\!\!\!\!\!\!\sum_{(x,r,y) \in S_{(x,r,y)}} \!\!\!\!\!\!\!\!\!\! \log P(y|x,r)\\
=\!\!\!\!\!\!\!\!\sum_{(u,v,i) \in D_{(u,v,i)}}&\!\!\!\!\!\!\!\!\!\!\log \sigma\big((\bm{u}+\!\bm{v})\!\cdot \!\bm{i}\big) + k\!\cdot\! \mathbb{E}_{i'\sim P_i}[\log\sigma\big(\!\!-\!(\bm{u}+\!\bm{v})\!\cdot\! \bm{i'}\big)] \\
+\!\!\!\!\!\!\!\!\!\!\!\!\sum_{(x,r,y) \in S_{(x,r,y)}}& \!\!\!\!\!\!\!\!\!\!\log \sigma\big((\bm{x}+\bm{r})\!\cdot\! \bm{y}\big) + k\!\cdot\! \mathbb{E}_{y'\sim P_r}[\log\sigma\big(\!\!-\!(\bm{x}+\bm{r})\!\cdot\! \bm{y'}\big)]\\
=\!\!\!\!\!\!\!\!\sum_{(u,q,i) \in D_{(u,q,i)}}&\!\!\!\!\!\!\!\!\!\!\log \sigma\big((\bm{u}+\!f(q))\!\cdot \!\bm{i}\big) + k\!\cdot\! \mathbb{E}_{i'\sim P_i}[\log\sigma\big(\!\!-\!(\bm{u}+\!f(q))\!\cdot\! \bm{i'}\big)] \\
+\!\!\!\!\!\!\!\!\!\!\!\!\sum_{(x,r,y) \in S_{(x,r,y)}}& \!\!\!\!\!\!\!\!\!\!\log \sigma\big((\bm{x}+\bm{r})\!\cdot\! \bm{y}\big) + k\!\cdot\! \mathbb{E}_{y'\sim P_r}[\log\sigma\big(\!\!-\!(\bm{x}+\bm{r})\!\cdot\! \bm{y'}\big)]\\
\end{split}
\label{equ:aggregated_loss}
\end{equation}
where \textit{Search\&Purchase} $(u,q,i)$ is the only dynamic relation in $D_{(u,v,i)}$. 
In DREM, $\bm{u}$, $\bm{i}$ and embeddings for all other entities are jointly learned with the parameters $W$ and $b$ in Equation (\ref{equ:fs}).
To conduct product search for a specific user $u$ with query $q$, we simply rank products $i\in I$ with their estimated purchase probability $P(i|u,\bm{v})$.

Empirically, the weight of static and dynamic relationships do not need to be equal in the model optimization.
To explicitly control their relative importance in the final entity representations, we add a hyper-parameter $\lambda$ in Equation~(\ref{equ:aggregated_loss}) as 
\begin{equation}
\begin{split}
\mathcal{L} =& \lambda\!\!\!\!\!\!\!\!\sum_{(u,q,i) \in D_{(u,q,i)}}\!\!\!\!\!\!\!\!\!\!\log \sigma\big((\bm{u}+\!f(q))\!\cdot \!\bm{i}\big) + k\!\cdot\! \mathbb{E}_{i'\sim P_i}[\log\sigma\big(\!\!-\!(\bm{u}+\!f(q))\!\cdot\! \bm{i'}\big)] \\
&+(1-\lambda)\!\!\!\!\!\!\!\!\!\!\!\!\sum_{(x,r,y) \in S_{(x,r,y)}} \!\!\!\!\!\!\!\!\!\!\log \sigma\big((\bm{x}+\bm{r})\!\cdot\! \bm{y}\big) + k\!\cdot\! \mathbb{E}_{y'\sim P_r}[\log\sigma\big(\!\!-\!(\bm{x}+\bm{r})\!\cdot\! \bm{y'}\big)]\\
\end{split}
\label{equ:final_loss}
\end{equation}
For simplicity, we assign equal weights for all relationships in most cases ($\lambda = 0.5$), but we discuss the results of DREM with respect to different $\lambda$ in Section~\ref{sec:parameter_sensitivity}.
Also, in this paper, we assume that all users and items have appeared in $D_{(u,v,i)}$ or $S_{(x,r,y)}$ at least once. 
We leave the exploration of cold-start product search for future studies.


\subsection{Time Complexity}

The construction of DREM includes two phases: the offline training of entity/relation embeddings and the online testing on unobserved user-query pairs.
The time complexity in the training phase mainly depends on the dimensionality of embedding vectors and the size of training data.
For each static relationship triple, the computation of local loss  (Equation~(\ref{equ:static_negative_sample})) is $\mathcal{O}(k\alpha)$, where $k$ is the number of negative samples, and $\alpha$ is the size of each embedding vector.
For each dynamic relationship triple, the computation of the relation embedding $\bm{v}$ (Equation~(\ref{equ:fs})) is $\mathcal{O}\big((|q| + \alpha)\alpha\big)$, and the computation of local loss (Equation~(\ref{equ:dynamic_loss})) is $\mathcal{O}\big((|q| + \alpha + k)\alpha\big)$.  
Thus, the time complexity of training DREM in one epoch is $\mathcal{O}\big((\mathbb{E}_{q}[|q|] + \alpha+k)\alpha|D_{(u,v,i)}|+k\alpha|S_{(x,r,y)}|\big)$, where $\mathbb{E}_{q}[|q|]$ is the average number of words in each query, and $|D_{(u,v,i)}|$ and $|S_{(x,r,y)}|$ is the number of observed dynamic and static relation triples, respectively.
Because $k$ and $\alpha$ are hyper-parameters, the computation cost of DREM is linear to the size of the training data, which is considered to be efficient in general.

For online testing, each item must be assigned with a score to generate the ranked list for a given user-query pair. 
As discussed in Section~\ref{sec:dynamic_relation}, we rank items according to the estimated purchase probability $P(i|u,\bm{v})$ in Equation~(\ref{equ:dynamic_softmax}).
Because $\exp(x)$ is a monotone increasing function and the denominator of the softmax function is equal for all items, we can directly rank items based on the dot product between $\bm{i}$ and $\bm{u} + \bm{v}$, which has $\mathcal{O}\big((|q| + \alpha)\alpha\big)$ complexity.
Because we only need to compute $\bm{v}$ once per query, the computation cost for each testing user-query pair is $\mathcal{O}\big((|q| + \alpha+|I|)\alpha\big)$, where $|I|$ is the total number of items in the product collection.
Since $|I|$ is much larger than $|q| + \alpha$, the overall complexity is approximately $\mathcal{O}(|I|\alpha)$.
To further improve the efficiency, one can reduce $|I|$ by adding additional retrieval phases to filter out irrelevant documents before applying DREM.
We leave these for future studies.




%% file: explanation.tex

\section{Explanation Extraction}\label{sec:explanation}

An important advantage of DREM is its support for explainable product search. 
With the knowledge graph, we can directly infer entity relationships and provide explanations of why retrieved items should be interesting to the users.
In this section, we discuss how to construct explanation paths in DREM and extract possible explanations for search results in product search.


\subsection{Explanation Path}

\begin{figure}[t]
	\centering
	\includegraphics[width=3.5in]{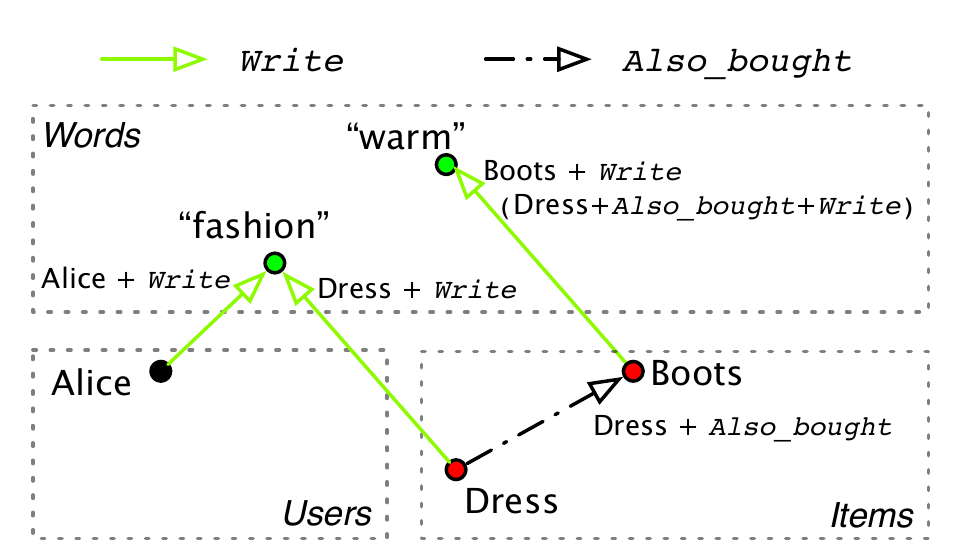}
	\caption{An example explanation path from user \textit{Alice} to item \textit{Dress} through word ``fashion'' in DREM. }
	\label{fig:explanation_path}
	\vspace{5pt}
\end{figure}

We formulate the problem of explaining why item $i$ is retrieved for user $u$ as finding an explanation path between $i$ and $u$ in the knowledge graph.
Figure~\ref{fig:explanation_path} shows an example search session where we retrieve a dress for user \textit{Alice}.
As shown in the figure, both the dress and \textit{Alice} are linked with the word ``fashion'' by the relationship \textit{Write} in the knowledge graph.
Based on this observation, we can say that ``we retrieve this dress for \textit{Alice} because she often writes about \textit{fashion} in her reviews and \textit{fashion} is frequently used to describe the dress by other users".

Formally, let $\Omega_x^r$ and $\Omega_y^r$ be the subspaces of $\Omega$ for the head and tail entities of relationship $r$, respectively. 
Then we define an explanation path from $u$ to $i$ as two lists of relationships $\{r_u^k\}$ (size $n$) and $\{r_i^j\}$ (size $m$) that connects $u$ and $i$:
\begin{equation}
\begin{split}
\bm{u} + \sum_{k=1}^{n}\bm{r_u^k} &= \bm{e} = \bm{i} + \sum_{j=1}^{m}\bm{r_i^j} 
\end{split}
\label{equ:explanation_path}
\end{equation} 
where the head entity space of $r_u^{1}$ is same to the entity space of user $u$ ($\Omega_x^{r_u^{1}} \!\!= \Omega_u$), the head entity space of $r_i^{1}$ is same to the entity space of item $i$ ($\Omega_x^{r_i^{1}} \!\!= \Omega_i$), the tail entity space of $r_u^{n}$ is same to the tail entity space of $r_i^{m}$ ($\Omega_y^{r_u^{n}} \!\!= \Omega_y^{r_i^{m}}$), and $\Omega_y^{r_u^{k-1}} \!\!\!\!= \Omega_x^{r_u^{k}}$ ($k\in[2,n]$), $\Omega_y^{r_i^{j-1}} \!\!\!\!= \Omega_x^{r_i^{j}}$ ($j\in[2,m]$). 
The relationship $r_u^k$ and $r_i^j$ can either be an identity relationship $\Phi$ (which projects an entity to itself) or any relationship in the observed data $S_{(x,r,y)}$ and $ D_{(u,v,i)}$. 
Here, $e$ is an entity in $\Omega_y^{r_i^{m}}$ ($\Omega_y^{r_u^{n}}$) that links $u$ and $i$ with $\{r_u^k\}$ and $\{r_i^j\}$.
Given this explanation path, we can generate a search explanation as ``we retrieve item $i$ for user $u$ because $u$ has relationships $\{r_u^k\}$ with $e$, and $i$ is also linked with $e$ through $\{r_i^j\}$''.
Therefore, the key of explainable product search is the finding of $\{r_u^k\}$ and $\{r_i^j\}$ given the user $u$ and the retrieved item $i$.

\subsection{Extraction Algorithm}

Finding an explanation path, however, is difficult for an arbitrary $(u,i)$ pair.
Because we only observe a limited number of relationship triples in the training data, the knowledge graph built on product data usually is sparse~\cite{zhang2016collaborative,ai2017learning,zhang2017joint} .
In most cases, it is impossible to find two sets of relationships $\{r_u^k\}$ and $\{r_i^j\}$ that directly link the user $u$ to the item $i$.
To tackle this problem, we propose a Soft Matching Algorithm (SMA) to extract explanation paths in DREM.

Let $\Omega_e$ be a subspace of $\Omega$ that contains all entities with the type of $e$, and $e_u$, $e_i$ be the projections of $u$ and $i$ in $\Omega_e$ given particular relation paths, then we define the soft matching score for $u$ and $i$ through $e\in \Omega_e$ as
\begin{equation}
S(e | u, i) = \log \big(P(e|e_u)P(e|e_i)\big) = \log P(e|e_u) + \log P(e|e_i)
\label{equ:soft_matching}
\end{equation}
where $P(e|e_u)$ and $P(e|e_i)$ are the probability of observing $e$ given $e_u$ and $e_i$. 
Intuitively, $P(e|e_u)$ and $P(e|e_i)$ can be model with any functions that measure the similarity between $e$, $e_u$, and $e_i$. 
In DREM, a straightforward method to compute $P(e|e_u)$ and $P(e|e_i)$ is to adopt the embedding-based generative framework as described in Equation~(\ref{equ:softmax}).  
This, however, ignores the length of the path from $u$ to $i$, and could potentially favor long and less meaningful search explanations in practice.
To explicitly encourage short explanation paths, we add a decay factor $\beta$ and define $P(e|e_u)$ and $P(e|e_i)$ as 
\begin{equation}
\begin{split}
P(e|e_u) \!\!=\!\! \frac{\exp(\bm{e_u}\!\cdot\! \bm{e} - \beta n)}{\sum_{e'\in \Omega_e}\!\!\exp(\bm{e_u}\!\cdot\! \bm{e'})}, P(e|e_i) \!\!=\!\! \frac{\exp(\bm{e_i}\!\cdot\! \bm{e} - \beta m)}{\sum_{e'\in \Omega_e}\!\!\exp(\bm{e_i}\!\cdot\! \bm{e'})} 
\end{split}
\label{equ:decay}
\end{equation}
where $\beta$ is a hyper-parameter that controls the effect of probability decay, and $n$, $m$ are the length of path $p_u = \{r_u^k\}$, $p_i = \{r_i^j\}$ that translate $u$, $i$ to $e_u$, $e_i$.
In this work, we set $\beta$ as 1.

\begin{algorithm}[t]
	\caption{Soft Matching Algorithm (SMA)}
	\SetAlgoLined
	\label{alg:SMA}
	\SetKwFunction{algo}{Main} \SetKwFunction{proc}{Dijkstra}
	\KwIn{$u$, $i$, $\beta$, $G = \{\Omega_e, r\}$}
	\KwOut{$path\_set, score\_set$}
	\SetKwProg{myalg}{Procedure}{}{}
	\myalg{\algo{}}{
	\nl Initialize $p_u \leftarrow \{\}, p_i \leftarrow \{\}, path\_set\leftarrow \{\}, score\_set \leftarrow \{\}$ \\
	\nl \For{$\Omega_e \in G$}{
		\nl $p_u[e] = Dijkstra(\Omega_u,\Omega_e, G)$\\
		\nl $p_i[e] = Dijkstra(\Omega_i,\Omega_e, G)$ 
	}
	\nl \For{$\Omega_e \in G$}{
		\nl \For{$e \in \Omega_e$}{
			\nl $score\_set[e] = S(e|u,i)$	// Equation~(\ref{equ:soft_matching}) and (\ref{equ:decay}). \\
			\nl $path\_set[e] = \{p_u[e], e, p_i[e]\}$. 
		}
	}
	\nl \Return $path\_set, score\_set$
	}
	\SetKwProg{myproc}{Function}{}{}
	\myproc{\proc{$\Omega_x,\Omega_y, G$}}{
		\nl $p_{xy} \leftarrow$ the shortest path from $\Omega_x$ to $\Omega_y$ in $G$; \\
		\nl \KwRet $p_{xy}$;}
\end{algorithm}
A summary of SMA for explanation extraction is shown in Algorithm~\ref{alg:SMA}.
Let $G = \{\Omega_e, r\}$ be a graph where each node $\Omega_e$ denotes a subspace of entity type $e$, and each edge $r$ denotes a relationship that connects two nodes with weight 1. 
First, given an arbitrary pair $(u,i)$, we find the shortest path from $\Omega_u$ to $\Omega_e$ and $\Omega_i$ to $\Omega_e$ as $p_u[e]$ and $p_i[e]$ with the Dijkstra algorithm~\cite{dijkstra1959note}.
Second, we compute their translations $e_u$, $e_i$ in $\Omega_e$ with $p_u[e]$, $p_i[e]$ and their soft matching scores with $e \in \Omega_e$ using Equation~(\ref{equ:soft_matching}) and (\ref{equ:decay}). 
Finally, we sort entity $e$ from each $\Omega_e \subset \Omega$ with their matching probabilities in descending order, and select the best path $\{p_u[e], e, p_i[e]\}$ to generate search explanations.
We manually ignore the path that only contains \textit{Search\&Purchase} because it does not provide any information for search explanation.


\revised{
Because the time complexity of the Dijkstra algorithm~\cite{dijkstra1959note} is $\mathcal{O}(|R| + |E|\log |E|)$ where $|E|$ is the number of entities (or nodes) in the knowledge graph and $|R|$ is the number of relations (or edges) between the entities, the computational cost of SMA could be large if we want to explore all possible paths in the explanation generation process. 
Thanks to the decay factor in Equation~(\ref{equ:decay}), however, we could ignore the paths with more than certain number of hoops (e.g., 4 in our experiments) and limit the Dijkstra algorithm~\cite{dijkstra1959note} to a relatively small graph for efficiency. 
Also, because search explanations can be extracted separately with the retrieval process, one could apply an asynchronous web service that shows the search results first while waiting for the generation of explanations to avoid hurting user experience.
We leave the exploration of how to efficiently generate search result explanations for future studies.
}

%% file: experiment_setup.tex

\section{Experimental Setup}\label{sec:setup}

In this section, we describe the details of our experiment settings.
We conduct experiments with Amazon product datasets and compare our method with state-of-the-art product search systems including both text-based models~\cite{ponte1998language} and latent space models~\cite{van2016learning,ai2017learning}.

\subsection{Datasets}\label{sec:datasets}

The Amazon product dataset\footnote{\url{http://jmcauley.ucsd.edu/data/amazon/}} is a well-known benchmark for product search and recommendation~\cite{van2016learning,ai2017learning,zhang2017joint}.
It contains information for millions of customers, products and associated metadata including descriptions, reviews, brands, and categories.
In our experiments, we used four subsets of the Amazon product data, which are \textit{Electronics}, \textit{Kinde Store}, \textit{CDs \& Vinyl}, and \textit{Cell Phones \& Accessories}.
We use the 5-core data provided by McAuley et al.~\cite{mcauley2015inferring} where each user and product has at least 5 purchases and 5 reviews.


\textbf{Query Extraction}.
To the best of our knowledge, no large-scale query log is available on the Amazon dataset.
A common paradigm used by previous studies is to extract queries from the category information of each product.
Similar to Van Gysel et al.~\cite{van2016learning}, we adopt a two-step process to extract search queries for each user.
First, given an arbitrary user and his purchase history, we extract the hierarchical category information of each item with more than two levels.
Second, we remove duplicated words and stopwords from a single hierarchy of categories and concatenate the terms to form a topic string.
The topic string is then treated as a query submitted by the user which leads to a purchase of the corresponding item.
Because users often search for ``a producer's name, a brand or a set of terms which describe the category of the product" in e-shopping~\cite{rowley2000product}, queries extracted with this paradigm are usually sufficient to simulate real-world product search queries~\cite{van2016learning,ai2017learning}. 

\textbf{Entities and Relationships}.
In this work, we consider five types of entities and their relationships in product search.
The entities we used are \textit{user}, \textit{item}, \textit{word}, \textit{brand} and \textit{category}.
We ignore words that have appeared for less than 5 times in the corresponding corpus.
Also, we split hierarchical category information of each product into multiple distinct categories and replace each category as a unique symbol in the training data.
For example, a two-level category hierarchy \textit{Camera, Photo $\rightarrow$ Digital Camera Lences} will be considered as two separate entities and anonymized as foo$_1$ and foo$_2$. 
An item that belongs to this category hierarchy will be connected to both foo$_1$ and foo$_2$.

The relationships used in our experiments include
\begin{itemize}
\item \textit{Write} : Word $w$ was written by user $u$ in their reviews ($u\rightarrow w$) or written for item $i$ in the item's reviews ($i\rightarrow w$). 
\item \textit{Also\_bought}: Users who purchased item $i_1$ previously also purchased item $i_2$ ($i_1\rightarrow i_2$). 
\item \textit{Also\_viewed}: Users who viewed item $i_1$ previously also viewed item $i_2$ ($i_1\rightarrow i_2$). 
\item \textit{Bought\_together}: Item $i_1$ was purchased with item $i_2$ in a single transaction ($i_1\rightarrow i_2$). 
\item \textit{Is\_brand}: Item $i$ belongs to brand $b$ ($i\rightarrow b$). 
\item \textit{Is\_category}: Item $i$ belongs to category $c$ ($i\rightarrow c$). 
\end{itemize}
The statistics of entities and relationships in the Amazon product datasets are summarized in Table~\ref{tab:dataset_statistics}. 
Similar to previous studies~\cite{zhang2016collaborative,ai2017learning,zhang2017joint}, the observed relation triples in our data are highly sparse.

\begin{table*}[t]
	\centering
	\caption{Statistics for the 5-core datasets for \textit{Electronics}, \textit{Kindle Store}, \textit{CDs \& Vinyl} and \textit{Cell Phones \& Accessories}~\cite{mcauley2015inferring}. 
	}
	\scalebox{0.80}{
		\begin{tabular}{ l  c  c  c  c   } 
			\toprule
			& \textit{Electronics}& \textit{Kindle Store}& \textit{CDs \& Vinyl} & \textit{Cell Phones \& Accessories}\\
			\midrule
			\textbf{Corpus}\\
			~~~~Vocabulary size & 142,922 & 95,729 & 202,959 & 22,493\\ %
			~~~~Number of reviews & 1,689,188 & 982,618 & 1,097,591 & 194,439\\ %
			~~~~Number of users & 192,403 & 68,223 & 75,258 & 27,879\\ %
			~~~~Number of items & 63,001 & 61,934 & 64,443 & 10,429\\ %
			~~~~Number of brands & 3,525 & 1 & 1,414 & 955\\ %
			~~~~Number of categories & 983 & 2,523 & 770 & 206\\ %
			
			\midrule
			\textbf{Relationships}\\
			~~~~\textit{Write} per user & 777.23$_{\pm 1748.6}$ & 1174.23$_{\pm 3682.39}$ & 1846.88$_{\pm 7667.51}$ & 500.01$_{\pm 979.78}$\\ %
			~~~~\textit{Write} per item & 2373.62$_{\pm 5860.33}$ & 1293.47$_{\pm 1916.72}$ & 2156.83$_{\pm 4024.15}$ & 1336.64$_{\pm 2698.30}$\\ %
			~~~~\textit{Also\_bought} per item & 36.70$_{\pm 38.56}$ & 82.56$_{\pm 29.92}$ & 57.28$_{\pm 39.22}$ & 56.53$_{\pm 35.82}$\\ %
			~~~~\textit{Also\_viewed} per item & 4.36$_{\pm 9.44}$ & 0.16$_{\pm 1.66}$ & 0.27$_{\pm 1.86}$ & 1.24$_{\pm 4.29}$ \\ %
			~~~~\textit{Bought\_together} per item & 0.59$_{\pm 0.72}$ & 0.00$_{\pm 0.04}$ & 0.68$_{\pm 0.80}$ & 0.81$_{\pm 0.77}$\\ %
			~~~~\textit{Is\_brand} per item & 0.47$_{\pm 0.50}$ & 0.00$_{\pm 0.00}$ & 0.21$_{\pm 0.41}$ & 0.52$_{\pm 0.50}$\\ %
			~~~~\textit{Is\_category} per item & 4.39$_{\pm 0.95}$ & 9.85$_{\pm 2.61}$ & 7.25$_{\pm 3.13}$ & 3.49$_{\pm 1.08}$\\ %
			
			\midrule
			\textbf{Train/Test}\\
			~~~~Number of reviews & 1,275,432/413,756 & 720,006/262,612 & 804,090/293,501 & 150,048/44,391\\ %
			~~~~Number of queries & 904/85 & 3313/1290 & 534/160 & 134/31\\
			~~~~Number of user-query pairs & 1,204,928/5,505 & 1,490,349/232,668 & 1,287,214/45,490 & 114,177/665\\ 
			~~~~Relevant items per pair & 1.12$_{\pm 0.48}$/1.01$_{\pm 0.09}$ & 1.87$_{\pm 3.30}$/1.48$_{\pm 1.94}$ & 2.57$_{\pm 6.59}$/1.30$_{\pm 1.19}$ & 1.52$_{\pm 1.13}$/1.00$_{\pm 0.05}$\\ 
			\bottomrule
		\end{tabular}
	}
	\label{tab:dataset_statistics}
\end{table*}

\subsection{Baselines}
To demonstrate the effectiveness of the DREM as a product search model, 
we incorporate five baselines in our experiments: the language modeling approach for IR~\cite{ponte1998language}, a probabilistic retrieval model (BM25)~\cite{robertson1994some}, a ensemble learning-to-rank model (LambdaMART)~\cite{wu2017ensemble}, the Latent Semantic Entity model~\cite{van2016learning}, and the Hierarchical Embedding Model~\cite{ai2017learning}.

\subsubsection{QL}: 
The language modeling approach for IR, which is often referred to as the Query Likelihood model (QL), is first proposed by Ponte and Croft~\cite{ponte1998language}.
It is a unigram model that ranks documents based on the posterior probability of observing the query words given a document's language model estimated with maximum likelihood estimation. 
In this paper, we concatenate the title, description and reviews of an item in the training data to form a document for it, and compute its ranking scores with respect to a query $q$ based on the language modeling approach with Dirichlet smoothing~\cite{zhai2001study} as:
$$
\log(P(q|d)) = \sum_{w\in q} \#(w,q)\log\frac{\#(w,d) + \mu \frac{\#(w,C)}{|C|}}{|d| + \mu }
$$
where $\#(w,q)$, $\#(w,d)$, and $\#(w,C)$ are the frequencies of word $w$ in the query $q$, document $d$, and the corpus $C$, respectively; $|d|$ and $|C|$ are the lengths of $d$ and $C$; and $\mu$ is a hyper-parameter that controls the effect of Dirichlet smoothing. 

\subsubsection{BM25}
Built on the bag-of-words representations of queries and documents, BM25 is a classic probabilistic retrieval model  proposed by Robertson and Walker.~\cite{robertson1994some}.
It assumes a 2-Poisson distribution for observed words in the corpus, and ranks documents with a statistical scoring function as
$$
BM25(q,d) = \sum_{w\in q}IDF(w, C) \cdot \frac{\#(w,q)\cdot (k_1+1)}{\#(w,q) + k_1 \cdot (1-b+b\cdot \frac{|d|}{avg(|d|)})}
$$ 
where $IDF(w, C)$ is the inverse document frequency of word $w$ in the corpus $C$, $avg(|d|)$ is the average document length, and $k_1$ and $b$ are two hyper-parameters for the ranking function.
Similar to QL, we concatenate the title, description, and reviews in the training data to form a document for each product.

\begin{table}
	\caption{A summary of the ranking features extracted for constructing a learning-to-rank model on the Amazon product search dataset.}
	\begin{tabular}
		{ l | p{0.75\textwidth}} \hline 
		\hline
		\multicolumn{2}{c}{Global Statistic Features} \\ \hline 
		\hline
		Length & The length of product title, descriptions, reviews.   \\\hline
		Purchase & The total number of purchases on each item in the training set. \\ \hline
		Distinct Purchase & The distinct number of users who have purchased a certain item in the training set. \\ \hline
		\hline
		\multicolumn{2}{c}{Query-item Features} \\ \hline 
		\hline
		TF & The average term frequency of query terms in product title, descriptions, reviews, and the whole document (title$+$description$+$reviews).   \\\hline
		IDF & The average inverse document frequency of query terms in product title, descriptions, reviews, and the whole document (title$+$description$+$reviews).   \\\hline
		TF-IDF & The average value of $tf\cdot idf$ of query terms in product title, descriptions, reviews, and the whole document (title$+$description$+$reviews).   \\\hline
		BM25 & The scores of BM25~\cite{robertson1994some} on product title, descriptions, reviews, and the whole document (title$+$description$+$reviews).   \\\hline
		LMABS & The scores of Language Model (LM)~\cite{ponte1998language} with absolute discounting~\cite{zhai2004study} on product titles, descriptions, reviews, and the whole document (title$+$description$+$reviews).   \\\hline
		LMDIR (QL) & The scores of LM with Dirichlet smoothing~\cite{zhai2004study} (which is same with QL) on product titles, descriptions, reviews, and the whole document (title$+$description$+$reviews).   \\\hline
		LMJM & The scores of LM with Jelinek-Mercer~\cite{zhai2004study} on product titles, descriptions, reviews, and the whole document (title$+$description$+$reviews).   \\\hline
		
	\end{tabular}\label{tab:ltr_features}
\end{table}

\subsubsection{LambdaMART}:
As a representative study on applying learning-to-rank techniques to product search, Wu et al.~\cite{wu2017ensemble} construct a LambdaMART model for product search by manually extracting a variety of ranking features for each item with their text data and user behavior logs.
In this paper, we construct a learning-to-rank baseline with LambdaMART following the same pipeline used by Wu et al.~\cite{wu2017ensemble}.
Due to the limits of Amazon review datasets, we cannot compute certain features such as session features and time features, but we manage to reproduce most global statistic features and query-item features proposed by Wu et al.~\cite{wu2017ensemble}.
Detailed feature descriptions are listed in Table~\ref{tab:ltr_features}.    

\subsubsection{LSE}: The Latent Semantic Entity model (LSE) is the first latent space model proposed for product search by Van Gysel et al.~\cite{van2016learning}.
It encodes queries and n-grams with a non-linear projection function similar to Equation~(\ref{equ:fs}).
It also learns the embedding representations of items by maximizing the similarity between an item and the encoded n-grams extracted from the corresponding item reviews.
Specifically, for each n-gram $s$ in the product review of an item $i$, the similarity between $s$ and $i$ in LSE is computed as
$$
P(i|s) = \sigma(\bm{i}\cdot f(s))
$$
where $\bm{i}$ is the representation of $i$ in the latent space, $f(x)$ is the non-linear projection function in Equation~(\ref{equ:fs}), and $\sigma (x)$ is a sigmoid function $\sigma (x) = \frac{1}{1 + e^{-x}}$. 
Products are retrieved based on their similarity with the query in the latent space.

\subsubsection{HEM}: 
The Hierarchical Embedding Model (HEM) proposed by Ai et al.~\cite{ai2017learning} is a state-of-the-art retrieval model for personalized product search. 
It is constructed based on a generative framework which assumes that reviews are generated by the language model of users/items and purchases are generated by the joint model of users and queries.
Similar to DREM, HEM learns the embeddings of users, items, and queries by maximizing the likelihood of observed review data, and ranks items based on their posterior probability given the user and the query.
However, HEM only considers the information of users, items, and product reviews, and does not differentiate the relations between different types of entities in the optimization process.

\subsection{Evaluation Methodology}
To train and test different product search models, we partitioned each dataset into a training set and a test set.
Following the methodology used by Ai et al.~\cite{ai2017learning}, we randomly hide 30\% of the user reviews from the training data and use their corresponding purchase information as the ground truth for testing.
We randomly select 30\% queries as test queries, and if all queries for an item were selected as test queries, we randomly pick one from the test query set and put it back to the training data.
After that, we match the queries with user-item pairs in the test set to construct the final test data.
An item is relevant to a user-query pair if and only if it is relevant to the query and has been purchased by the user.
In this setting, all query-user-item triples in the test set are unobserved in the training process.
More statistics about our data partitions are shown in Table~\ref{tab:dataset_statistics}.

To evaluate retrieval performance in our experiments, we adopt three metrics, which include the mean average precision (MAP), the mean reciprocal rank (MRR) and the normalized discounted cumulative gain (NDCG). 
For each user-query pair, we only retrieve 100 items to generate the rank list.
Both MAP and MRR are computed based on the whole rank list, while NDCG is computed only based on the top 10 items.  
Significant differences are measured by the Fisher randomization test~\cite{smucker2007comparison} with $p<0.01$.


\subsection{Implementation Details}

For QL and BM25, we used galago\footnote{\url{https://sourceforge.net/p/lemur/wiki/Galago/}} to index and retrieve items.
For LambdaMART, we manually extract features from raw data and build the model with an open-source learning-to-rank package  ranklib\footnote{\url{https://sourceforge.net/p/lemur/wiki/RankLib/}}. 
And for LSE and HEM, we used the implementation provided by Ai et al.~\cite{ai2017learning}\footnote{\url{https://github.com/QingyaoAi/Hierarchical-Embedding-Model-for-Personalized-Product-Search}}.
QL, BM25, LSE, and HEM conduct product search based on the text information of items, which is the same as DREM built with the \textit{Write} relationship only.
To further analyze the usefulness of other relationships, we tested DREM built on \textit{Write} together with other relationships.
We refer to DREM with \textit{Also\_bought}, \textit{Also\_viewed}, \textit{Bought\_together}, \textit{Is\_brand} and \textit{Is\_category} as DREM$_{AB}$, DREM$_{AV}$, DREM$_{BT}$, DREM$_{Bnd}$ and DREM$_{Cat}$, respectively.
DREM with \textit{Write} only and the DREM with all relationships are referred to as DREM$_{NoMeta}$ and DREM$_{All}$.

The latent space models (LSE, HEM, and DREM) are trained with stochastic gradient descent with batch size 64.
We manually clip the norm of batch gradients with 5 to avoid unstable parameter updates.
We train each model with 20 epochs and gradually decrease the learning rate from 0.5 to 0 in the training process.
For baselines, we tuned the Dirichlet smoothing parameter $\mu$ of QL from 1000 to 3000, and the BM25 scoring parameter $k_1$ and $b$ from 0.5 to 4 and 0.25 to 1, respectively.
The number of trees and leaf nodes in LambdaMART are set as 1000 and 10, respectively, and we tuned the personalization weight $\eta$ of HEM from 0 to 1.
We also tuned the embedding size $\alpha$ for LSE, HEM, and DREM from 100 to 500.
In order to better illustrate the importance of different product relationships in different datasets, we fix the dynamic relation weight $\lambda$ in Equation~(\ref{equ:final_loss}) as 0.5 for most experiments, but we will discuss its effect in Section~\ref{sec:parameter_sensitivity}.
We will release our data and code upon the publication of this manuscript.

%% file: experiment_results.tex

\begin{table*}[t]
	\centering
	\caption{Comparison of baselines and DREM on the Amazon product search datasets. 
		$*$, $+$ and $\dagger$ denote significant differences to all baselines (QL, BM25, LambdaMART, LSE, and HEM), 	DREM$_{NoMeta}$, and all tested models, respectively, in Fisher randomization test~\cite{smucker2007comparison} with $p \leq 0.01$. The best performance is highlighted in boldface.}
	\begin{tabular}{ c || l | l | l || l | l | l  } 
		\hline
		\multicolumn{1}{c||}{ } & \multicolumn{3}{c||}{\textit{Electronics}} & \multicolumn{3}{c}{\textit{Kindle Store}} \\ \hline 
		Model & MAP & MRR & NDCG & MAP & MRR & NDCG  \\\hline
		\hline
		QL & 0.289 & 0.289 & 0.316 & 0.011 & 0.012 & 0.013 \\ \hline
		BM25 & 0.283 & 0.280 & 0.304 & 0.021 & 0.013 & 0.014 \\ \hline
		LambdaMART & 0.180 & 0.181 & 0.237 & 0.028 & 0.029 & 0.018 \\ \hline
		LSE & 0.233 & 0.234 & 0.239 & 0.006 & 0.007 & 0.007 \\ \hline
		HEM & 0.308$^{*+}$ & 0.309$^{*+}$ & 0.329$^{*+}$ & 0.029 & 0.035$^{*}$ & 0.033$^{*}$ \\ \hline 
		\hline
		DREM$_{NoMeta}$ & 0.291 & 0.291 & 0.319 & 0.036$^{*}$ & 0.044$^{*}$ & 0.042$^{*}$ \\ \hline
		DREM$_{AB}$ & 0.283 & 0.283 & 0.312 & 0.043$^{*+}$ & 0.052$^{*+}$ & 0.050$^{*+}$ \\ \hline
		DREM$_{AV}$ & 0.318$^{*+}$ & 0.319$^{*+}$ & 0.349$^{*+}$ & 0.035$^{*}$ & 0.043$^{*}$ & 0.041$^{*}$ \\ \hline
		DREM$_{BT}$ & 0.320$^{*+}$ & 0.321$^{*+}$ & 0.346$^{*+}$ & 0.037$^{*}$ & 0.045$^{*}$ & 0.042$^{*}$ \\ \hline
		DREM$_{Bnd}$ & 0.314$^{*+}$ & 0.315$^{*+}$ & 0.340$^{*+}$ & 0.037$^{*}$ & 0.044$^{*}$ & 0.043$^{*}$ \\ \hline
		DREM$_{Cat}$ & 0.299$^{+}$ & 0.300$^{+}$ & 0.360$^{*+}$ & 0.048$^{*+}$ & 0.056$^{*+}$ & 0.056$^{*+}$ \\ \hline \hline
		DREM$_{All}$ & \textbf{0.366}$^{*+\dagger}$ & \textbf{0.367}$^{*+\dagger}$ & \textbf{0.408}$^{*+\dagger}$ & \textbf{0.057}$^{*+\dagger}$ & \textbf{0.067}$^{*+\dagger}$ & \textbf{0.067}$^{*+\dagger}$ \\ \hline
		\hline
		\multicolumn{1}{c||}{ } & \multicolumn{3}{c||}{\textit{CDs \& Vinyl}} & \multicolumn{3}{c}{\textit{Cell Phones \& Accessories}} \\ \hline
		Model & MAP & MRR & NDCG& MAP & MRR & NDCG \\ \hline
		\hline
		QL & 0.009 & 0.011 & 0.010 & 0.081 & 0.081 & 0.092 \\ \hline
		BM25 & 0.027 & 0.018 & 0.016 & 0.083 & 0.081 & 0.115 \\ \hline
		LambdaMART & 0.054$^{*+}$ & 0.057$^{*+}$ & 0.051$^{*+}$ & 0.121 & 0.121 & 0.148 \\ \hline
		LSE& 0.018 & 0.022 & 0.020 & 0.098 & 0.098 & 0.084 \\ \hline
		HEM & 0.034 & 0.040 & 0.040 & 0.124$^{*+}$ & 0.124$^{*+}$ & 0.153$^{*+}$ \\ \hline
		\hline
		DREM$_{NoMeta}$ & 0.034 & 0.041 & 0.040 & 0.107 & 0.107 & 0.127 \\ \hline
		DREM$_{AB}$ & 0.046$^{+}$ & 0.054$^{+}$ & 0.054$^{+}$ & 0.098 & 0.098 & 0.120 \\ \hline
		DREM$_{AV}$ & 0.034 & 0.041 & 0.040 & 0.095 & 0.096 & 0.096 \\ \hline
		DREM$_{BT}$ & 0.037$^{+}$ & 0.044$^{+}$ & 0.042$^{+}$ & 0.089 & 0.089 & 0.096 \\ \hline
		DREM$_{Bnd}$ & 0.035 & 0.041 & 0.040 & 0.134$^{*+}$ & 0.134$^{*+}$ & 0.152$^{+}$ \\ \hline
		DREM$_{Cat}$ & 0.059$^{*+}$ & 0.068$^{*+}$ & 0.070$^{*+}$ & 0.193$^{*+}$ & 0.193$^{*+}$ & 0.229$^{*+}$ \\ \hline
		\hline
		DREM$_{All}$ & \textbf{0.074}$^{*+\dagger}$ & \textbf{0.084}$^{*+\dagger}$ & \textbf{0.086}$^{*+\dagger}$ & \textbf{0.249}$^{*+\dagger}$ & \textbf{0.249}$^{*+\dagger}$ & \textbf{0.282}$^{*+\dagger}$ \\ \hline
		
	\end{tabular}
	\label{tab:overall_results}
\end{table*}

\section{Results and Discussions}\label{sec:results}

In this section, we report the results of our experiments.
We first present and discuss the retrieval performance of DREM and baseline models.
Then we provide a case study to analyze the effectiveness of DREM for explainable product search.

\subsection{Retrieval Performance}

Table~\ref{tab:overall_results} summarizes the results of our product search experiments on the four subsets of Amazon product data.
We group the models into three groups -- the baseline models (QL, BM25, LambdaMART, LSE, HEM); DREM with \textit{Write} and another relationship among \textit{Also\_bought}, \textit{Also\_viewed}, \textit{Bought\_together}, \textit{Is\_brand} and \textit{Is\_category} (DREM$_{AB}$, DREM$_{AV}$, DREM$_{BT}$, DREM$_{Bnd}$, DREM$_{Cat}$); and the DREM with \textit{Write} only or with all the relationships (DREM$_{NoMeta}$, DREM$_{All}$) 

\subsubsection{Overall Results}
As we can see from the table, the relative performance of bag-of-words models (QL, BM25) and latent space models  without personalization (LSE) varies across different datasets. 
While QL and BM25 have comparable performance on all datasets, LSE outperformed them on \textit{CDs \& Vinyl} but performed worse than them on \textit{Electronics} and \textit{Kindle Store}.
As shown by previous studies~\cite{guo2016deep,guo2016semantic}, the main difference between unigram models and latent space models is their ability to conduct semantic matching.
The latter performs well when vocabulary mismatch between queries and documents is severe, while the former works better in other cases.
Our results indicate that the severity of vocabulary mismatch is low on \textit{Electronics} or \textit{Kindle Store}, but high on \textit{CDs \& Vinyl}. 
By incorporating personalization, HEM consistently outperformed QL and LSE on all the datasets tested in our experiments.
Because purchasing is a highly personalized behavior, incorporating user information can help HEM better understand the search intents of each user and retrieve items that suits different individuals.
The results for LambdaMART are more complicated.
While it achieved superior performance on \textit{CDs \& Vinyl}, it also produced bad results on \textit{Electronics}.  
Further analysis on ranking features are needed in order to understand why learning-to-rank models perform differently on different product categories, which is beyond the scope of this paper and we will leave it for future studies.  

After incorporating other product knowledge information discussed in Section~\ref{sec:datasets}, DREM$_{All}$ significantly outperformed all baseline models on all datasets.
Its obtained 19\%, 97\%, 118\% and 101\% improvements with respect to MAP over HEM on \textit{Electronics}, \textit{Kinde Store}, \textit{CDs \& Vinyl} and \textit{Cell Phones \& Accessories}, respectively. 
This demonstrate the usefulness of multi-relational product data and the effectiveness of DREM as a product retrieval model.

In Table~\ref{tab:overall_results}, the performance of HEM and DREM$_{NoMeta}$ is competitive in most cases.
HEM and DREM$_{NoMeta}$ are both constructed based on users, items and their associated reviews.
The only difference between them is the method they used to model entity relationships.
HEM directly uses user embeddings to predict both review words and purchased items, while DREM$_{NoMeta}$ uses relationship embeddings to project users into the space of words and items separately.
According to our results, the two paradigms are equally effective for product search and neither of them is consistently better than the other.
However, DREM is more powerful in terms of extendability because it creates a knowledge graph that can integrate different kinds of product information for retrieval tasks.


\subsubsection{Usefulness of Different Relationships}
In our experiments, we analyze the importance of different relationships by training DREMs with each of the relationships separately.
As shown in Table~\ref{tab:overall_results}, the importance of relationships varies considerably on different datasets.
On \textit{Electronics}, nearly all types of product knowledge brought benefits to DREM except DREM$_{AB}$, which is built on the \textit{Write} and \textit{Also\_bought} relationships. 
As shown by Ai et al.~\cite{ai2017learning}, the importance of personalization for product search is less significant on \textit{Electronics} than on other datasets.
Two co-purchased items in \textit{Electronics} are less likely to satisfy the same type of user preference or search intent.
For example, users may not intend to buy a keyboard when they search for ``mouse", despite that they often buy keyboards before or after the purchase of a mouse.  
Therefore, the relationship \textit{Also\_brought} introduces less information but more noise for DREM on \textit{Electronics}.

In contrast to \textit{Electronics}, the incorporation of \textit{Also\_bought} significantly improved the retrieval performance of DREM on \textit{Kinde Store} and \textit{CDs \& Vinyl}.
DREM$_{AB}$ outperformed DREM$_{NoMeta}$ by 19\% and 35\% with respect to MAP on \textit{Kinde Store} and \textit{CDs \& Vinyl}, respectively.
This indicates that co-purchased items often fit the same need of users in \textit{Kinde Store} and \textit{CDs \& Vinyl}.
This is reasonable because \textit{Kinde Store} and \textit{CDs \& Vinyl} consist of books and music, on which people usually have consistent tastes.
If a CD is relevant to a query, then other frequently co-purchased CDs are also likely to be relevant. 

As shown in Table~\ref{tab:overall_results}, we observed that \textit{Is\_brand} is more useful for product search on \textit{Cell Phones \& Accessories} than on other datasets.
On \textit{Cell Phones \& Accessories}, DREM$_{Bnd}$ significantly outperformed DREM$_{NoMeta}$ with a 25\% improvement on MAP.
According to a recent report\footnote{\url{https://www.statista.com/statistics/716086/smartphone-brand-loyalty-in-us/}}, most people have high loyalty to the manufacturer of their phones and 56\% of people who currently possess a smartphone used to own a phone from the same manufacturer.
Thus, it is not surprising to see that \textit{Is\_brand} exhibits high correlations with user purchases in \textit{Cell Phones \& Accessories}.



Although we have split each hierarchical category into distinct categories and anonymized them in model construction, there might be a concern that incorporating category entities in DREM may hurt the fairness of the evaluation since the test queries are generated based on the hierarchy of categories.
In our experiment, we indeed observed that DREM with \textit{Is\_category} (DREM$_{Cat}$) performed better than the DREM with other relationships on \textit{Kinde Store}, \textit{CDs \& Vinyl} and \textit{Cell Phones \& Accessories}. 
However, it's worth noticing that DREMs without \textit{Is\_category} also significantly outperformed the state-of-the-art baseline methods. 
In Table~\ref{tab:overall_results}, DREM$_{BT}$ on \textit{Electronics}, DREM$_{AB}$ on \textit{Kinde Store}, DREM$_{AB}$ on \textit{CDs \& Vinyl} and DREM$_{Bnd}$ on \textit{Cell Phones \& Accessories} obtained 4\%, 48\%, 35\% and 8\% improvements on MAP over the best baseline (HEM), respectively.
Again, these results indicate the effectiveness of DREM and the usefulness of multi-relational product data for product search.







\begin{figure*}
	\centering
	\begin{subfigure}{.5\textwidth}
		\centering
		\includegraphics[width=2.7in]{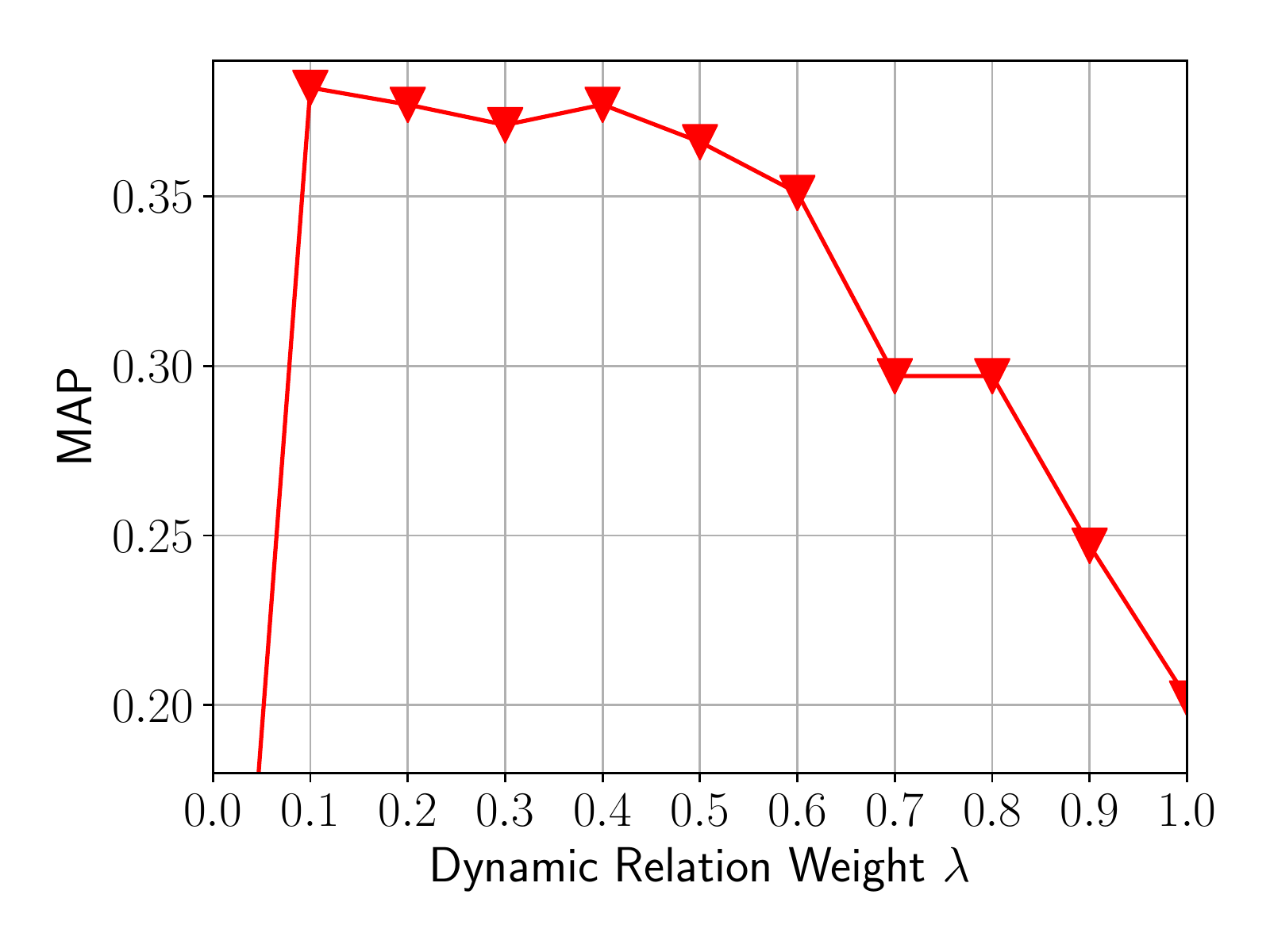}
		\caption{\textit{Electronics}}
		\label{fig:e_dynamic}
	\end{subfigure}%
	\begin{subfigure}{.5\textwidth}
		\centering
		\includegraphics[width=2.7in]{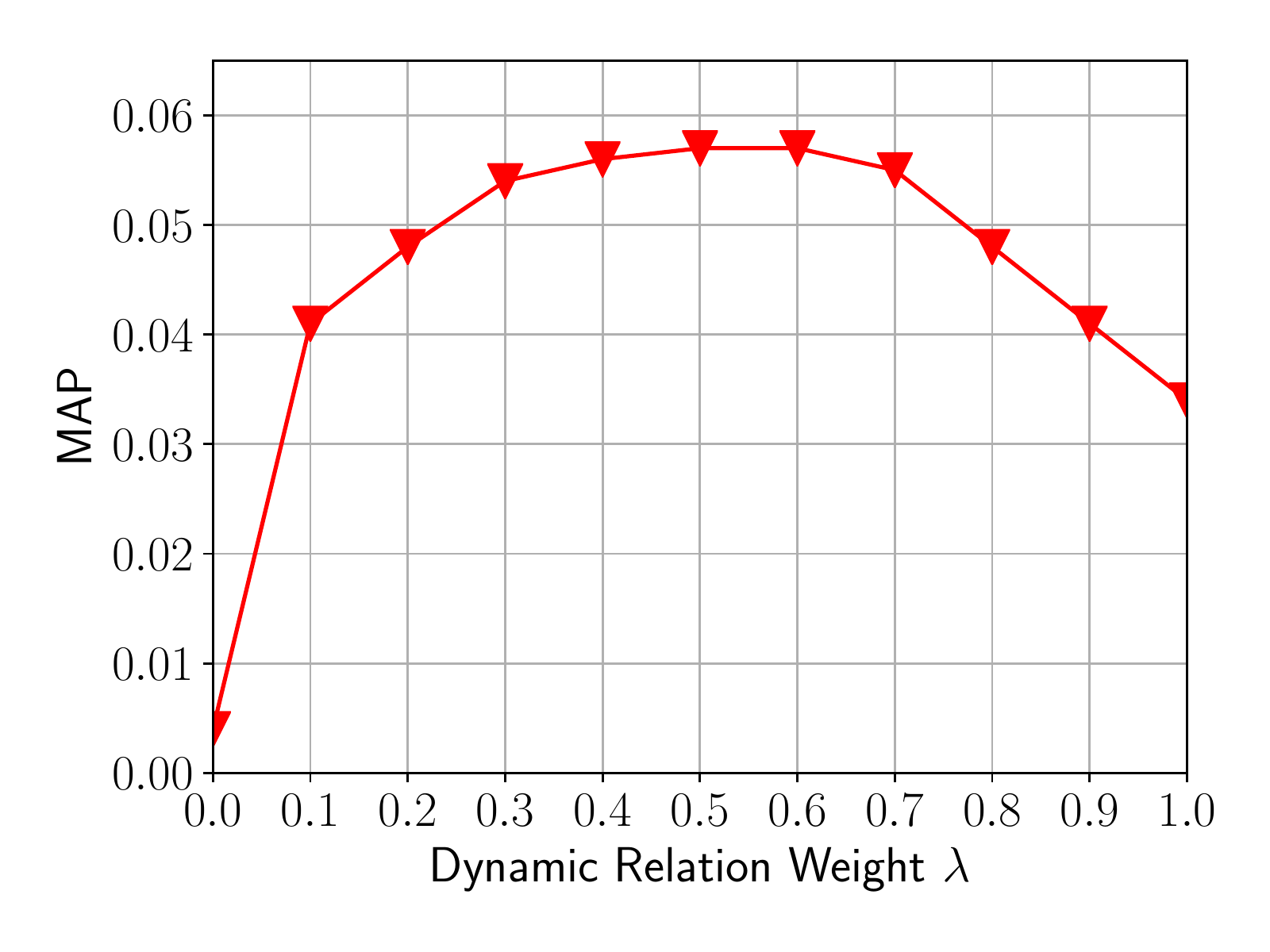}
		\caption{\textit{Kindle Store}}
		\label{fig:k_dynamic}
	\end{subfigure}
	\\
	\begin{subfigure}{.5\textwidth}
		\centering
		\includegraphics[width=2.7in]{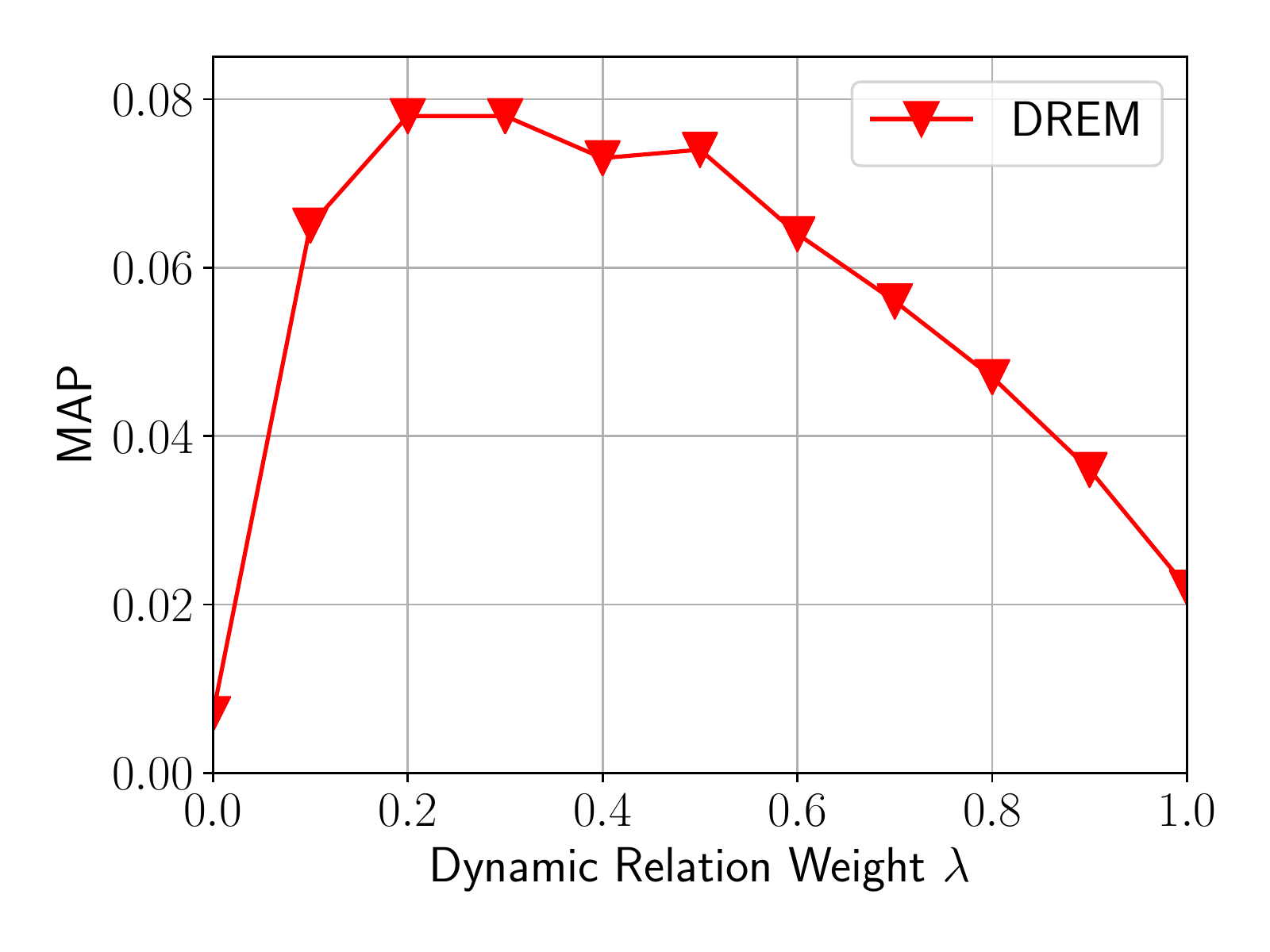}
		\caption{\textit{CDs \& Vinyl}}
		\label{fig:cd_dynamic}
	\end{subfigure}%
	\begin{subfigure}{.5\textwidth}
		\centering
		\includegraphics[width=2.7in]{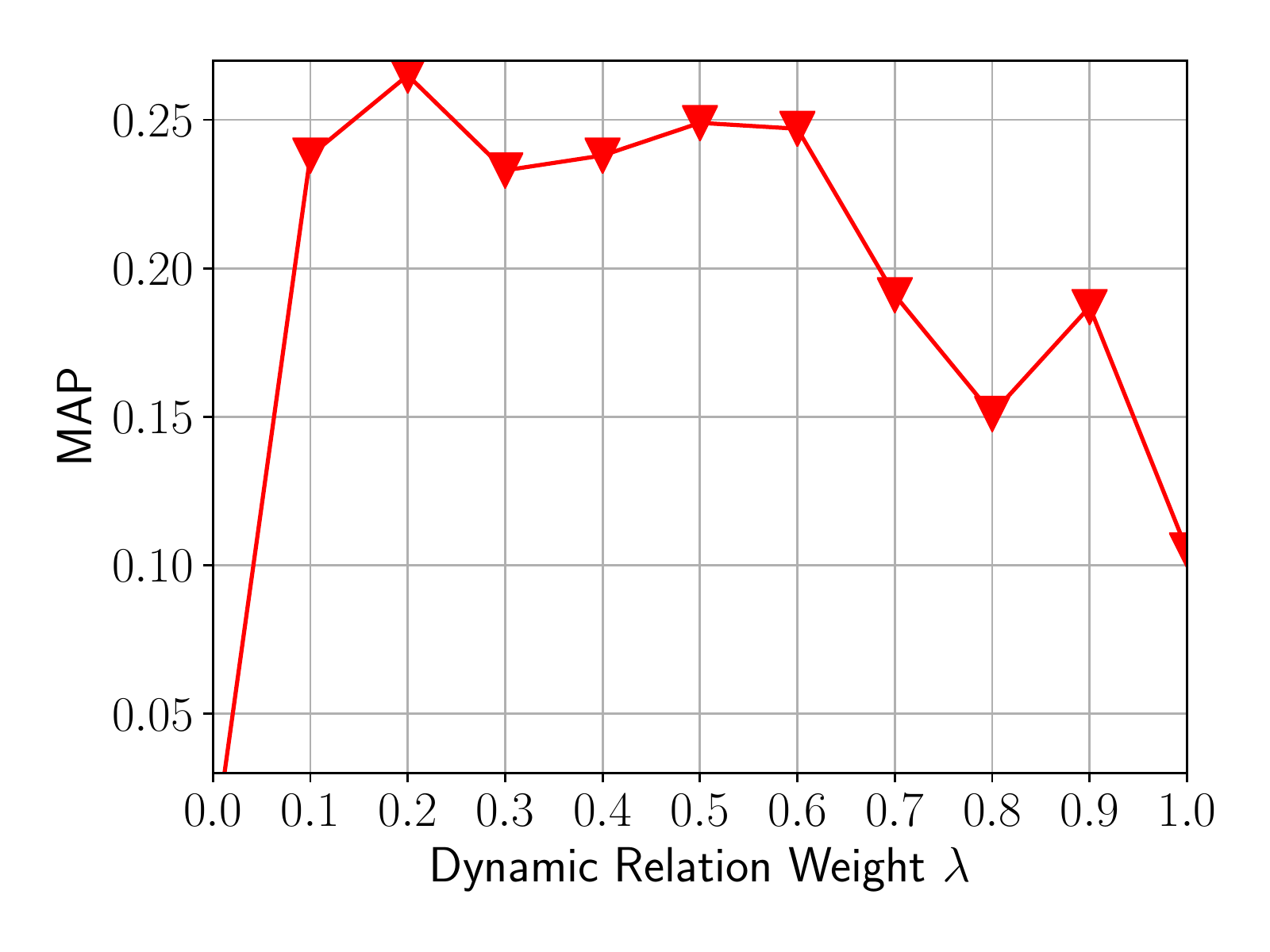}
		\caption{\textit{Cell Phones \& Accessories}}
		\label{fig:ce_dynamic}
	\end{subfigure}%
	\caption{The performance of DREM with different dynamic relation weight $\lambda$.}
	\label{fig:dynamic}
\end{figure*}

\begin{figure*}
	\centering
	\begin{subfigure}{.5\textwidth}
		\centering
		\includegraphics[width=2.7in]{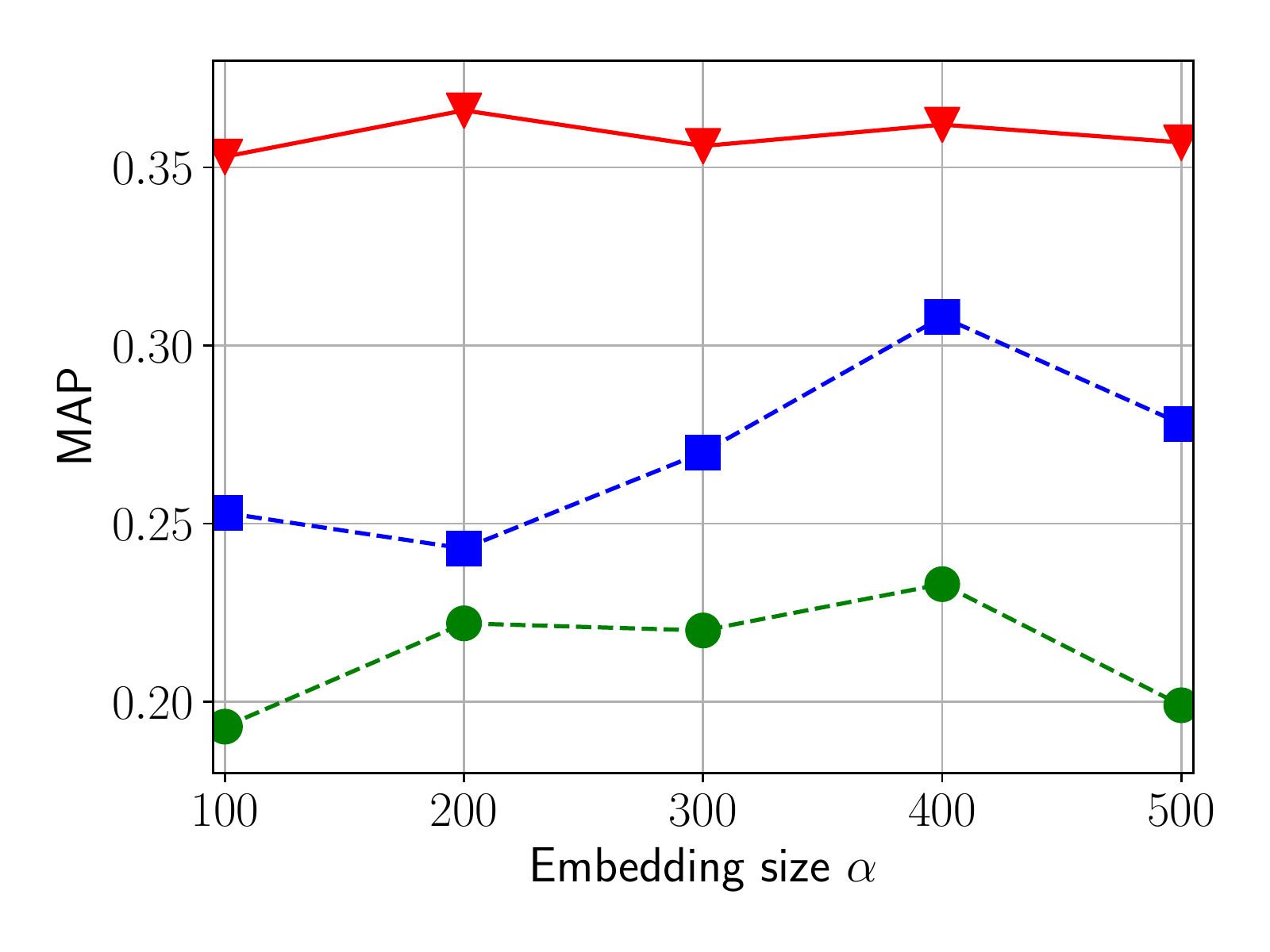}
		\caption{\textit{Electronics}}
		\label{fig:e_embed}
	\end{subfigure}%
	\begin{subfigure}{.5\textwidth}
		\centering
		\includegraphics[width=2.7in]{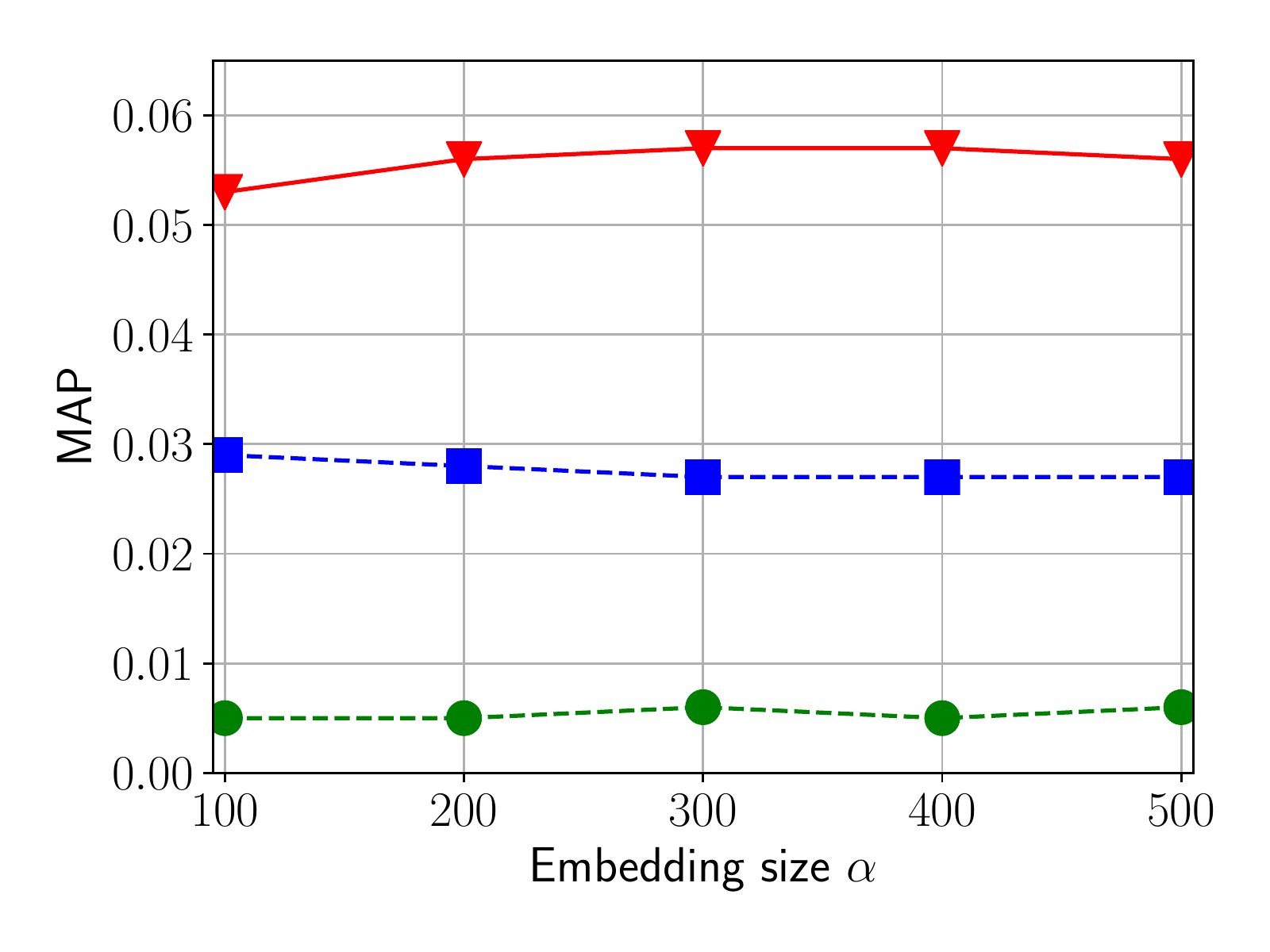}
		\caption{\textit{Kindle Store}}
		\label{fig:k_embed}
	\end{subfigure}
	\\
	\begin{subfigure}{.5\textwidth}
		\centering
		\includegraphics[width=2.7in]{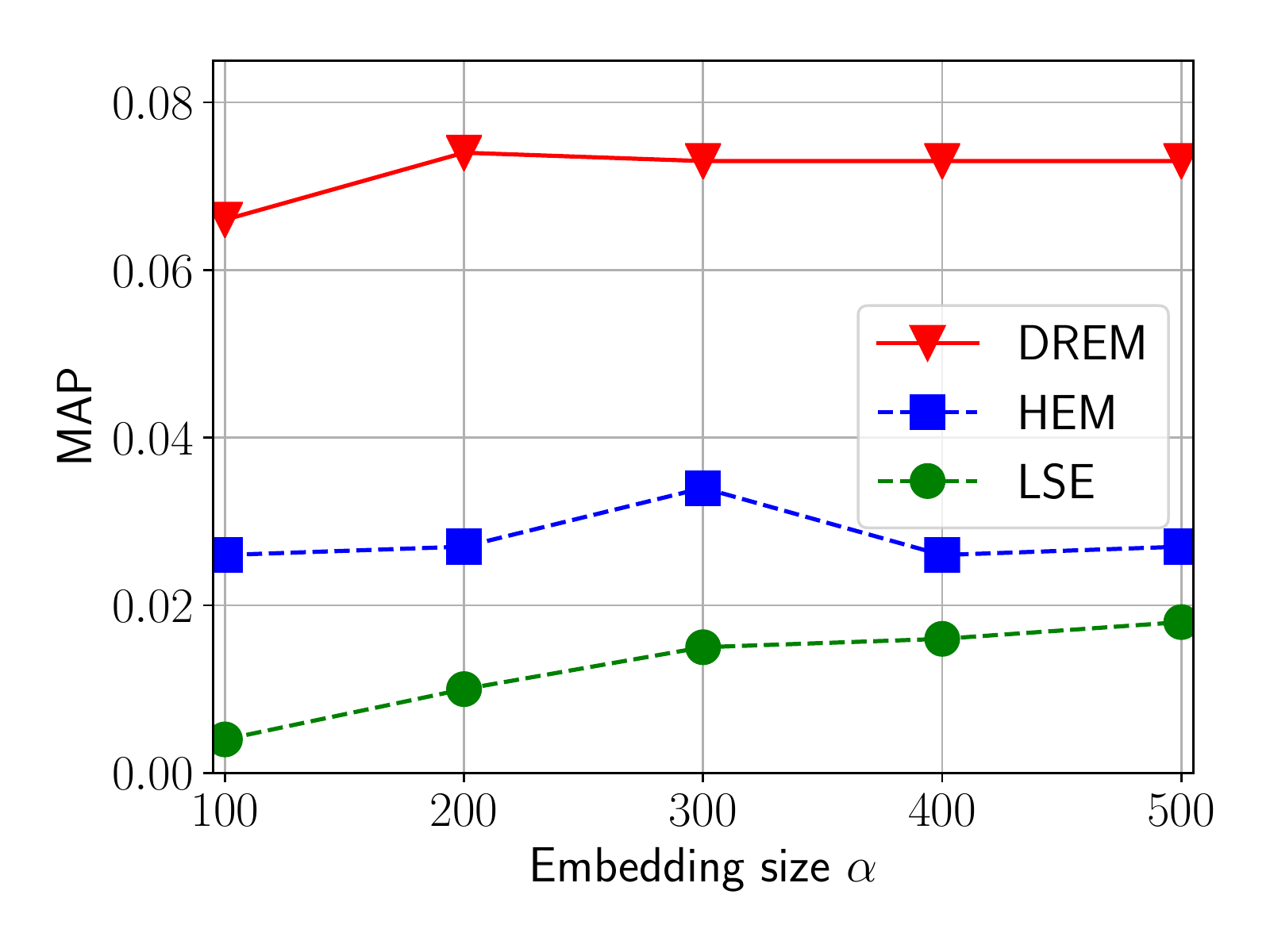}
		\caption{\textit{CDs \& Vinyl}}
		\label{fig:cd_embed}
	\end{subfigure}%
	\begin{subfigure}{.5\textwidth}
		\centering
		\includegraphics[width=2.7in]{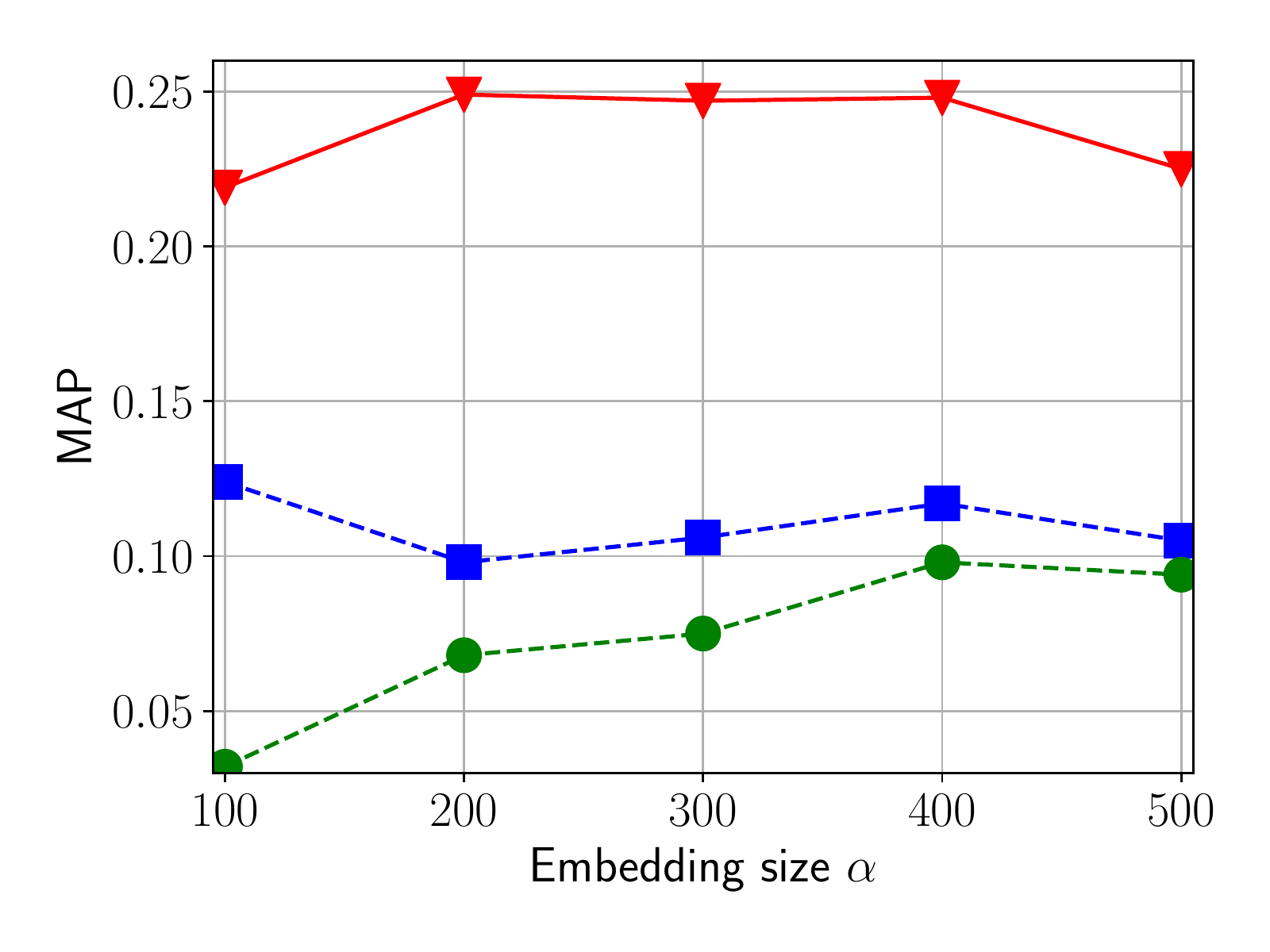}
		\caption{\textit{Cell Phones \& Accessories}}
		\label{fig:ce_embed}
	\end{subfigure}%
	\caption{The performance of DREM and baselines with different embedding size $\alpha$. The red solid line with triangles represents the numbers for DREM$_{All}$; the green and blue dashed lines with circles and squares are results for LSE and HEM, respectively.}
	\label{fig:embed}
\end{figure*}

\subsubsection{Parameter Sensitivity}\label{sec:parameter_sensitivity}
There are two hyper-parameters used in the training of DREM -- the dynamic relation weight $\lambda$ in Equation~(\ref{equ:final_loss}) and the embedding size $\alpha$. 
To analyze the parameter sensitivity of DREM, we plot the MAP of DREM$_{All}$ with different parameter settings in Figure~\ref{fig:dynamic} and Figure~\ref{fig:embed}.

Figure~\ref{fig:dynamic} shows the performance of DREM$_{All}$ on different product categories with respect to the dynamic relation weight $\lambda$ ranged from 0 to 1.
When $\lambda=0$, DREM$_{All}$  learns nothing on the dynamic relationships, and search queries would have no influence on the final search results, which means that the model will be degraded from a search model to recommendation model. 
As expected, the retrieval performance of DREM$_{All}$ with $\lambda=0$ is significantly worse than other models.
When $\lambda=1$, DREM$_{All}$ does not incorporate any information from static relationships.
While it performs reasonable well compared to the text-based baseline models such as QL and LSE, it produces inferior performance compared to DREM$_{All}$ with smaller $\lambda$.
As shown in Figure~\ref{fig:dynamic}, DREM$_{All}$ usually achieves the best performance when $\lambda$ is larger than 0.1 but less than 0.7. 
This demonstrates that both dynamic and static relationship information are valuable for product search. 

Figure~\ref{fig:embed} plots the retrieval performance of both baseline methods (LSE and HEM) and DREM$_{All}$.
As we can see, the size of embeddings has minor effect on the performance of DREM. 
DREM$_{All}$ obtained similar results with different $\alpha$ and outperformed LSE and HEM with large margins.
Therefore, in practice, we advise to start with a small $\alpha$ and increasing it when necessary.


\subsection{Case Study}

\begin{figure*}[t]
	\centering
	\includegraphics[width=5.5in]{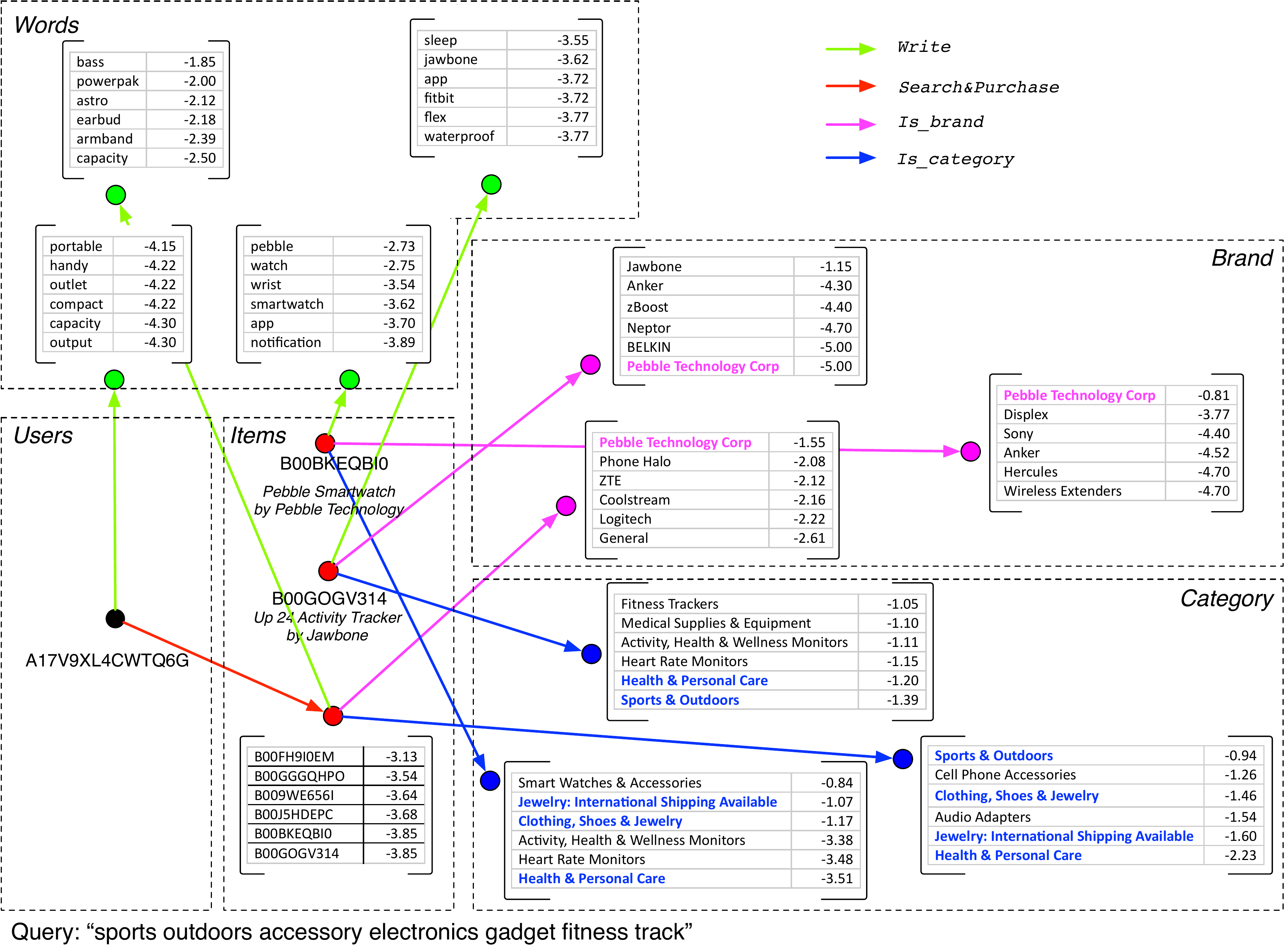}
	\caption{The knowledge graph created by DREM$_{All}$ on \textit{Cell Phones \& Accessories} for query ``sports outdoors accessory electronics gadget fitness track". 
	Six top retrieved entities and the corresponding probabilities (Equation~(\ref{equ:decay})) are shown for each node.
	}
	\label{fig:case_study}
\end{figure*}

To show the effectiveness of DREM as an explainable product search model, \revised{we show an example knowledge graph created by DREM in the experiments and conduct a laboratory study to analyze the performance of its search result explanations.}

\revised{
	\subsubsection{Example Knowledge Graph and Result Explanation Generation}
	Figure~\ref{fig:case_study} depicts 
}the knowledge graph created by DREM$_{All}$ on \textit{Cell Phones \& Accessories} for query ``sports outdoors accessory electronics gadget fitness track".
Here we show the nodes and translations of user ``A17V9XL4CWTQ6G", item ``B00GOGV314" (\textit{Up 24 Activity Tracker} by \textit{Jawbone}), and item ``B00BKEQBI0" (\textit{Pebble Smartwatch} by \textit{Pebble Technology}) in different entity subspace.
Each edge in the graph represents a particular type of relationship.
Entities are connected with their translations through edges with solid arrows.  
For clarity, we hide the item-item relationships (\textit{Also\_bought}, \textit{Also\_viewed} and \textit{Bought\_together}) in the graph.
On each node, we show a list of six results selected from the top retrieved entities with soft matching (Equation~(\ref{equ:soft_matching}) and (\ref{equ:decay})).
Entities shared by multiple lists in the same subspace are highlighted with colors.
We use $u$, $i_j$, $i_p$ to denote the node of the user, \textit{Up 24 Activity Tracker} and \textit{Pebble Smartwatch}, and use $\vec{SP}$, $\vec{B}$, $\vec{C}$ to denote the relationships of \textit{Search\&Purchase}, \textit{Is\_brand} and \textit{Is\_category}.

As shown in Figure~\ref{fig:case_study}, given the Soft Matching Algorithm, we can find the following explanation paths from user ``A17V9XL4CWTQ6G" to \textit{Pebble Smartwatch} ``B00BKEQBI0":  
\begin{itemize}
	\item $u\!+\!\vec{SP}\!+\!\vec{B} \!\!\rightarrow\!\! \textbf{Pebble Technology} \!\!\leftarrow\!\! i_p \!+\! \vec{B}$ with  $S(e|u,i) = -2.36$.
	\item $u\!+\!\vec{SP}\!+\!\vec{C} \!\!\rightarrow\!\! \textbf{Clothing, Shoes, Jewelry} \!\!\leftarrow\!\! i_p \!+\! \vec{C}$ with  $S(e|u,i) = -2.63$.
	\item $u\!+\!\vec{SP}\!+\!\vec{C} \!\!\rightarrow\!\! \textbf{Jewelry:International Ship} \!\!\leftarrow\!\! i_p \!+\! \vec{C}$ with  $S(e|u,i) = -2.67$.
	\item $u\!+\!\vec{SP}\!+\!\vec{C} \!\!\rightarrow\!\! \textbf{Health\&Personal Care} \!\!\leftarrow\!\! i_p \!+\! \vec{C}$ with $S(e|u,i) = -5.84$.
\end{itemize}

With simple templates, we can create four explanations for why the user should be interested in \textit{Pebble Smartwatch} as
\begin{itemize}
	\item ``Based on your profile and query, you may like to see somethings by \textit{Pebble Technology}, and \textit{Pebble Smartwatch} is a top product of this brand." ($S(e|u,i) = -2.36$)
	\item ``Based on your profile and query, you may like to see somethings in \textit{Clothing, Shoes, Jewelry}, and \textit{Pebble Smartwatch} is a top product in this category." ($S(e|u,i) = -2.63$)
	\item ``Based on your profile and query, you may like to see somethings in \textit{Jewelry:International Ship}, and \textit{Pebble Smartwatch by Pebble Technology} is a top product in this category."   ($S(e|u,i) = -2.67$)
	\item ``Based on your profile and query, you may like to see somethings in \textit{Health\&Personal Care}, and \textit{Pebble Smartwatch} is a top product in this category." ($S(e|u,i) = -5.84$)
\end{itemize}

Similarly, for \textit{Up 24 Activity Tracker} ``B00GOGV314", we have the following explanation paths that connect it to the search user ``A17V9XL4CWTQ6G": 
\begin{itemize}
	\item $u\!+\!\vec{SP}\!+\!\vec{C} \!\!\rightarrow\!\! \textbf{Sports\&Outdoors} \!\!\leftarrow\!\! i_j \!+\! \vec{C}$ with  $S(e|u,i) = -2.33$: \\ ``Based on your profile and query, you may like to see somethings in \textit{Sports\&Outdoors}, and \textit{Up 24 Activity Tracker} is a top product in this category."
	\item $u\!+\!\vec{SP}\!+\!\vec{C} \!\!\rightarrow\!\! \textbf{Health\&Personal Care} \!\!\leftarrow\!\! i_j \!+\! \vec{C}$ with $S(e|u,i) = -3.43$:\\ ``Based on your profile and query, you may like to see somethings in \textit{Health\&Personal Care}, and \textit{Up 24 Activity Tracker} is a top product in this category." 
	\item $u\!+\!\vec{SP}\!+\!\vec{B} \!\!\rightarrow\!\! \textbf{Pebble Technology} \!\!\leftarrow\!\! i_j \!+\! \vec{B}$ with $S(e|u,i) = -5.81$\\``Based on your profile and query, you may like to see somethings by \textit{Pebble Technology}, which is a top brand related to \textit{Up 24 Activity Tracker by Jawbone}." 
\end{itemize}

Given more information on the query and corresponding products, we find that most explanations above are actually reasonable.
According to the query, the user is looking for electronic fitness trackers. 
\textit{Pebble Technology} is a company famous for its fitness tracking devices, while \textit{Pebble Smartwatch} is one of its bestsellers.
Also, when the user is searching for fitness trackers within the domain of \textit{Cell Phones \& Accessories}, it is likely that he or she is interested in wearable devices with health tracking functions.
\textit{Pebble Smartwatch} is a wearable device well-known for its multi-functionality and stylish design, while \textit{Up 24 Activity Tracker} is one of its competitors that focuses on health tracking functions and has a cheaper price. 
It is reasonable to recommend the former based on its popularity in \textit{Clothing, Shoes, Jewelry}, while recommend the latter based on its relationship with \textit{Health\&Personal Care}.
In fact, the query word ``fitness" is more related to \textit{Health\&Personal Care}, and the user purchased \textit{Up 24 Activity Tracker} in the end.

\revised{
\subsubsection{Laboratory User Study}
Due to the limit of our experimental environment, we cannot access the original users of the Amazon product dataset to evaluate DREM in terms of explanation generation.
Instead, we conduct a laboratory study and recruit graduate students to analyze the performance of result explanations created by DREM.
For each of our experimental datasets (i.e., \textit{Electronics}, \textit{Kindle Store}, \textit{CDs \& Vinyl} and \textit{Cell Phones \& Accessories}), we randomly sampled 50 test search sessions and the top three explanations generated by the DREM$_{All}$ for the first item retrieved in each session.
Because our annotators are not the original users who conducted the search, it is neither reasonable nor reliable to let them judge whether the sampled result explanations could satisfy the personal need of each user. 
Thus, in our experiments, we focus on evaluating whether the generated result explanations can (1) provide more information for the users and (2) attract people to purchase the item in general. 
Specifically, we follow the methodology proposed by Wang et al.~\cite{wang2018explainable} and create three questions for the annotations of search result explanations in our laboratory study:
\begin{itemize}
	\item Q1: \textit{Item Information}. After reading the search result explanations, have you learned more about the item we retrieved?
	\item Q2: \textit{Query Information}. After reading the search result explanations, have you learned more about the query and what the relevant items may look like?
	\item Q3: \textit{Usefulness}. Overall, do you think people would be more likely to purchase the item after they read the search result explanations?
\end{itemize}   
Annotators are asked to do a 3-level judgments for each question: 0 for \textit{irrelevant}, 1 for \textit{fair}, and 2 for \textit{excellent}.
In total, we have 200 annotated sessions in which each session has been judged by at least two graduate students with master or Ph.D. degrees in Computer Science.
To the best of our knowledge, DREM is the first model that can generate search result explanations for product search, so we only conduct the laboratory study on DREM in this paper.
}




\begin{table}[t]
	\centering
	\caption{The performance of search result explanations generated by DREM in the laboratory user study. \textit{Q1}, \textit{Q2}, and \textit{Q3} refer to the questions of whether the explantion provide more information about the item, whether the explanation provide more information about the query, and whether the explanation is useful in persuading the user to purchase the item. Each question are answered with three-level annotations: 0 (\textit{irrelevant}), 1 (\textit{fair}), and 2 (\textit{excellent}). For each question and each dataset, we show both the mean and the variance of their evaluation scores.}
	\begin{tabular}{ c || c | c | c | } 
		\hline
		Dataset & Q1: Item information & Q2: Query information & Q3: Usefulness \\ \hline \hline
		\textit{Electronics} & 1.08$_{\pm 0.58}$ & 0.65$_{\pm 0.68}$ & 0.48$_{\pm 0.69}$ \\ \hline
		\textit{Kindle Store} & 0.88$_{\pm 0.60}$ & 0.26$_{\pm 0.46}$ & 0.40$_{\pm 0.54}$ \\ \hline
		\textit{CDs \& Vinyl} & 0.93$_{\pm 0.55}$ & 0.42$_{\pm 0.53}$ & 0.44$_{\pm 0.57}$ \\ \hline
		\textit{Cell Phones \& Accessories} & 0.85$_{\pm 0.62}$ & 0.27$_{\pm 0.47}$ & 0.19$_{\pm 0.46}$ \\ \hline
	\end{tabular}
	\label{tab:explanation_eval}
\end{table}

\begin{figure*}
	\centering
	\begin{subfigure}{.5\textwidth}
		\centering
		\includegraphics[width=2.7in]{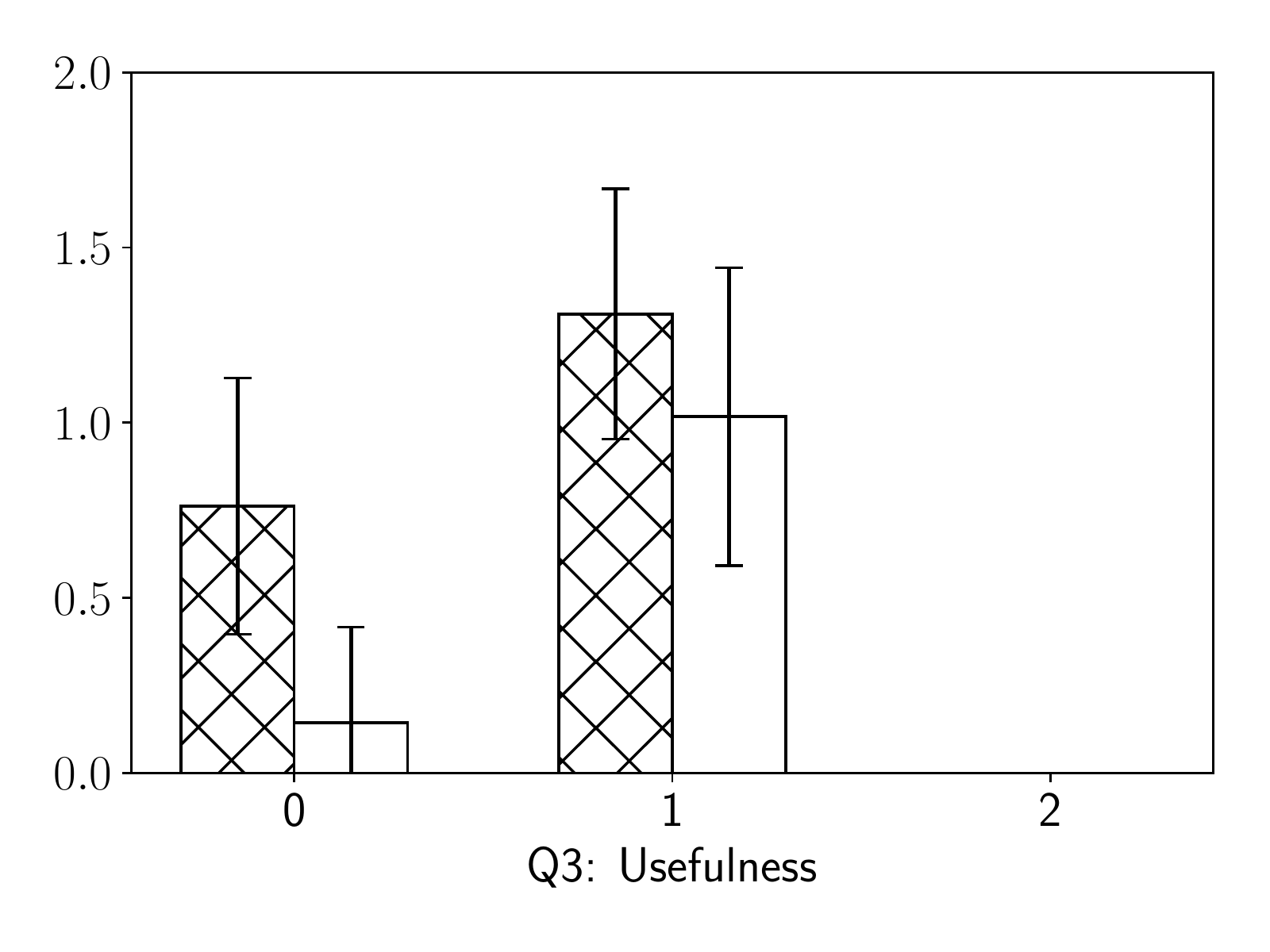}
		\caption{\textit{Electronics}}
		\label{fig:e_user_score}
	\end{subfigure}%
	\begin{subfigure}{.5\textwidth}
		\centering
		\includegraphics[width=2.7in]{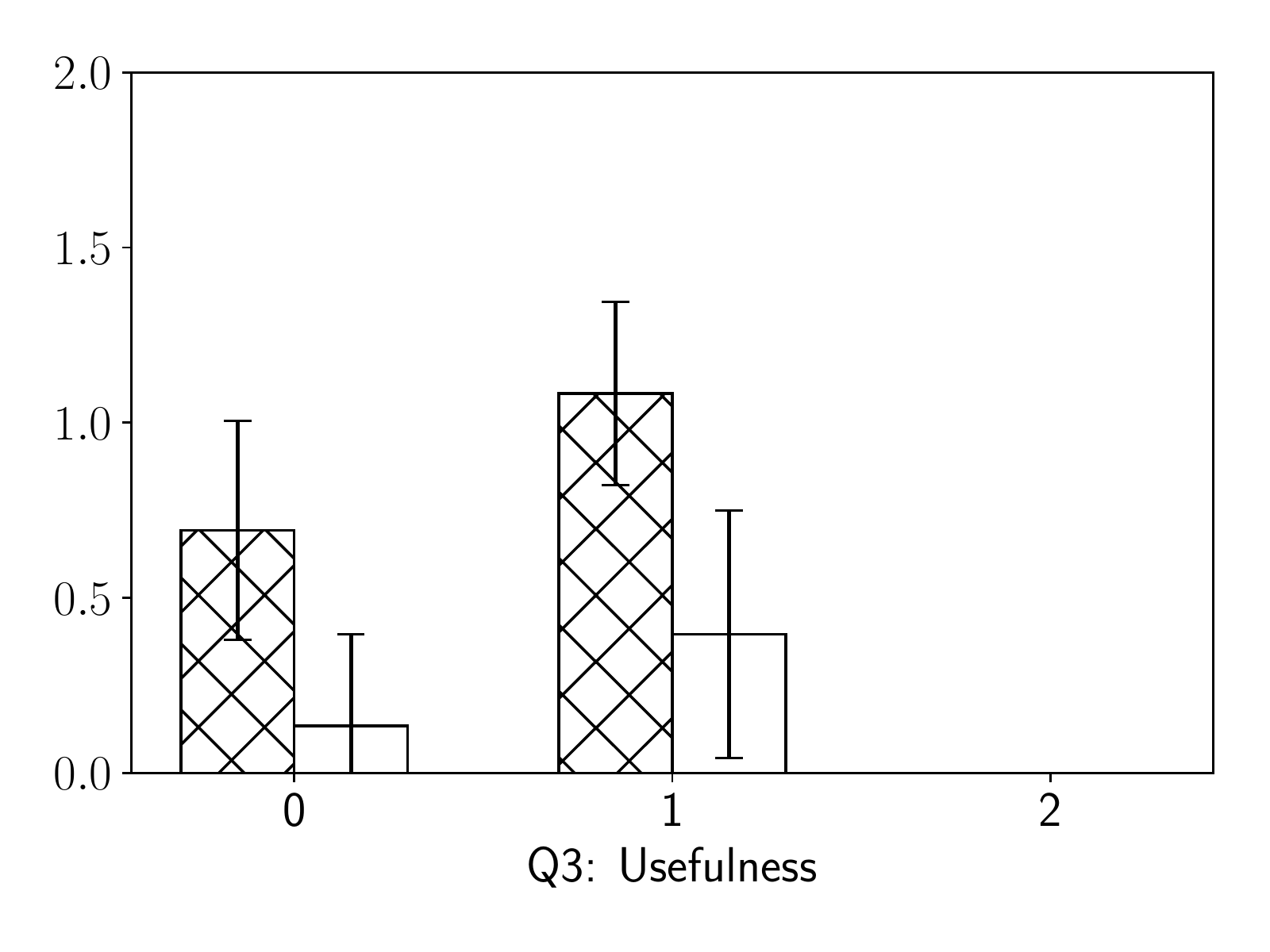}
		\caption{\textit{Kindle Store}}
		\label{fig:k_user_score}
	\end{subfigure}
	\\
	\begin{subfigure}{.5\textwidth}
		\centering
		\includegraphics[width=2.7in]{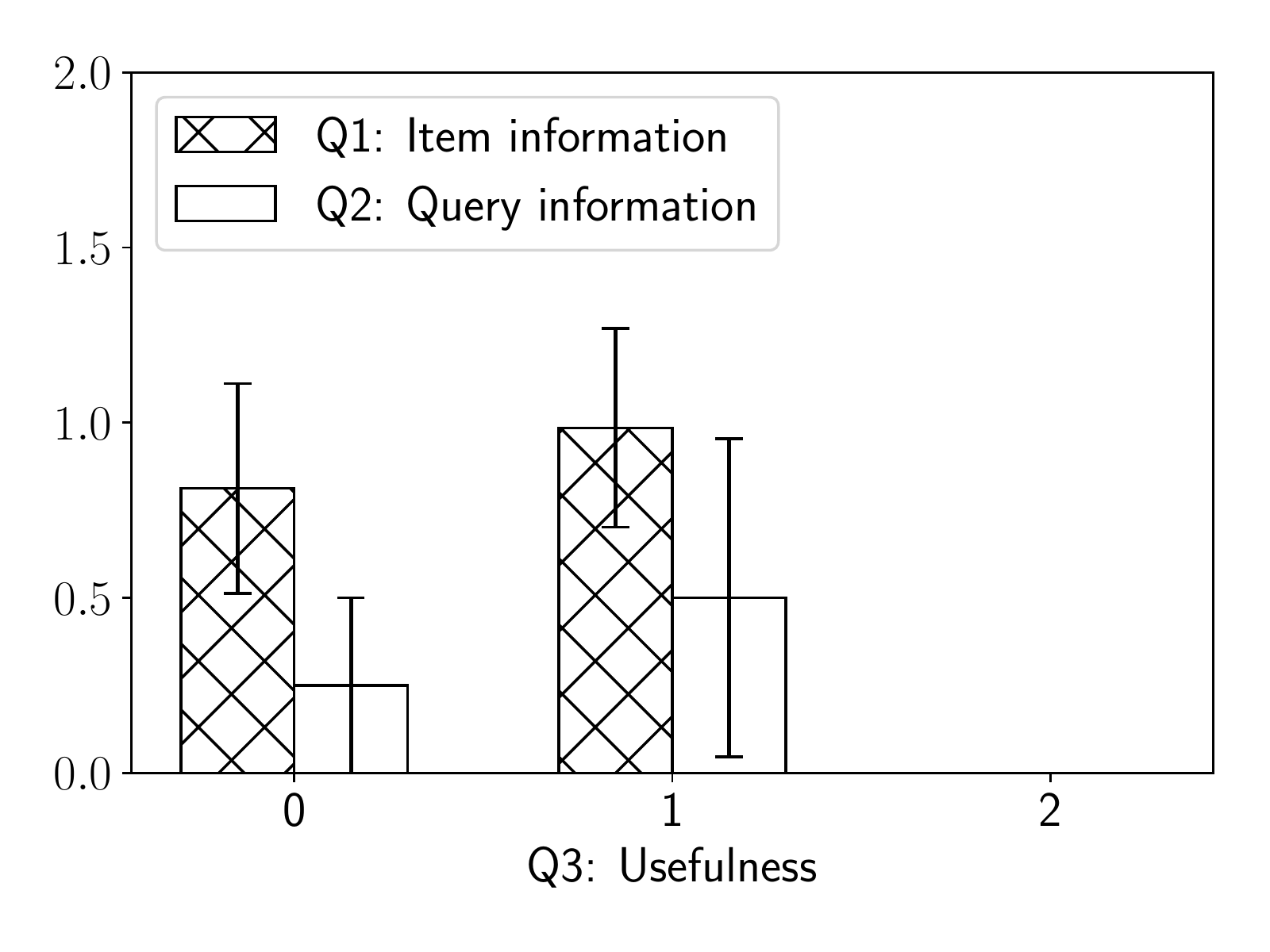}
		\caption{\textit{CDs \& Vinyl}}
		\label{fig:cd_user_score}
	\end{subfigure}%
	\begin{subfigure}{.5\textwidth}
		\centering
		\includegraphics[width=2.7in]{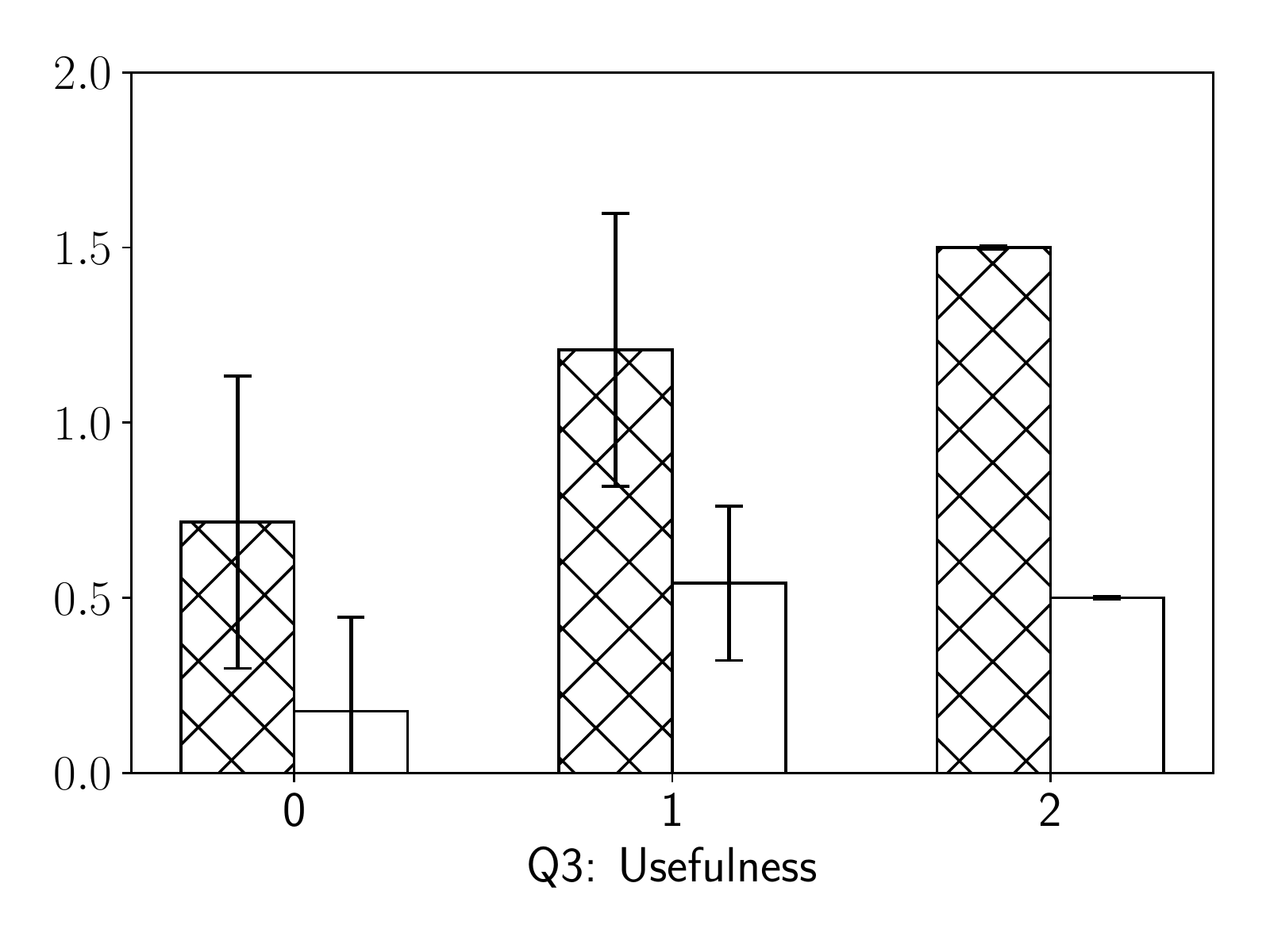}
		\caption{\textit{Cell Phones \& Accessories}}
		\label{fig:ce_user_score}
	\end{subfigure}%
	\caption{The average Q1/Q2 scores with respect to different usefulness (i.e., Q3 scores) for the result explanations created by DREM in the laboratory user study. The usefulness of the explanations in each session are discretized as ``0'' (\textit{irrelevant}), ``1'' (\textit{fair}), and ``2'' (\textit{excellent}).}
	\label{fig:user_score}
\end{figure*}

\begin{figure*}
	\centering
	\begin{subfigure}{.5\textwidth}
		\centering
		\includegraphics[width=2.7in]{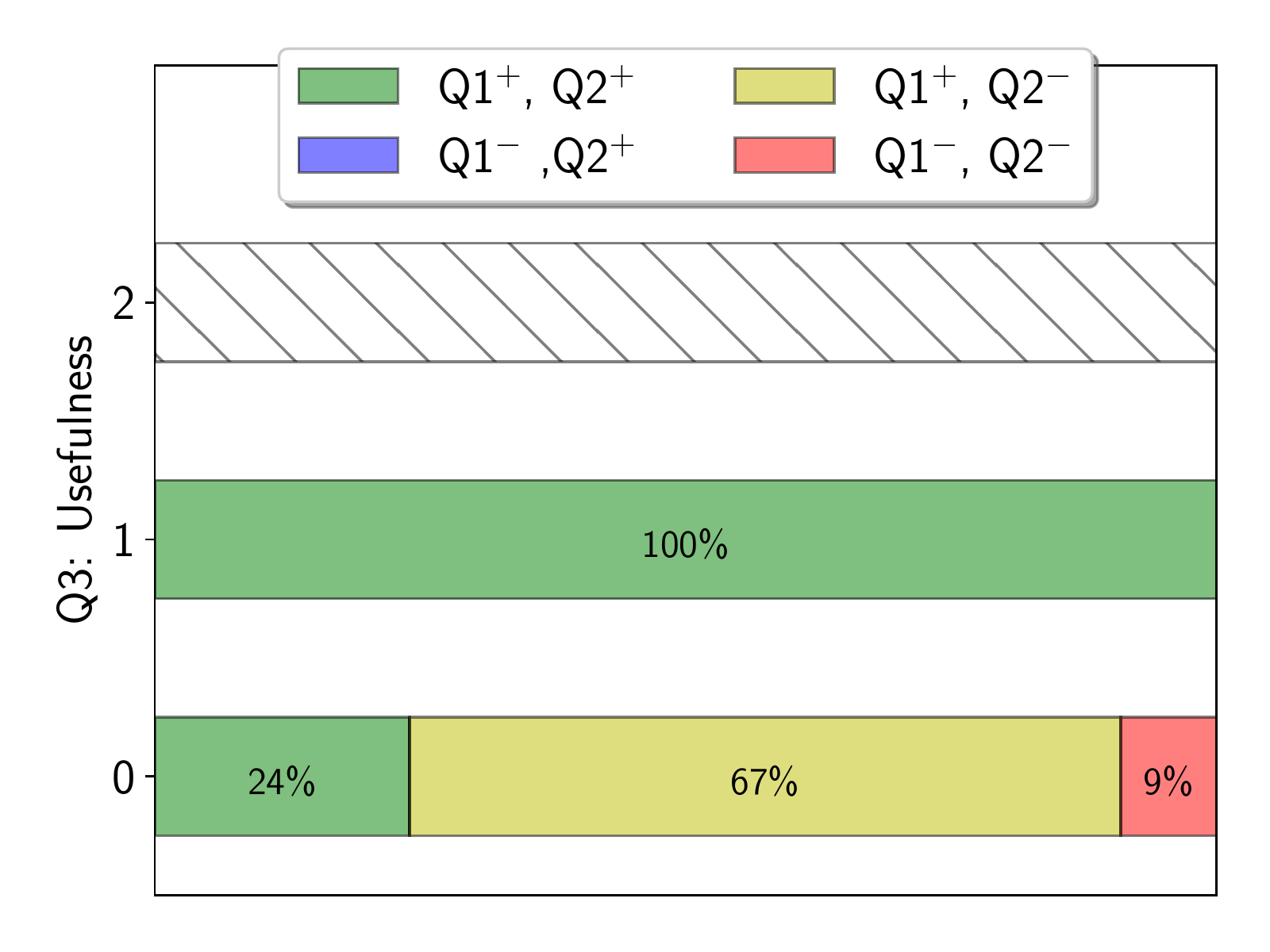}
		\caption{\textit{Electronics}}
		\label{fig:e_user_count}
	\end{subfigure}%
	\begin{subfigure}{.5\textwidth}
		\centering
		\includegraphics[width=2.7in]{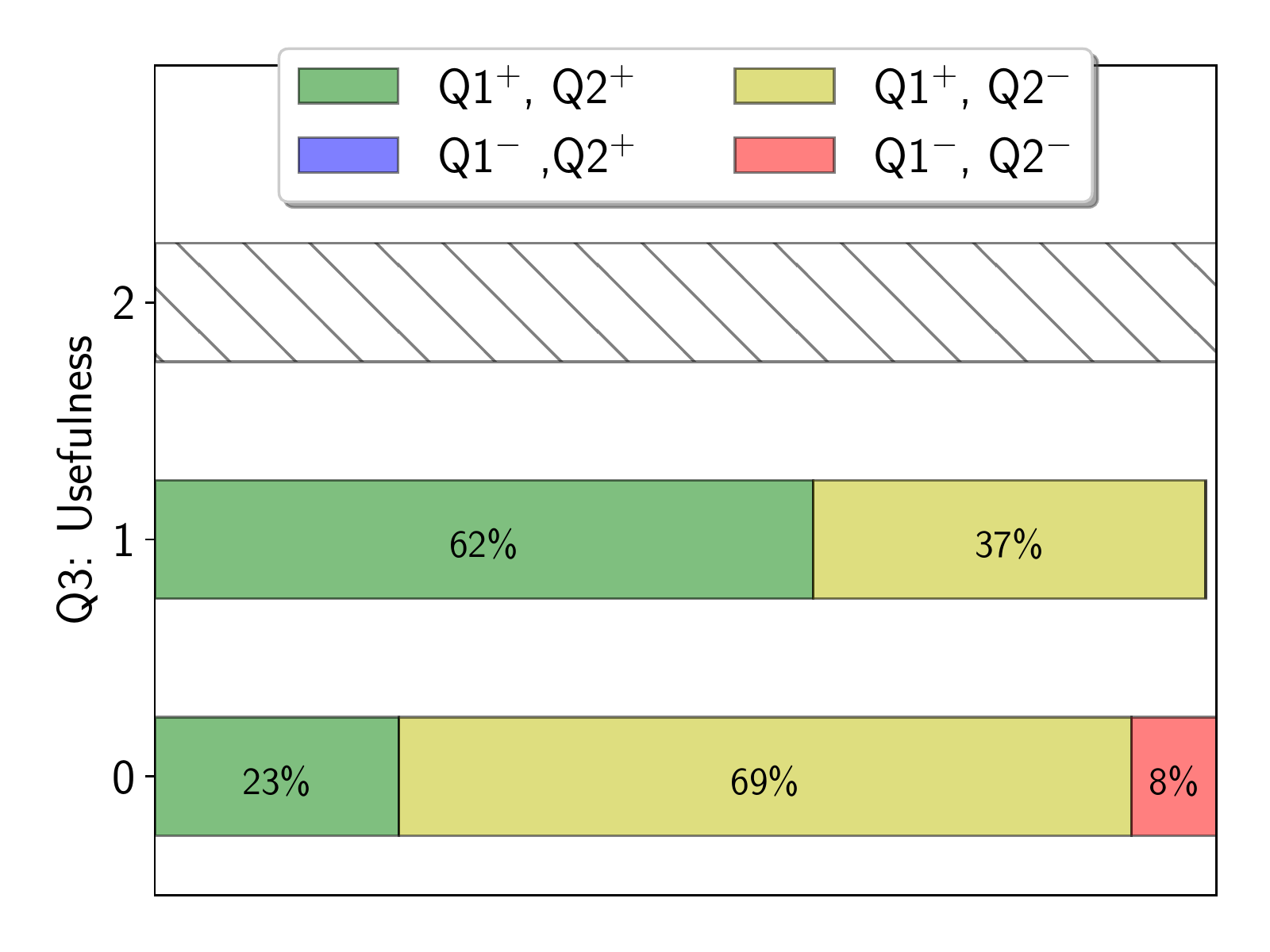}
		\caption{\textit{Kindle Store}}
		\label{fig:k_user_count}
	\end{subfigure}
	\\
	\begin{subfigure}{.5\textwidth}
		\centering
		\includegraphics[width=2.7in]{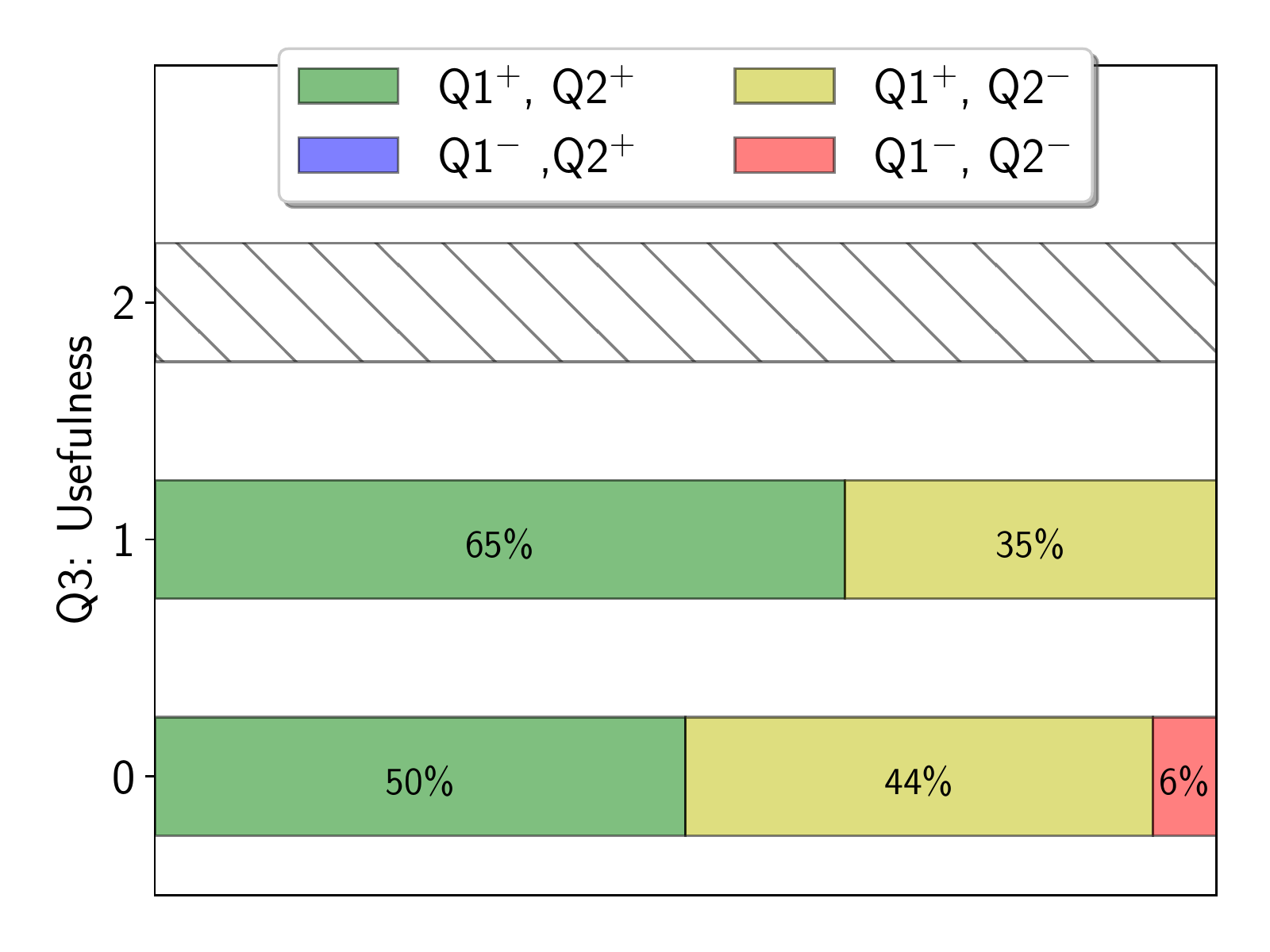}
		\caption{\textit{CDs \& Vinyl}}
		\label{fig:cd_user_count}
	\end{subfigure}%
	\begin{subfigure}{.5\textwidth}
		\centering
		\includegraphics[width=2.7in]{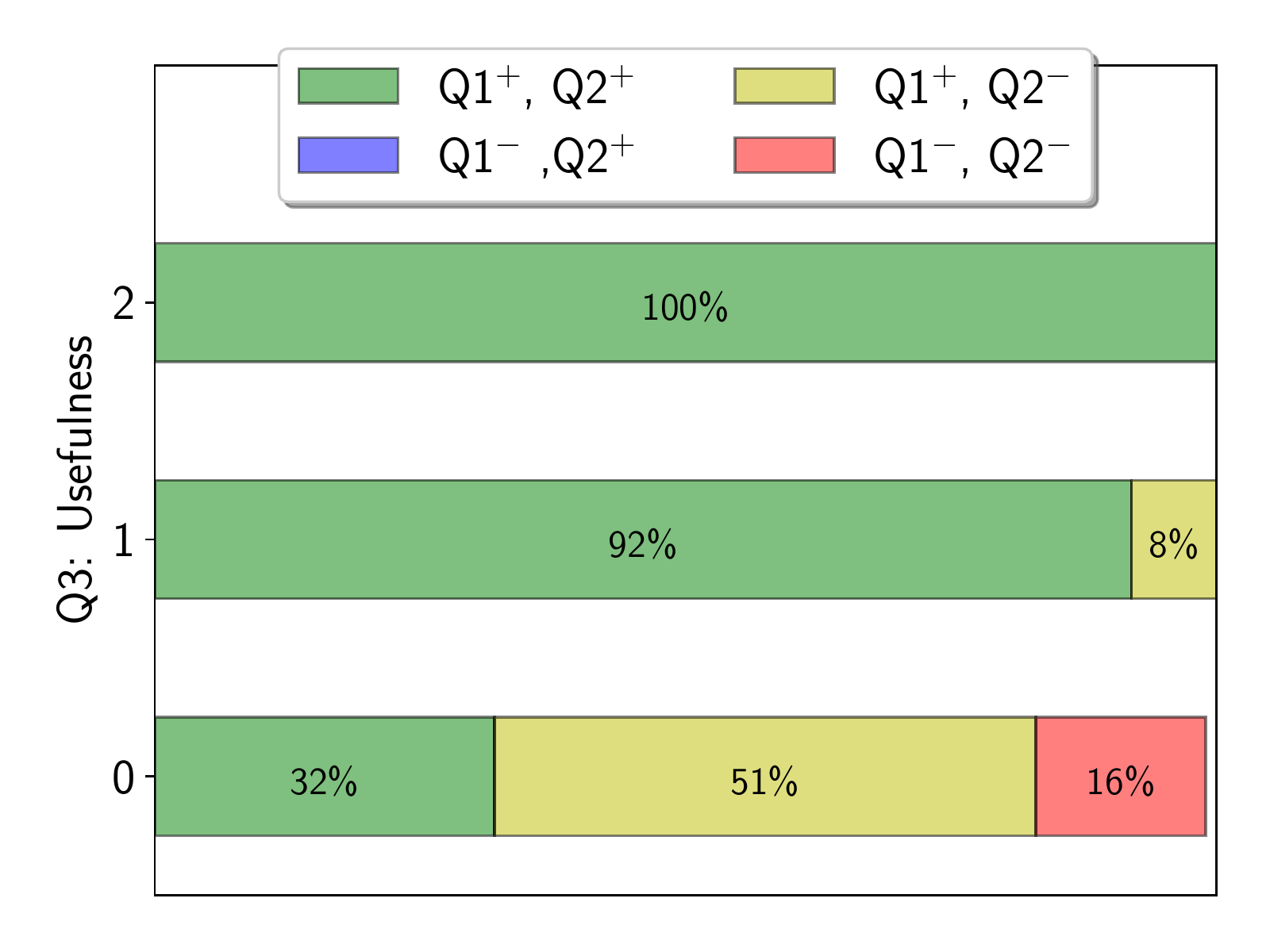}
		\caption{\textit{Cell Phones \& Accessories}}
		\label{fig:ce_user_count}
	\end{subfigure}%
	\caption{The distribution of sessions with different Q1 and Q2 scores with respect to explanation usefulness (i.e., Q3 scores) in the laboratory user study. The cases where Q1/Q2 scores are larger than 0.5 are treated as positive (i.e., Q1$^+$/Q2$^+$), while the cases where Q1/Q2 scores are less than 0.5 are treated as negative (i.e., Q1$^-$/Q2$^-$).}
	\label{fig:user_count}
\end{figure*}

\revised{
Table~\ref{tab:explanation_eval} shows the overall results of our laboratory study on DREM. 
As we can see, the average scores of DREM in Q1 (\textit{item information}) are approximately 1 on all datasets, which means that the result explanations provided by DREM usually provide relevant information about the retrieved items.   
The average scores of DREM in Q2 (\textit{query information}), on the other hand, are relatively low comparing to the scores of Q1.
In DREM, we create result explanations by finding paths between users and items, which are modeled as entities, on a knowledge graph where queries are simply treated as a dynamic relationship.
This approach may impose an emphasis on item modeling and make the model more likely to retrieve relevant information about the item rather than the query. 
As for Q3 (\textit{usefulness}), DREM achieved a score of 0.48 on \textit{Electronics}, 0.40 on \textit{Kindle Store}, 0.44 on \textit{CDs \& Vinyl}, and 0.19 on \textit{Cell Phones \& Accessories}.
While these scores are far from perfect, they indicate that the result explanations provided by DREM is useful in persuading people to purchase the corresponding items in general.
}

\revised{
To understand the relationship between different evaluation metrics, we further analyze the distribution of the DREM's scores in Q1, Q2 and Q3.
Figure~\ref{fig:user_score} shows the average Q1 and Q2 scores of DREM in sessions with different explanation usefulness (i.e., the Q3 scores).
Specifically, we discretize the Q3 scores by labeling the sessions with $[0, 0.5)$ as ``0'' (\textit{irrelevant}), $[0.5, 1.5)$ as ``1'' (\textit{fair}), and $[1.5,2]$ as ``2'' (\textit{excellent}).
The columns of ``2'' in \textit{Electronics}, \textit{Kindle Store} and \textit{CDs \& Vinyl} are missing because there is no session with \textit{excellent} result explanations in these datasets.
As we can see, there is a correlation between both the scores of Q1, Q3 and the scores of Q2 and Q3.
Useful result explanations (i.e., Q3 scores ``1'' or ``2'' ) usually provide more information on the retrieved item (i.e., higher Q1 scores) and the search query (i.e., higher Q2 scores) than unuseful explanations (i.e., Q3 scores ``0'').
Figure~\ref{fig:user_count} depicts the percentage of annotated sessions with different $<$Q1, Q2$>$ pairs with respect to Q3 scores.
Here, we apply the same discretization strategy to Q1/Q2 scores and treat \textit{fair} or \textit{excellent} explanations as poistive instances (i.e, Q1$^+$ and Q2$^+$) and \textit{irrelevant} explanations as negative instances (i.e, Q1$^-$ and Q2$^-$).
As shown in Figure~\ref{fig:user_count}, useful result explanations (i.e., Q3 scores equal or greater than 1) almost always provide relevant item information (i.e., Q1$^+$) or query information (i.e., Q2$^+$).
This supports the hypothesis that result explanations can be useful only when its provide additional information that helps the queries or the retrieved items.
Also, even when their usefulness in persuading people to purchase the items is not good (i.e., Q3 scores are 0), the percentage of result explanations with $<$Q1$^-$, Q2$^-\!\!>$ is lower than 16\% in all datasets.
This indicates that the result explanations created by DREM can provide relevant information about the items or the queries in most cases.
}

\revised{
In our experiments, we haven't observed significant correlation between the retrieval performance and the result explanation usefulness on different datasets, but there are some interesting comments provided by the annotators.
First, the annotators find that different entity relationships could have different value for the generation of result explanations in each dataset. 
For example, the relationships of \textit{Also\_viewed} and \textit{Also\_bought} tend to be less useful in \textit{Kindle Store} and \textit{CDs \& Vinyl} because, when the annotators are not familiar with the topic of the retrieved item, they are also not familiar with the items that are viewed or bought together with the retrieved item.
Also, the annotators note that the relevance of retrieved items could affect their judgment process.
In our laboratory study, we only sample the result explanations for the first retrieved items in each sampled search session.
In many cases, however, the first retrieved item of DREM are irrelevant to the query or not purchased by the user.
The annotators find it particularly annoying when the result explanations are relevant while the actual retrieved items are not.
This raises an interesting question of whether search result explanations should be created according to how much we believe that the retrieved items are relevant to the users.
We leave these topics for future studies.
}


%

%% file: conclusion.tex
\section{Conclusion and Future Work}\label{sec:conclusion}

In this paper, we present our initial attempt to tackle the problem of explainable product search.
We propose a Dynamic Relation Embedding Model that jointly learns embedding representations for entities/relationships and creates session-dependent knowledge graphs.
Empirical experiments show that our approach significantly outperforms the state-of-the-art product retrieval methods and has the ability to produce reasonable explanations for search results.
This indicates that the construction of dynamic knowledge graph with multi-relational product data is beneficial for the effectiveness and explainability of product retrieval models.

We believe that the study of explainable retrieval models is in an early stage and could be fruitful for product search. 
In DREM, we generate search explanations based on each explanation path separately.
As shown in our laboratory user study, these result explanations are far from perfect. 
In practice, it may be preferable to organize and combine multiple explanation paths to create a single but more persuasive explanation.
Also, in our framework, the explanation path extracted by the Soft Matching Algorithm are scored by heuristic functions (Equation~(\ref{equ:soft_matching}) and (\ref{equ:decay})), which are empirically effective but not theoretically principled.
How to unify product retrieval and search explanation in terms of the model design is still an open question, and we believe that one promising approach is to combine the power of neural embedding models with rule-based knowledge systems.
\revised{Last but not least, although it seems intuitive that good result explanations could improve the transaction rate of e-commerce search, their actual effects on online systems are mostly unknown.
Understanding the real impact of expainable search systems and developing effective evaluation metircs for result explanations are important research problems for the future of e-commerce search engines.
We will explore these topics in future studies.
}


%% file: TOIS2018.bbl

\begin{thebibliography}{00}


\ifx \showCODEN    \undefined \def \showCODEN     #1{\unskip}     \fi
\ifx \showDOI      \undefined \def \showDOI       #1{{\tt DOI:}\penalty0{#1}\ }
  \fi
\ifx \showISBNx    \undefined \def \showISBNx     #1{\unskip}     \fi
\ifx \showISBNxiii \undefined \def \showISBNxiii  #1{\unskip}     \fi
\ifx \showISSN     \undefined \def \showISSN      #1{\unskip}     \fi
\ifx \showLCCN     \undefined \def \showLCCN      #1{\unskip}     \fi
\ifx \shownote     \undefined \def \shownote      #1{#1}          \fi
\ifx \showarticletitle \undefined \def \showarticletitle #1{#1}   \fi
\ifx \showURL      \undefined \def \showURL       #1{#1}          \fi

\bibitem[\protect\citeauthoryear{Ai, Azizi, Chen, and Zhang}{Ai
  et~al\mbox{.}}{2018a}]%
        {ai2018learninghe}
{Qingyao Ai}, {Vahid Azizi}, {Xu Chen}, {and} {Yongfeng Zhang}. 2018a.
\newblock \showarticletitle{Learning heterogeneous knowledge base embeddings
  for explainable recommendation}. {\em Algorithms\/} {11}, 9 (2018), 137.
\newblock


\bibitem[\protect\citeauthoryear{Ai, Bi, Guo, and Croft}{Ai
  et~al\mbox{.}}{2018b}]%
        {ai2018learning}
{Qingyao Ai}, {Keping Bi}, {Jiafeng Guo}, {and} {W~Bruce Croft}. 2018b.
\newblock \showarticletitle{Learning a deep listwise context model for ranking
  refinement}. In {\em The 41st International ACM SIGIR Conference on Research
  \& Development in Information Retrieval}. ACM, 135--144.
\newblock


\bibitem[\protect\citeauthoryear{Ai, Hill, Vishwanathan, and Croft}{Ai
  et~al\mbox{.}}{2019}]%
        {ai2019zero}
{Qingyao Ai}, {Daniel~N Hill}, {SVN Vishwanathan}, {and} {W~Bruce Croft}. 2019.
\newblock \showarticletitle{A Zero Attention Model for Personalized Product
  Search}. In {\em Proceedings of the 28th ACM International Conference on
  Information and Knowledge Management}. ACM.
\newblock


\bibitem[\protect\citeauthoryear{Ai, Yang, Guo, and Croft}{Ai
  et~al\mbox{.}}{2016a}]%
        {ai2016analysis}
{Qingyao Ai}, {Liu Yang}, {Jiafeng Guo}, {and} {W~Bruce Croft}. 2016a.
\newblock \showarticletitle{Analysis of the paragraph vector model for
  information retrieval}. In {\em Proceedings of the ACM ICTIR'16}. ACM,
  133--142.
\newblock


\bibitem[\protect\citeauthoryear{Ai, Yang, Guo, and Croft}{Ai
  et~al\mbox{.}}{2016b}]%
        {ai2016improving}
{Qingyao Ai}, {Liu Yang}, {Jiafeng Guo}, {and} {W~Bruce Croft}. 2016b.
\newblock \showarticletitle{Improving language estimation with the paragraph
  vector model for ad-hoc retrieval}. In {\em Proceedings of the 39th
  International ACM SIGIR conference}. ACM, 869--872.
\newblock


\bibitem[\protect\citeauthoryear{Ai, Zhang, Bi, Chen, and Croft}{Ai
  et~al\mbox{.}}{2017}]%
        {ai2017learning}
{Qingyao Ai}, {Yongfeng Zhang}, {Keping Bi}, {Xu Chen}, {and} {W~Bruce Croft}.
  2017.
\newblock \showarticletitle{Learning a hierarchical embedding model for
  personalized product search}. In {\em Proceedings of the 40th International
  ACM SIGIR Conference}. ACM, 645--654.
\newblock


\bibitem[\protect\citeauthoryear{Aryafar, Guillory, and Hong}{Aryafar
  et~al\mbox{.}}{2017}]%
        {aryafar2017ensemble}
{Kamelia Aryafar}, {Devin Guillory}, {and} {Liangjie Hong}. 2017.
\newblock \showarticletitle{An Ensemble-based Approach to Click-Through Rate
  Prediction for Promoted Listings at Etsy}. In {\em Proceedings of the
  ADKDD'17}. ACM, 10.
\newblock


\bibitem[\protect\citeauthoryear{Bello, Kulkarni, Jain, Boutilier, Chi, Eban,
  Luo, Mackey, and Meshi}{Bello et~al\mbox{.}}{2018}]%
        {bello2018seq2slate}
{Irwan Bello}, {Sayali Kulkarni}, {Sagar Jain}, {Craig Boutilier}, {Ed Chi},
  {Elad Eban}, {Xiyang Luo}, {Alan Mackey}, {and} {Ofer Meshi}. 2018.
\newblock \showarticletitle{Seq2Slate: Re-ranking and Slate Optimization with
  RNNs}.
\newblock {\em arXiv preprint arXiv:1810.02019\/} (2018).
\newblock


\bibitem[\protect\citeauthoryear{Bi, Ai, Zhang, and Croft}{Bi
  et~al\mbox{.}}{2019a}]%
        {bi2019conversational}
{Keping Bi}, {Qingyao Ai}, {Yongfeng Zhang}, {and} {W~Bruce Croft}. 2019a.
\newblock \showarticletitle{Conversational Product Search Based on Negative
  Feedback}. In {\em Proceedings of the 28th ACM International Conference on
  Information and Knowledge Management}. ACM.
\newblock


\bibitem[\protect\citeauthoryear{Bi, Teo, Dattatreya, Mohan, and Croft}{Bi
  et~al\mbox{.}}{2019b}]%
        {bi2019leverage}
{Keping Bi}, {Choon~Hui Teo}, {Yesh Dattatreya}, {Vijai Mohan}, {and} {W~Bruce
  Croft}. 2019b.
\newblock \showarticletitle{Leverage Implicit Feedback for Context-aware
  Product Search}. In {\em Proceedings of the 28th ACM International Conference
  on Information and Knowledge Management}. ACM.
\newblock


\bibitem[\protect\citeauthoryear{Bilgic and Mooney}{Bilgic and Mooney}{2005}]%
        {bilgic2005explaining}
{Mustafa Bilgic} {and} {Raymond~J Mooney}. 2005.
\newblock \showarticletitle{Explaining recommendations: Satisfaction vs.
  promotion}. In {\em Beyond Personalization Workshop, IUI}, Vol.~5. 153.
\newblock


\bibitem[\protect\citeauthoryear{Bordes, Usunier, Garcia-Duran, Weston, and
  Yakhnenko}{Bordes et~al\mbox{.}}{2013}]%
        {bordes2013translating}
{Antoine Bordes}, {Nicolas Usunier}, {Alberto Garcia-Duran}, {Jason Weston},
  {and} {Oksana Yakhnenko}. 2013.
\newblock \showarticletitle{Translating embeddings for modeling
  multi-relational data}. In {\em Advances in neural information processing
  systems}. 2787--2795.
\newblock


\bibitem[\protect\citeauthoryear{Bordes, Weston, Collobert, Bengio,
  et~al\mbox{.}}{Bordes et~al\mbox{.}}{2011}]%
        {bordes2011learning}
{Antoine Bordes}, {Jason Weston}, {Ronan Collobert}, {Yoshua Bengio}, {and}
  {others}. 2011.
\newblock \showarticletitle{Learning Structured Embeddings of Knowledge
  Bases.}. In {\em AAAI}, Vol.~6. 6.
\newblock


\bibitem[\protect\citeauthoryear{Borisov, Markov, de~Rijke, and
  Serdyukov}{Borisov et~al\mbox{.}}{2016}]%
        {borisov2016neural}
{Alexey Borisov}, {Ilya Markov}, {Maarten de Rijke}, {and} {Pavel Serdyukov}.
  2016.
\newblock \showarticletitle{A neural click model for web search}. In {\em
  Proceedings of the 25th International Conference on World Wide Web}.
  International World Wide Web Conferences Steering Committee, 531--541.
\newblock


\bibitem[\protect\citeauthoryear{Cramer, Evers, Ramlal, Van~Someren, Rutledge,
  Stash, Aroyo, and Wielinga}{Cramer et~al\mbox{.}}{2008}]%
        {cramer2008effects}
{Henriette Cramer}, {Vanessa Evers}, {Satyan Ramlal}, {Maarten Van~Someren},
  {Lloyd Rutledge}, {Natalia Stash}, {Lora Aroyo}, {and} {Bob Wielinga}. 2008.
\newblock \showarticletitle{The effects of transparency on trust in and
  acceptance of a content-based art recommender}.
\newblock {\em User Modeling and User-Adapted Interaction\/} {18}, 5 (2008),
  455.
\newblock


\bibitem[\protect\citeauthoryear{Dijkstra}{Dijkstra}{1959}]%
        {dijkstra1959note}
{Edsger~W Dijkstra}. 1959.
\newblock \showarticletitle{A note on two problems in connexion with graphs}.
\newblock {\em Numerische mathematik\/} {1}, 1 (1959), 269--271.
\newblock


\bibitem[\protect\citeauthoryear{Duan, Zhai, Cheng, and Gattani}{Duan
  et~al\mbox{.}}{2013}]%
        {duan2013supporting}
{Huizhong Duan}, {ChengXiang Zhai}, {Jinxing Cheng}, {and} {Abhishek Gattani}.
  2013.
\newblock \showarticletitle{Supporting keyword search in product database: a
  probabilistic approach}.
\newblock {\em Proceedings of the VLDB Endowment\/} {6}, 14 (2013), 1786--1797.
\newblock


\bibitem[\protect\citeauthoryear{Guo, Fan, Ai, and Croft}{Guo
  et~al\mbox{.}}{2016a}]%
        {guo2016deep}
{Jiafeng Guo}, {Yixing Fan}, {Qingyao Ai}, {and} {W~Bruce Croft}. 2016a.
\newblock \showarticletitle{A deep relevance matching model for ad-hoc
  retrieval}. In {\em Proceedings of the 25th ACM CIKM}. ACM, 55--64.
\newblock


\bibitem[\protect\citeauthoryear{Guo, Fan, Ai, and Croft}{Guo
  et~al\mbox{.}}{2016b}]%
        {guo2016semantic}
{Jiafeng Guo}, {Yixing Fan}, {Qingyao Ai}, {and} {W~Bruce Croft}. 2016b.
\newblock \showarticletitle{Semantic Matching by Non-Linear Word Transportation
  for Information Retrieval}. In {\em Proceedings of the 25th ACM International
  on Conference on Information and Knowledge Management}. ACM, 701--710.
\newblock


\bibitem[\protect\citeauthoryear{Guo, Cheng, Nie, Xu, and Kankanhalli}{Guo
  et~al\mbox{.}}{2018}]%
        {guo2018multi}
{Yangyang Guo}, {Zhiyong Cheng}, {Liqiang Nie}, {Xin-Shun Xu}, {and} {Mohan
  Kankanhalli}. 2018.
\newblock \showarticletitle{Multi-modal preference modeling for product
  search}. In {\em 2018 ACM Multimedia Conference on Multimedia Conference}.
  ACM, 1865--1873.
\newblock


\bibitem[\protect\citeauthoryear{Harshman and Lundy}{Harshman and
  Lundy}{1994}]%
        {harshman1994parafac}
{Richard~A Harshman} {and} {Margaret~E Lundy}. 1994.
\newblock \showarticletitle{PARAFAC: Parallel factor analysis}.
\newblock {\em Computational Statistics \& Data Analysis\/} {18}, 1 (1994),
  39--72.
\newblock


\bibitem[\protect\citeauthoryear{He, Kang, and McAuley}{He
  et~al\mbox{.}}{2018}]%
        {he2018translation}
{Ruining He}, {Wang-Cheng Kang}, {and} {Julian McAuley}. 2018.
\newblock \showarticletitle{Translation-based Recommendation: A Scalable Method
  for Modeling Sequential Behavior.}. In {\em IJCAI}. 5264--5268.
\newblock


\bibitem[\protect\citeauthoryear{He, Chen, Kan, and Chen}{He
  et~al\mbox{.}}{2015}]%
        {he2015trirank}
{Xiangnan He}, {Tao Chen}, {Min-Yen Kan}, {and} {Xiao Chen}. 2015.
\newblock \showarticletitle{Trirank: Review-aware explainable recommendation by
  modeling aspects}. In {\em Proceedings of the 24th ACM International on
  Conference on Information and Knowledge Management}. ACM, 1661--1670.
\newblock


\bibitem[\protect\citeauthoryear{He, Liao, Zhang, Nie, Hu, and Chua}{He
  et~al\mbox{.}}{2017}]%
        {he2017neural}
{Xiangnan He}, {Lizi Liao}, {Hanwang Zhang}, {Liqiang Nie}, {Xia Hu}, {and}
  {Tat-Seng Chua}. 2017.
\newblock \showarticletitle{Neural collaborative filtering}. In {\em
  Proceedings of the 26th International Conference on World Wide Web}.
  International World Wide Web Conferences Steering Committee, 173--182.
\newblock


\bibitem[\protect\citeauthoryear{Hechtlinger}{Hechtlinger}{2016}]%
        {hechtlinger2016interpretation}
{Yotam Hechtlinger}. 2016.
\newblock \showarticletitle{Interpretation of prediction models using the input
  gradient}.
\newblock {\em arXiv preprint arXiv:1611.07634\/} (2016).
\newblock


\bibitem[\protect\citeauthoryear{Herlocker, Konstan, and Riedl}{Herlocker
  et~al\mbox{.}}{2000}]%
        {herlocker2000explaining}
{Jonathan~L Herlocker}, {Joseph~A Konstan}, {and} {John Riedl}. 2000.
\newblock \showarticletitle{Explaining collaborative filtering
  recommendations}. In {\em Proceedings of the 2000 ACM conference on Computer
  supported cooperative work}. ACM, 241--250.
\newblock


\bibitem[\protect\citeauthoryear{Hinton, Vinyals, and Dean}{Hinton
  et~al\mbox{.}}{2015}]%
        {hinton2015distilling}
{Geoffrey Hinton}, {Oriol Vinyals}, {and} {Jeff Dean}. 2015.
\newblock \showarticletitle{Distilling the knowledge in a neural network}.
\newblock {\em arXiv preprint arXiv:1503.02531\/} (2015).
\newblock


\bibitem[\protect\citeauthoryear{Hu, Da, Zeng, Yu, and Xu}{Hu
  et~al\mbox{.}}{2018}]%
        {Hu:2018:RLR:3219819.3219846}
{Yujing Hu}, {Qing Da}, {Anxiang Zeng}, {Yang Yu}, {and} {Yinghui Xu}. 2018.
\newblock \showarticletitle{Reinforcement Learning to Rank in E-Commerce Search
  Engine: Formalization, Analysis, and Application}. In {\em Proceedings of the
  24th ACM SIGKDD International Conference on Knowledge Discovery \&\#38; Data
  Mining} {\em (KDD '18)}. ACM, New York, NY, USA, 368--377.
\newblock
\showISBNx{978-1-4503-5552-0}
\showDOI{%
\url{http://dx.doi.org/10.1145/3219819.3219846}}


\bibitem[\protect\citeauthoryear{Huang, He, Gao, Deng, Acero, and Heck}{Huang
  et~al\mbox{.}}{2013}]%
        {huang2013learning}
{Po-Sen Huang}, {Xiaodong He}, {Jianfeng Gao}, {Li Deng}, {Alex Acero}, {and}
  {Larry Heck}. 2013.
\newblock \showarticletitle{Learning deep structured semantic models for web
  search using clickthrough data}. In {\em Proceedings of the 22nd ACM
  international conference on Conference on information \& knowledge
  management}. ACM, 2333--2338.
\newblock


\bibitem[\protect\citeauthoryear{Karmaker~Santu, Sondhi, and
  Zhai}{Karmaker~Santu et~al\mbox{.}}{2017}]%
        {karmaker2017application}
{Shubhra~Kanti Karmaker~Santu}, {Parikshit Sondhi}, {and} {ChengXiang Zhai}.
  2017.
\newblock \showarticletitle{On application of learning to rank for e-commerce
  search}. In {\em Proceedings of the 40th International ACM SIGIR Conference
  on Research and Development in Information Retrieval}. ACM, 475--484.
\newblock


\bibitem[\protect\citeauthoryear{Krakovna and Doshi-Velez}{Krakovna and
  Doshi-Velez}{2016}]%
        {krakovna2016increasing}
{Viktoriya Krakovna} {and} {Finale Doshi-Velez}. 2016.
\newblock \showarticletitle{Increasing the interpretability of recurrent neural
  networks using hidden Markov models}.
\newblock {\em arXiv preprint arXiv:1606.05320\/} (2016).
\newblock


\bibitem[\protect\citeauthoryear{Le and Mikolov}{Le and Mikolov}{2014}]%
        {le2014distributed}
{Quoc~V Le} {and} {Tomas Mikolov}. 2014.
\newblock \showarticletitle{Distributed Representations of Sentences and
  Documents.}. In {\em ICML}, Vol.~14. 1188--1196.
\newblock


\bibitem[\protect\citeauthoryear{Levy and Goldberg}{Levy and Goldberg}{2014}]%
        {levy2014neural}
{Omer Levy} {and} {Yoav Goldberg}. 2014.
\newblock \showarticletitle{Neural word embedding as implicit matrix
  factorization}. In {\em Advances in neural information processing systems}.
  2177--2185.
\newblock


\bibitem[\protect\citeauthoryear{Liang, Altosaar, Charlin, and Blei}{Liang
  et~al\mbox{.}}{2016}]%
        {liang2016factorization}
{Dawen Liang}, {Jaan Altosaar}, {Laurent Charlin}, {and} {David~M Blei}. 2016.
\newblock \showarticletitle{Factorization meets the item embedding:
  Regularizing matrix factorization with item co-occurrence}. In {\em
  Proceedings of the 10th ACM conference on recommender systems}. ACM, 59--66.
\newblock


\bibitem[\protect\citeauthoryear{Lim, Liu, and Lee}{Lim et~al\mbox{.}}{2010}]%
        {lim2010multi}
{Soon Chong~Johnson Lim}, {Ying Liu}, {and} {Wing~Bun Lee}. 2010.
\newblock \showarticletitle{Multi-facet product information search and
  retrieval using semantically annotated product family ontology}.
\newblock {\em Information Processing \& Management\/} {46}, 4 (2010),
  479--493.
\newblock


\bibitem[\protect\citeauthoryear{Loyola, Liu, and Hirate}{Loyola
  et~al\mbox{.}}{2017}]%
        {loyola2017modeling}
{Pablo Loyola}, {Chen Liu}, {and} {Yu Hirate}. 2017.
\newblock \showarticletitle{Modeling User Session and Intent with an
  Attention-based Encoder-Decoder Architecture}. In {\em Proceedings of the
  Eleventh ACM Conference on Recommender Systems}. ACM, 147--151.
\newblock


\bibitem[\protect\citeauthoryear{McAuley, Pandey, and Leskovec}{McAuley
  et~al\mbox{.}}{2015}]%
        {mcauley2015inferring}
{Julian McAuley}, {Rahul Pandey}, {and} {Jure Leskovec}. 2015.
\newblock \showarticletitle{Inferring networks of substitutable and
  complementary products}. In {\em Proceedings of the 21th ACM SIGKDD}. ACM,
  785--794.
\newblock


\bibitem[\protect\citeauthoryear{McLachlan and Krishnan}{McLachlan and
  Krishnan}{2007}]%
        {mclachlan2007algorithm}
{Geoffrey McLachlan} {and} {Thriyambakam Krishnan}. 2007.
\newblock {\em The EM algorithm and extensions}. Vol. 382.
\newblock John Wiley \& Sons.
\newblock


\bibitem[\protect\citeauthoryear{Mikolov, Chen, Corrado, and Dean}{Mikolov
  et~al\mbox{.}}{2013a}]%
        {mikolov2013efficient}
{Tomas Mikolov}, {Kai Chen}, {Greg Corrado}, {and} {Jeffrey Dean}. 2013a.
\newblock \showarticletitle{Efficient estimation of word representations in
  vector space}.
\newblock {\em arXiv preprint arXiv:1301.3781\/} (2013).
\newblock


\bibitem[\protect\citeauthoryear{Mikolov, Sutskever, Chen, Corrado, and
  Dean}{Mikolov et~al\mbox{.}}{2013b}]%
        {mikolov2013distributed}
{Tomas Mikolov}, {Ilya Sutskever}, {Kai Chen}, {Greg~S Corrado}, {and} {Jeff
  Dean}. 2013b.
\newblock \showarticletitle{Distributed representations of words and phrases
  and their compositionality}. In {\em Advances in neural information
  processing systems}. 3111--3119.
\newblock


\bibitem[\protect\citeauthoryear{Miller, Jordan, and Griffiths}{Miller
  et~al\mbox{.}}{2009}]%
        {miller2009nonparametric}
{Kurt Miller}, {Michael~I Jordan}, {and} {Thomas~L Griffiths}. 2009.
\newblock \showarticletitle{Nonparametric latent feature models for link
  prediction}. In {\em Advances in neural information processing systems}.
  1276--1284.
\newblock


\bibitem[\protect\citeauthoryear{Montavon, Lapuschkin, Binder, Samek, and
  M{\"u}ller}{Montavon et~al\mbox{.}}{2017}]%
        {montavon2017explaining}
{Gr{\'e}goire Montavon}, {Sebastian Lapuschkin}, {Alexander Binder}, {Wojciech
  Samek}, {and} {Klaus-Robert M{\"u}ller}. 2017.
\newblock \showarticletitle{Explaining nonlinear classification decisions with
  deep taylor decomposition}.
\newblock {\em Pattern Recognition\/}  {65} (2017), 211--222.
\newblock


\bibitem[\protect\citeauthoryear{Nickel, Tresp, and Kriegel}{Nickel
  et~al\mbox{.}}{2011}]%
        {nickel2011three}
{Maximilian Nickel}, {Volker Tresp}, {and} {Hans-Peter Kriegel}. 2011.
\newblock \showarticletitle{A Three-Way Model for Collective Learning on
  Multi-Relational Data.}. In {\em ICML}, Vol.~11. 809--816.
\newblock


\bibitem[\protect\citeauthoryear{Nurmi, Lagerspetz, Buntine, Flor{\'e}en, and
  Kukkonen}{Nurmi et~al\mbox{.}}{2008}]%
        {nurmi2008product}
{Petteri Nurmi}, {Eemil Lagerspetz}, {Wray Buntine}, {Patrik Flor{\'e}en},
  {and} {Joonas Kukkonen}. 2008.
\newblock \showarticletitle{Product retrieval for grocery stores}. In {\em
  Proceedings of the 31st ACM SIGIR}. ACM, 781--782.
\newblock


\bibitem[\protect\citeauthoryear{Palangi, Deng, Shen, Gao, He, Chen, Song, and
  Ward}{Palangi et~al\mbox{.}}{2016}]%
        {palangi2016deep}
{Hamid Palangi}, {Li Deng}, {Yelong Shen}, {Jianfeng Gao}, {Xiaodong He},
  {Jianshu Chen}, {Xinying Song}, {and} {Rabab Ward}. 2016.
\newblock \showarticletitle{Deep sentence embedding using long short-term
  memory networks: Analysis and application to information retrieval}.
\newblock {\em IEEE/ACM Transactions on ASLP\/} {24}, 4 (2016), 694--707.
\newblock


\bibitem[\protect\citeauthoryear{Ponte and Croft}{Ponte and Croft}{1998}]%
        {ponte1998language}
{Jay~M Ponte} {and} {W~Bruce Croft}. 1998.
\newblock \showarticletitle{A language modeling approach to information
  retrieval}. In {\em Proceedings of the 21st ACM SIGIR}. ACM, 275--281.
\newblock


\bibitem[\protect\citeauthoryear{Robertson and Walker}{Robertson and
  Walker}{1994}]%
        {robertson1994some}
{Stephen~E Robertson} {and} {Steve Walker}. 1994.
\newblock \showarticletitle{Some simple effective approximations to the
  2-poisson model for probabilistic weighted retrieval}. In {\em Proceedings of
  the 17th annual international ACM SIGIR conference on Research and
  development in information retrieval}. Springer-Verlag New York, Inc.,
  232--241.
\newblock


\bibitem[\protect\citeauthoryear{Rowley}{Rowley}{2000}]%
        {rowley2000product}
{Jennifer Rowley}. 2000.
\newblock \showarticletitle{Product search in e-shopping: a review and research
  propositions}.
\newblock {\em Journal of consumer marketing\/} {17}, 1 (2000), 20--35.
\newblock


\bibitem[\protect\citeauthoryear{Sarwar, Karypis, Konstan, and Riedl}{Sarwar
  et~al\mbox{.}}{2001}]%
        {sarwar2001item}
{Badrul Sarwar}, {George Karypis}, {Joseph Konstan}, {and} {John Riedl}. 2001.
\newblock \showarticletitle{Item-based collaborative filtering recommendation
  algorithms}. In {\em Proceedings of the 10th international conference on
  World Wide Web}. ACM, 285--295.
\newblock


\bibitem[\protect\citeauthoryear{Sharma and Cosley}{Sharma and Cosley}{2013}]%
        {sharma2013social}
{Amit Sharma} {and} {Dan Cosley}. 2013.
\newblock \showarticletitle{Do social explanations work?: studying and modeling
  the effects of social explanations in recommender systems}. In {\em
  Proceedings of the 22nd international conference on World Wide Web}. ACM,
  1133--1144.
\newblock


\bibitem[\protect\citeauthoryear{Shen, Tan, and Zhai}{Shen
  et~al\mbox{.}}{2005}]%
        {shen2005context}
{Xuehua Shen}, {Bin Tan}, {and} {ChengXiang Zhai}. 2005.
\newblock \showarticletitle{Context-sensitive information retrieval using
  implicit feedback}. In {\em Proceedings of the 28th annual international ACM
  SIGIR conference}. ACM, 43--50.
\newblock


\bibitem[\protect\citeauthoryear{Singh and Gordon}{Singh and Gordon}{2008}]%
        {singh2008relational}
{Ajit~P Singh} {and} {Geoffrey~J Gordon}. 2008.
\newblock \showarticletitle{Relational learning via collective matrix
  factorization}. In {\em Proceedings of the 14th ACM SIGKDD international
  conference on Knowledge discovery and data mining}. ACM, 650--658.
\newblock


\bibitem[\protect\citeauthoryear{Smucker, Allan, and Carterette}{Smucker
  et~al\mbox{.}}{2007}]%
        {smucker2007comparison}
{Mark~D Smucker}, {James Allan}, {and} {Ben Carterette}. 2007.
\newblock \showarticletitle{A comparison of statistical significance tests for
  information retrieval evaluation}. In {\em Proceedings of the sixteenth ACM
  CIKM}. ACM, 623--632.
\newblock


\bibitem[\protect\citeauthoryear{Socher, Chen, Manning, and Ng}{Socher
  et~al\mbox{.}}{2013}]%
        {socher2013reasoning}
{Richard Socher}, {Danqi Chen}, {Christopher~D Manning}, {and} {Andrew Ng}.
  2013.
\newblock \showarticletitle{Reasoning with neural tensor networks for knowledge
  base completion}. In {\em Advances in neural information processing systems}.
  926--934.
\newblock


\bibitem[\protect\citeauthoryear{Tintarev and Masthoff}{Tintarev and
  Masthoff}{2007}]%
        {tintarev2007survey}
{Nava Tintarev} {and} {Judith Masthoff}. 2007.
\newblock \showarticletitle{A survey of explanations in recommender systems}.
  In {\em Data Engineering Workshop, 2007 IEEE 23rd International Conference
  on}. IEEE, 801--810.
\newblock


\bibitem[\protect\citeauthoryear{Tintarev and Masthoff}{Tintarev and
  Masthoff}{2011}]%
        {tintarev2011designing}
{Nava Tintarev} {and} {Judith Masthoff}. 2011.
\newblock \showarticletitle{Designing and evaluating explanations for
  recommender systems}.
\newblock {\em Recommender Systems Handbook\/} (2011), 479--510.
\newblock


\bibitem[\protect\citeauthoryear{Van~Gysel, de~Rijke, and Kanoulas}{Van~Gysel
  et~al\mbox{.}}{2016}]%
        {van2016learning}
{Christophe Van~Gysel}, {Maarten de Rijke}, {and} {Evangelos Kanoulas}. 2016.
\newblock \showarticletitle{Learning latent vector spaces for product search}.
  In {\em Proceedings of the 25th ACM CIKM}. ACM, 165--174.
\newblock


\bibitem[\protect\citeauthoryear{Vidovic, G{\"o}rnitz, M{\"u}ller, and
  Kloft}{Vidovic et~al\mbox{.}}{2016}]%
        {vidovic2016feature}
{Marina M-C Vidovic}, {Nico G{\"o}rnitz}, {Klaus-Robert M{\"u}ller}, {and}
  {Marius Kloft}. 2016.
\newblock \showarticletitle{Feature importance measure for non-linear learning
  algorithms}.
\newblock {\em arXiv preprint arXiv:1611.07567\/} (2016).
\newblock


\bibitem[\protect\citeauthoryear{Vuli{\'c} and Moens}{Vuli{\'c} and
  Moens}{2015}]%
        {vulic2015monolingual}
{Ivan Vuli{\'c}} {and} {Marie-Francine Moens}. 2015.
\newblock \showarticletitle{Monolingual and cross-lingual information retrieval
  models based on (bilingual) word embeddings}. In {\em Proceedings of the 38th
  ACM SIGIR}. ACM, 363--372.
\newblock


\bibitem[\protect\citeauthoryear{Wang, Wang, Jia, and Yin}{Wang
  et~al\mbox{.}}{2018}]%
        {wang2018explainable}
{Nan Wang}, {Hongning Wang}, {Yiling Jia}, {and} {Yue Yin}. 2018.
\newblock \showarticletitle{Explainable recommendation via multi-task learning
  in opinionated text data}. In {\em The 41st International ACM SIGIR
  Conference on Research \& Development in Information Retrieval}. ACM,
  165--174.
\newblock


\bibitem[\protect\citeauthoryear{Wu, Yan, and Si}{Wu et~al\mbox{.}}{2017}]%
        {wu2017ensemble}
{Chen Wu}, {Ming Yan}, {and} {Luo Si}. 2017.
\newblock \showarticletitle{Ensemble Methods for Personalized E-Commerce Search
  Challenge at CIKM Cup 2016}.
\newblock {\em arXiv preprint arXiv:1708.04479\/} (2017).
\newblock


\bibitem[\protect\citeauthoryear{Wu, Hu, Hong, and Liu}{Wu
  et~al\mbox{.}}{2018}]%
        {wu2018turning}
{Liang Wu}, {Diane Hu}, {Liangjie Hong}, {and} {Huan Liu}. 2018.
\newblock \showarticletitle{Turning Clicks into Purchases: Revenue Optimization
  for Product Search in E-Commerce}.
\newblock  (2018).
\newblock


\bibitem[\protect\citeauthoryear{Yang, Liu, Wang, and Hu}{Yang
  et~al\mbox{.}}{2018}]%
        {yang2018towards}
{Fan Yang}, {Ninghao Liu}, {Suhang Wang}, {and} {Xia Hu}. 2018.
\newblock \showarticletitle{Towards Interpretation of Recommender Systems with
  Sorted Explanation Paths}. In {\em 2018 IEEE International Conference on Data
  Mining (ICDM)}. IEEE, 667--676.
\newblock


\bibitem[\protect\citeauthoryear{Yang, Ai, Guo, and Croft}{Yang
  et~al\mbox{.}}{2016}]%
        {yang2016anmm}
{Liu Yang}, {Qingyao Ai}, {Jiafeng Guo}, {and} {W~Bruce Croft}. 2016.
\newblock \showarticletitle{aNMM: Ranking short answer texts with
  attention-based neural matching model}. In {\em Proceedings of the 25th ACM
  International on Conference on Information and Knowledge Management}. ACM,
  287--296.
\newblock


\bibitem[\protect\citeauthoryear{Yu, Mohan, Putthividhya, and Wong}{Yu
  et~al\mbox{.}}{2014}]%
        {yu2014latent}
{Jun Yu}, {Sunil Mohan}, {Duangmanee~Pew Putthividhya}, {and} {Weng-Keen Wong}.
  2014.
\newblock \showarticletitle{Latent dirichlet allocation based diversified
  retrieval for e-commerce search}. In {\em Proceedings of the 7th ACM
  international conference on Web search and data mining}. ACM, 463--472.
\newblock


\bibitem[\protect\citeauthoryear{Zamani and Croft}{Zamani and Croft}{2016}]%
        {zamani2016estimating}
{Hamed Zamani} {and} {W~Bruce Croft}. 2016.
\newblock \showarticletitle{Estimating embedding vectors for queries}. In {\em
  Proceedings of the ACM ICTIR'16}. ACM, 123--132.
\newblock


\bibitem[\protect\citeauthoryear{Zhai and Lafferty}{Zhai and Lafferty}{2001}]%
        {zhai2001study}
{Chengxiang Zhai} {and} {John Lafferty}. 2001.
\newblock \showarticletitle{A study of smoothing methods for language models
  applied to ad hoc information retrieval}. In {\em Proceedings of the 24th ACM
  SIGIR}. ACM, 334--342.
\newblock


\bibitem[\protect\citeauthoryear{Zhai and Lafferty}{Zhai and Lafferty}{2004}]%
        {zhai2004study}
{Chengxiang Zhai} {and} {John Lafferty}. 2004.
\newblock \showarticletitle{A study of smoothing methods for language models
  applied to information retrieval}.
\newblock {\em ACM Transactions on Information Systems (TOIS)\/} {22}, 2
  (2004), 179--214.
\newblock


\bibitem[\protect\citeauthoryear{Zhang, Yuan, Lian, Xie, and Ma}{Zhang
  et~al\mbox{.}}{2016}]%
        {zhang2016collaborative}
{Fuzheng Zhang}, {Nicholas~Jing Yuan}, {Defu Lian}, {Xing Xie}, {and} {Wei-Ying
  Ma}. 2016.
\newblock \showarticletitle{Collaborative knowledge base embedding for
  recommender systems}. In {\em Proceedings of the 22nd ACM SIGKDD
  international conference on knowledge discovery and data mining}. ACM,
  353--362.
\newblock


\bibitem[\protect\citeauthoryear{Zhang}{Zhang}{2017}]%
        {zhang2017explainable}
{Yongfeng Zhang}. 2017.
\newblock \showarticletitle{Explainable Recommendation: Theory and
  Applications}.
\newblock {\em arXiv preprint arXiv:1708.06409\/} (2017).
\newblock


\bibitem[\protect\citeauthoryear{Zhang, Ai, Chen, and Croft}{Zhang
  et~al\mbox{.}}{2017}]%
        {zhang2017joint}
{Yongfeng Zhang}, {Qingyao Ai}, {Xu Chen}, {and} {W~Bruce Croft}. 2017.
\newblock \showarticletitle{Joint representation learning for top-n
  recommendation with heterogeneous information sources}. In {\em Proceedings
  of the 2017 ACM on Conference on Information and Knowledge Management}. ACM,
  1449--1458.
\newblock


\bibitem[\protect\citeauthoryear{Zhang, Ai, Chen, and Wang}{Zhang
  et~al\mbox{.}}{2018a}]%
        {zhang2018learning}
{Yongfeng Zhang}, {Qingyao Ai}, {Xu Chen}, {and} {Pengfei Wang}. 2018a.
\newblock \showarticletitle{Learning over knowledge-base embeddings for
  recommendation}. {\em arXiv preprint arXiv:1803.06540\/} (2018).
\newblock


\bibitem[\protect\citeauthoryear{Zhang, Chen, Ai, Yang, and Croft}{Zhang
  et~al\mbox{.}}{2018b}]%
        {zhang2018towards}
{Yongfeng Zhang}, {Xu Chen}, {Qingyao Ai}, {Liu Yang}, {and} {W~Bruce Croft}.
  2018b.
\newblock \showarticletitle{Towards conversational search and recommendation:
  System ask, user respond}. In {\em Proceedings of the 27th ACM International
  Conference on Information and Knowledge Management}. ACM, 177--186.
\newblock


\bibitem[\protect\citeauthoryear{Zhang, Lai, Zhang, Zhang, Liu, and Ma}{Zhang
  et~al\mbox{.}}{2014}]%
        {zhang2014explicit}
{Yongfeng Zhang}, {Guokun Lai}, {Min Zhang}, {Yi Zhang}, {Yiqun Liu}, {and}
  {Shaoping Ma}. 2014.
\newblock \showarticletitle{Explicit factor models for explainable
  recommendation based on phrase-level sentiment analysis}. In {\em Proceedings
  of the 37th international ACM SIGIR conference}. ACM, 83--92.
\newblock


\bibitem[\protect\citeauthoryear{Zhang, Mao, and Ai}{Zhang
  et~al\mbox{.}}{2019}]%
        {zhang2019sigir}
{Yongfeng Zhang}, {Jiaxin Mao}, {and} {Qingyao Ai}. 2019.
\newblock \showarticletitle{SIGIR 2019 Tutorial on Explainable Recommendation
  and Search}. In {\em Proceedings of the 42nd International ACM SIGIR
  Conference on Research and Development in Information Retrieval}. ACM,
  1417--1418.
\newblock


\bibitem[\protect\citeauthoryear{Zhang, Zhang, Zhang, Liu, and Ma}{Zhang
  et~al\mbox{.}}{2014}]%
        {zhang2014users}
{Yongfeng Zhang}, {Haochen Zhang}, {Min Zhang}, {Yiqun Liu}, {and} {Shaoping
  Ma}. 2014.
\newblock \showarticletitle{Do users rate or review?: Boost phrase-level
  sentiment labeling with review-level sentiment classification}. In {\em
  Proceedings of the 37th international ACM SIGIR conference}. ACM, 1027--1030.
\newblock


\bibitem[\protect\citeauthoryear{Zhu, Song, and Chen}{Zhu
  et~al\mbox{.}}{2016}]%
        {zhu2016max}
{Jun Zhu}, {Jiaming Song}, {and} {Bei Chen}. 2016.
\newblock \showarticletitle{Max-margin nonparametric latent feature models for
  link prediction}.
\newblock {\em arXiv preprint arXiv:1602.07428\/} (2016).
\newblock


\end{thebibliography}


\begin{thebibliography}{10}

\bibitem{TeXFAQ}
{UK \TeX{} Users Group}.
\newblock {UK} list of {\TeX} frequently asked questions.
\newblock \url{https://texfaq.org}, 2019.

\bibitem{Downes04:amsart}
Michael Downes and Barbara Beeton.
\newblock {\em The \textsf{amsart}, \textsf{amsproc}, and \textsf{amsbook}
  document~classes}.
\newblock American Mathematical Society, August 2004.
\newblock \url{http://www.ctan.org/pkg/amslatex}.

\bibitem{Fiorio15}
Cristophe Fiorio.
\newblock {\em {a}lgorithm2e.sty---package for algorithms}, October 2015.
\newblock \url{http://www.ctan.org/pkg/algorithm2e}.

\bibitem{Brito09}
Rog\'erio Brito.
\newblock {\em The algorithms bundle}, August 2009.
\newblock \url{http://www.ctan.org/pkg/algorithms}.

\bibitem{Heinz15}
Carsten Heinz, Brooks Moses, and Jobst Hoffmann.
\newblock {\em The Listings Package}, June 2015.
\newblock \url{http://www.ctan.org/pkg/listings}.

\bibitem{Fear05}
Simon Fear.
\newblock {\em Publication quality tables in {\LaTeX}}, April 2005.
\newblock \url{http://www.ctan.org/pkg/booktabs}.

\bibitem{ACMIdentityStandards}
Association for Computing Machinery.
\newblock {\em {ACM} Visual Identity Standards}, 2007.
\newblock \url{http://identitystandards.acm.org}.

\bibitem{Sommerfeldt13:Subcaption}
Axel Sommerfeldt.
\newblock {\em The subcaption package}, April 2013.
\newblock \url{http://www.ctan.org/pkg/subcaption}.

\bibitem{Nomencl}
Boris Veytsman, Bern Schandl, Lee Netherton, and C.~V. Radhakrishnan.
\newblock {\em A package to create a nomenclature}, September 2005.
\newblock \url{http://www.ctan.org/pkg/nomencl}.

\bibitem{Talbot16:Glossaries}
Nicola L.~C. Talbot.
\newblock {\em User Manual for glossaries.sty v4.25}, June 2016.
\newblock \url{http://www.ctan.org/pkg/subcaption}.

\bibitem{Carlisle04:Textcase}
David Carlisle.
\newblock {\em The \textsl{textcase} package}, October 2004.
\newblock \url{http://www.ctan.org/pkg/textcase}.

\end{thebibliography}



\begin{thebibliography}{35}


\ifx \showCODEN    \undefined \def \showCODEN     #1{\unskip}     \fi
\ifx \showDOI      \undefined \def \showDOI       #1{#1}\fi
\ifx \showISBNx    \undefined \def \showISBNx     #1{\unskip}     \fi
\ifx \showISBNxiii \undefined \def \showISBNxiii  #1{\unskip}     \fi
\ifx \showISSN     \undefined \def \showISSN      #1{\unskip}     \fi
\ifx \showLCCN     \undefined \def \showLCCN      #1{\unskip}     \fi
\ifx \shownote     \undefined \def \shownote      #1{#1}          \fi
\ifx \showarticletitle \undefined \def \showarticletitle #1{#1}   \fi
\ifx \showURL      \undefined \def \showURL       {\relax}        \fi
\providecommand\bibfield[2]{#2}
\providecommand\bibinfo[2]{#2}
\providecommand\natexlab[1]{#1}
\providecommand\showeprint[2][]{arXiv:#2}

\bibitem[\protect\citeauthoryear{Ablamowicz and Fauser}{Ablamowicz and
  Fauser}{2007}]%
        {Ablamowicz07}
\bibfield{author}{\bibinfo{person}{Rafal Ablamowicz} {and}
  \bibinfo{person}{Bertfried Fauser}.} \bibinfo{year}{2007}\natexlab{}.
\newblock \bibinfo{booktitle}{\emph{CLIFFORD: a Maple 11 Package for Clifford
  Algebra Computations, version 11}}.
\newblock
\urldef\tempurl%
\url{http://math.tntech.edu/rafal/cliff11/index.html}
\showURL{%
Retrieved February 28, 2008 from \tempurl}


\bibitem[\protect\citeauthoryear{Abril and Plant}{Abril and Plant}{2007}]%
        {Abril07}
\bibfield{author}{\bibinfo{person}{Patricia~S. Abril} {and}
  \bibinfo{person}{Robert Plant}.} \bibinfo{year}{2007}\natexlab{}.
\newblock \showarticletitle{The patent holder's dilemma: Buy, sell, or troll?}
\newblock \bibinfo{journal}{\emph{Commun. ACM}} \bibinfo{volume}{50},
  \bibinfo{number}{1} (\bibinfo{date}{Jan.} \bibinfo{year}{2007}),
  \bibinfo{pages}{36--44}.
\newblock
\urldef\tempurl%
\url{https://doi.org/10.1145/1188913.1188915}
\showDOI{\tempurl}


\bibitem[\protect\citeauthoryear{Andler}{Andler}{1979}]%
        {Andler79}
\bibfield{author}{\bibinfo{person}{Sten Andler}.}
  \bibinfo{year}{1979}\natexlab{}.
\newblock \showarticletitle{Predicate Path expressions}. In
  \bibinfo{booktitle}{\emph{Proceedings of the 6th. ACM SIGACT-SIGPLAN
  symposium on Principles of Programming Languages}}
  \emph{(\bibinfo{series}{POPL '79})}. \bibinfo{publisher}{ACM Press},
  \bibinfo{address}{New York, NY}, \bibinfo{pages}{226--236}.
\newblock
\urldef\tempurl%
\url{https://doi.org/10.1145/567752.567774}
\showDOI{\tempurl}


\bibitem[\protect\citeauthoryear{Anisi}{Anisi}{2003}]%
        {anisi03}
\bibfield{author}{\bibinfo{person}{David~A. Anisi}.}
  \bibinfo{year}{2003}\natexlab{}.
\newblock \emph{\bibinfo{title}{Optimal Motion Control of a Ground Vehicle}}.
\newblock \bibinfo{thesistype}{Master's\ thesis}. \bibinfo{school}{Royal
  Institute of Technology (KTH), Stockholm, Sweden}.
\newblock


\bibitem[\protect\citeauthoryear{Anzaroot and McCallum}{Anzaroot and
  McCallum}{2013}]%
        {UMassCitations}
\bibfield{author}{\bibinfo{person}{Sam Anzaroot} {and} \bibinfo{person}{Andrew
  McCallum}.} \bibinfo{year}{2013}\natexlab{}.
\newblock \bibinfo{booktitle}{\emph{{UMass} Citation Field Extraction
  Dataset}}.
\newblock
\urldef\tempurl%
\url{http://www.iesl.cs.umass.edu/data/data-umasscitationfield}
\showURL{%
Retrieved May 27, 2019 from \tempurl}


\bibitem[\protect\citeauthoryear{Clarkson}{Clarkson}{1985}]%
        {Clarkson85}
\bibfield{author}{\bibinfo{person}{Kenneth~L. Clarkson}.}
  \bibinfo{year}{1985}\natexlab{}.
\newblock \emph{\bibinfo{title}{Algorithms for Closest-Point Problems
  (Computational Geometry)}}.
\newblock \bibinfo{thesistype}{Ph.D. Dissertation}. \bibinfo{school}{Stanford
  University}, \bibinfo{address}{Palo Alto, CA}.
\newblock
\newblock
\shownote{UMI Order Number: AAT 8506171.}


\bibitem[\protect\citeauthoryear{Cohen}{Cohen}{1996}]%
        {JCohen96}
\bibfield{editor}{\bibinfo{person}{Jacques Cohen}} (Ed.).
  \bibinfo{year}{1996}\natexlab{}. \showarticletitle{Special issue: Digital
  Libraries}.
\newblock \bibinfo{journal}{\emph{Commun. {ACM}}} \bibinfo{volume}{39},
  \bibinfo{number}{11} (\bibinfo{date}{Nov.} \bibinfo{year}{1996}).

\bibitem[\protect\citeauthoryear{Cohen, Nutt, and Sagic}{Cohen
  et~al\mbox{.}}{2007}]%
        {Cohen07}
\bibfield{author}{\bibinfo{person}{Sarah Cohen}, \bibinfo{person}{Werner Nutt},
  {and} \bibinfo{person}{Yehoshua Sagic}.} \bibinfo{year}{2007}\natexlab{}.
\newblock \showarticletitle{Deciding equivalances among conjunctive aggregate
  queries}.
\newblock \bibinfo{journal}{\emph{J. ACM}} \bibinfo{volume}{54},
  \bibinfo{number}{2}, Article \bibinfo{articleno}{5} (\bibinfo{date}{April}
  \bibinfo{year}{2007}), \bibinfo{numpages}{50}~pages.
\newblock
\urldef\tempurl%
\url{https://doi.org/10.1145/1219092.1219093}
\showDOI{\tempurl}


\bibitem[\protect\citeauthoryear{Douglass, Harel, and Trakhtenbrot}{Douglass
  et~al\mbox{.}}{1998}]%
        {Douglass98}
\bibfield{author}{\bibinfo{person}{Bruce~P. Douglass}, \bibinfo{person}{David
  Harel}, {and} \bibinfo{person}{Mark~B. Trakhtenbrot}.}
  \bibinfo{year}{1998}\natexlab{}.
\newblock \showarticletitle{Statecarts in use: structured analysis and
  object-orientation}.
\newblock In \bibinfo{booktitle}{\emph{Lectures on Embedded Systems}},
  \bibfield{editor}{\bibinfo{person}{Grzegorz Rozenberg} {and}
  \bibinfo{person}{Frits~W. Vaandrager}} (Eds.). \bibinfo{series}{Lecture Notes
  in Computer Science}, Vol.~\bibinfo{volume}{1494}.
  \bibinfo{publisher}{Springer-Verlag}, \bibinfo{address}{London},
  \bibinfo{pages}{368--394}.
\newblock
\urldef\tempurl%
\url{https://doi.org/10.1007/3-540-65193-4_29}
\showDOI{\tempurl}


\bibitem[\protect\citeauthoryear{Editor}{Editor}{2007}]%
        {Editor00}
\bibfield{editor}{\bibinfo{person}{Ian Editor}} (Ed.).
  \bibinfo{year}{2007}\natexlab{}.
\newblock \bibinfo{booktitle}{\emph{The title of book one}
  (\bibinfo{edition}{1st.} ed.)}. \bibinfo{series}{The name of the series one},
  Vol.~\bibinfo{volume}{9}.
\newblock \bibinfo{publisher}{University of Chicago Press},
  \bibinfo{address}{Chicago}.
\newblock
\urldef\tempurl%
\url{https://doi.org/10.1007/3-540-09237-4}
\showDOI{\tempurl}


\bibitem[\protect\citeauthoryear{Editor}{Editor}{2008}]%
        {Editor00a}
\bibfield{editor}{\bibinfo{person}{Ian Editor}} (Ed.).
  \bibinfo{year}{2008}\natexlab{}.
\newblock \bibinfo{booktitle}{\emph{The title of book two}
  (\bibinfo{edition}{2nd.} ed.)}.
\newblock \bibinfo{publisher}{University of Chicago Press},
  \bibinfo{address}{Chicago}, Chapter 100.
\newblock
\urldef\tempurl%
\url{https://doi.org/10.1007/3-540-09237-4}
\showDOI{\tempurl}


\bibitem[\protect\citeauthoryear{Gundy, Balzarotti, and Vigna}{Gundy
  et~al\mbox{.}}{2007}]%
        {VanGundy07}
\bibfield{author}{\bibinfo{person}{Matthew~Van Gundy}, \bibinfo{person}{Davide
  Balzarotti}, {and} \bibinfo{person}{Giovanni Vigna}.}
  \bibinfo{year}{2007}\natexlab{}.
\newblock \showarticletitle{Catch me, if you can: Evading network signatures
  with web-based polymorphic worms}. In \bibinfo{booktitle}{\emph{Proceedings
  of the first USENIX workshop on Offensive Technologies}}
  \emph{(\bibinfo{series}{WOOT '07})}. \bibinfo{publisher}{USENIX Association},
  \bibinfo{address}{Berkley, CA}, Article \bibinfo{articleno}{7},
  \bibinfo{numpages}{9}~pages.
\newblock


\bibitem[\protect\citeauthoryear{Harel}{Harel}{1978}]%
        {Harel78}
\bibfield{author}{\bibinfo{person}{David Harel}.}
  \bibinfo{year}{1978}\natexlab{}.
\newblock \bibinfo{booktitle}{\emph{LOGICS of Programs: AXIOMATICS and
  DESCRIPTIVE POWER}}.
\newblock \bibinfo{type}{MIT Research Lab Technical Report} TR-200.
  \bibinfo{institution}{Massachusetts Institute of Technology},
  \bibinfo{address}{Cambridge, MA}.
\newblock


\bibitem[\protect\citeauthoryear{Harel}{Harel}{1979}]%
        {Harel79}
\bibfield{author}{\bibinfo{person}{David Harel}.}
  \bibinfo{year}{1979}\natexlab{}.
\newblock \bibinfo{booktitle}{\emph{First-Order Dynamic Logic}}.
  \bibinfo{series}{Lecture Notes in Computer Science},
  Vol.~\bibinfo{volume}{68}.
\newblock \bibinfo{publisher}{Springer-Verlag}, \bibinfo{address}{New York,
  NY}.
\newblock
\urldef\tempurl%
\url{https://doi.org/10.1007/3-540-09237-4}
\showDOI{\tempurl}


\bibitem[\protect\citeauthoryear{H{\"o}rmander}{H{\"o}rmander}{1985a}]%
        {MR781537}
\bibfield{author}{\bibinfo{person}{Lars H{\"o}rmander}.}
  \bibinfo{year}{1985}\natexlab{a}.
\newblock \bibinfo{booktitle}{\emph{The analysis of linear partial differential
  operators. {III}}}. \bibinfo{series}{Grundlehren der Mathematischen
  Wissenschaften [Fundamental Principles of Mathematical Sciences]},
  Vol.~\bibinfo{volume}{275}.
\newblock \bibinfo{publisher}{Springer-Verlag}, \bibinfo{address}{Berlin,
  Germany}. viii+525 pages.
\newblock
\showISBNx{3-540-13828-5}
\newblock
\shownote{Pseudodifferential operators.}


\bibitem[\protect\citeauthoryear{H{\"o}rmander}{H{\"o}rmander}{1985b}]%
        {MR781536}
\bibfield{author}{\bibinfo{person}{Lars H{\"o}rmander}.}
  \bibinfo{year}{1985}\natexlab{b}.
\newblock \bibinfo{booktitle}{\emph{The analysis of linear partial differential
  operators. {IV}}}. \bibinfo{series}{Grundlehren der Mathematischen
  Wissenschaften [Fundamental Principles of Mathematical Sciences]},
  Vol.~\bibinfo{volume}{275}.
\newblock \bibinfo{publisher}{Springer-Verlag}, \bibinfo{address}{Berlin,
  Germany}. vii+352 pages.
\newblock
\showISBNx{3-540-13829-3}
\newblock
\shownote{Fourier integral operators.}


\bibitem[\protect\citeauthoryear{IEEE}{IEEE}{2004}]%
        {2004:ITE:1009386.1010128}
IEEE \bibinfo{year}{2004}\natexlab{}.
\newblock \showarticletitle{IEEE TCSC Executive Committee}. In
  \bibinfo{booktitle}{\emph{Proceedings of the IEEE International Conference on
  Web Services}} \emph{(\bibinfo{series}{ICWS '04})}. \bibinfo{publisher}{IEEE
  Computer Society}, \bibinfo{address}{Washington, DC, USA},
  \bibinfo{pages}{21--22}.
\newblock
\showISBNx{0-7695-2167-3}
\urldef\tempurl%
\url{https://doi.org/10.1109/ICWS.2004.64}
\showDOI{\tempurl}


\bibitem[\protect\citeauthoryear{Kirschmer and Voight}{Kirschmer and
  Voight}{2010}]%
        {Kirschmer:2010:AEI:1958016.1958018}
\bibfield{author}{\bibinfo{person}{Markus Kirschmer} {and}
  \bibinfo{person}{John Voight}.} \bibinfo{year}{2010}\natexlab{}.
\newblock \showarticletitle{Algorithmic Enumeration of Ideal Classes for
  Quaternion Orders}.
\newblock \bibinfo{journal}{\emph{SIAM J. Comput.}} \bibinfo{volume}{39},
  \bibinfo{number}{5} (\bibinfo{date}{Jan.} \bibinfo{year}{2010}),
  \bibinfo{pages}{1714--1747}.
\newblock
\showISSN{0097-5397}
\urldef\tempurl%
\url{https://doi.org/10.1137/080734467}
\showDOI{\tempurl}


\bibitem[\protect\citeauthoryear{Knuth}{Knuth}{1997}]%
        {Knuth97}
\bibfield{author}{\bibinfo{person}{Donald~E. Knuth}.}
  \bibinfo{year}{1997}\natexlab{}.
\newblock \bibinfo{booktitle}{\emph{The Art of Computer Programming, Vol. 1:
  Fundamental Algorithms (3rd. ed.)}}.
\newblock \bibinfo{publisher}{Addison Wesley Longman Publishing Co., Inc.}
\newblock


\bibitem[\protect\citeauthoryear{Kosiur}{Kosiur}{2001}]%
        {Kosiur01}
\bibfield{author}{\bibinfo{person}{David Kosiur}.}
  \bibinfo{year}{2001}\natexlab{}.
\newblock \bibinfo{booktitle}{\emph{Understanding Policy-Based Networking}
  (\bibinfo{edition}{2nd.} ed.)}.
\newblock \bibinfo{publisher}{Wiley}, \bibinfo{address}{New York, NY}.
\newblock


\bibitem[\protect\citeauthoryear{Lamport}{Lamport}{1986}]%
        {Lamport:LaTeX}
\bibfield{author}{\bibinfo{person}{Leslie Lamport}.}
  \bibinfo{year}{1986}\natexlab{}.
\newblock \bibinfo{booktitle}{\emph{\it {\LaTeX}: A Document Preparation
  System}}.
\newblock \bibinfo{publisher}{Addison-Wesley}, \bibinfo{address}{Reading, MA.}
\newblock


\bibitem[\protect\citeauthoryear{Lee}{Lee}{2005}]%
        {Lee05}
\bibfield{author}{\bibinfo{person}{Newton Lee}.}
  \bibinfo{year}{2005}\natexlab{}.
\newblock \showarticletitle{Interview with Bill Kinder: January 13, 2005}.
\newblock \bibinfo{howpublished}{Video}.
\newblock \bibinfo{journal}{\emph{Comput. Entertain.}} \bibinfo{volume}{3},
  \bibinfo{number}{1}, Article \bibinfo{articleno}{4}
  (\bibinfo{date}{Jan.-March} \bibinfo{year}{2005}).
\newblock
\urldef\tempurl%
\url{https://doi.org/10.1145/1057270.1057278}
\showDOI{\tempurl}


\bibitem[\protect\citeauthoryear{Novak}{Novak}{2003}]%
        {Novak03}
\bibfield{author}{\bibinfo{person}{Dave Novak}.}
  \bibinfo{year}{2003}\natexlab{}.
\newblock \showarticletitle{Solder man}. \bibinfo{howpublished}{Video}. In
  \bibinfo{booktitle}{\emph{ACM SIGGRAPH 2003 Video Review on Animation theater
  Program: Part I - Vol. 145 (July 27--27, 2003)}}. \bibinfo{publisher}{ACM
  Press}, \bibinfo{address}{New York, NY}, \bibinfo{pages}{4}.
\newblock
\urldef\tempurl%
\url{https://doi.org/99.9999/woot07-S422}
\showDOI{\tempurl}


\bibitem[\protect\citeauthoryear{Obama}{Obama}{2008}]%
        {Obama08}
\bibfield{author}{\bibinfo{person}{Barack Obama}.}
  \bibinfo{year}{2008}\natexlab{}.
\newblock \bibinfo{title}{A more perfect union}.
\newblock \bibinfo{howpublished}{Video}.
\newblock
\urldef\tempurl%
\url{http://video.google.com/videoplay?docid=6528042696351994555}
\showURL{%
Retrieved March 21, 2008 from \tempurl}


\bibitem[\protect\citeauthoryear{Poker-Edge.Com}{Poker-Edge.Com}{2006}]%
        {Poker06}
\bibfield{author}{\bibinfo{person}{Poker-Edge.Com}.}
  \bibinfo{year}{2006}\natexlab{}.
\newblock \bibinfo{title}{Stats and Analysis}.
\newblock
\newblock
\urldef\tempurl%
\url{http://www.poker-edge.com/stats.php}
\showURL{%
Retrieved June 7, 2006 from \tempurl}


\bibitem[\protect\citeauthoryear{{R Core Team}}{{R Core Team}}{2019}]%
        {R}
\bibfield{author}{\bibinfo{person}{{R Core Team}}.}
  \bibinfo{year}{2019}\natexlab{}.
\newblock \bibinfo{booktitle}{\emph{R: A Language and Environment for
  Statistical Computing}}.
\newblock R Foundation for Statistical Computing, Vienna, Austria.
\newblock
\urldef\tempurl%
\url{https://www.R-project.org/}
\showURL{%
\tempurl}


\bibitem[\protect\citeauthoryear{Rous}{Rous}{2008}]%
        {rous08}
\bibfield{author}{\bibinfo{person}{Bernard Rous}.}
  \bibinfo{year}{2008}\natexlab{}.
\newblock \showarticletitle{The Enabling of Digital Libraries}.
\newblock \bibinfo{journal}{\emph{Digital Libraries}} \bibinfo{volume}{12},
  \bibinfo{number}{3}, Article \bibinfo{articleno}{5} (\bibinfo{date}{July}
  \bibinfo{year}{2008}).
\newblock
\newblock
\shownote{To appear.}


\bibitem[\protect\citeauthoryear{Saeedi, Zamani, and Sedighi}{Saeedi
  et~al\mbox{.}}{2010a}]%
        {SaeediMEJ10}
\bibfield{author}{\bibinfo{person}{Mehdi Saeedi},
  \bibinfo{person}{Morteza~Saheb Zamani}, {and} \bibinfo{person}{Mehdi
  Sedighi}.} \bibinfo{year}{2010}\natexlab{a}.
\newblock \showarticletitle{A library-based synthesis methodology for
  reversible logic}.
\newblock \bibinfo{journal}{\emph{Microelectron. J.}} \bibinfo{volume}{41},
  \bibinfo{number}{4} (\bibinfo{date}{April} \bibinfo{year}{2010}),
  \bibinfo{pages}{185--194}.
\newblock


\bibitem[\protect\citeauthoryear{Saeedi, Zamani, Sedighi, and Sasanian}{Saeedi
  et~al\mbox{.}}{2010b}]%
        {SaeediJETC10}
\bibfield{author}{\bibinfo{person}{Mehdi Saeedi},
  \bibinfo{person}{Morteza~Saheb Zamani}, \bibinfo{person}{Mehdi Sedighi},
  {and} \bibinfo{person}{Zahra Sasanian}.} \bibinfo{year}{2010}\natexlab{b}.
\newblock \showarticletitle{Synthesis of Reversible Circuit Using Cycle-Based
  Approach}.
\newblock \bibinfo{journal}{\emph{J. Emerg. Technol. Comput. Syst.}}
  \bibinfo{volume}{6}, \bibinfo{number}{4} (\bibinfo{date}{Dec.}
  \bibinfo{year}{2010}).
\newblock


\bibitem[\protect\citeauthoryear{Scientist}{Scientist}{2009}]%
        {JoeScientist001}
\bibfield{author}{\bibinfo{person}{Joseph Scientist}.}
  \bibinfo{year}{2009}\natexlab{}.
\newblock \bibinfo{title}{The fountain of youth}.
\newblock
\newblock
\newblock
\shownote{Patent No. 12345, Filed July 1st., 2008, Issued Aug. 9th., 2009.}


\bibitem[\protect\citeauthoryear{Smith}{Smith}{2010}]%
        {Smith10}
\bibfield{author}{\bibinfo{person}{Stan~W. Smith}.}
  \bibinfo{year}{2010}\natexlab{}.
\newblock \showarticletitle{An experiment in bibliographic mark-up: Parsing
  metadata for XML export}. In \bibinfo{booktitle}{\emph{Proceedings of the
  3rd. annual workshop on Librarians and Computers}}
  \emph{(\bibinfo{series}{LAC '10})},
  \bibfield{editor}{\bibinfo{person}{Reginald~N. Smythe} {and}
  \bibinfo{person}{Alexander Noble}} (Eds.), Vol.~\bibinfo{volume}{3}.
  \bibinfo{publisher}{Paparazzi Press}, \bibinfo{address}{Milan Italy},
  \bibinfo{pages}{422--431}.
\newblock
\urldef\tempurl%
\url{https://doi.org/99.9999/woot07-S422}
\showDOI{\tempurl}


\bibitem[\protect\citeauthoryear{Spector}{Spector}{1990}]%
        {Spector90}
\bibfield{author}{\bibinfo{person}{Asad~Z. Spector}.}
  \bibinfo{year}{1990}\natexlab{}.
\newblock \showarticletitle{Achieving application requirements}.
\newblock In \bibinfo{booktitle}{\emph{Distributed Systems}
  (\bibinfo{edition}{2nd.} ed.)}, \bibfield{editor}{\bibinfo{person}{Sape
  Mullender}} (Ed.). \bibinfo{publisher}{ACM Press}, \bibinfo{address}{New
  York, NY}, \bibinfo{pages}{19--33}.
\newblock
\urldef\tempurl%
\url{https://doi.org/10.1145/90417.90738}
\showDOI{\tempurl}


\bibitem[\protect\citeauthoryear{Thornburg}{Thornburg}{2001}]%
        {Thornburg01}
\bibfield{author}{\bibinfo{person}{Harry Thornburg}.}
  \bibinfo{year}{2001}\natexlab{}.
\newblock \bibinfo{booktitle}{\emph{Introduction to Bayesian Statistics}}.
\newblock
\urldef\tempurl%
\url{http://ccrma.stanford.edu/~jos/bayes/bayes.html}
\showURL{%
Retrieved March 2, 2005 from \tempurl}


\bibitem[\protect\citeauthoryear{TUG}{TUG}{2017}]%
        {TUGInstmem}
TUG \bibinfo{year}{2017}\natexlab{}.
\newblock \bibinfo{booktitle}{\emph{Institutional members of the {\TeX} Users
  Group}}.
\newblock
\urldef\tempurl%
\url{http://wwtug.org/instmem.html}
\showURL{%
Retrieved May 27, 2017 from \tempurl}


\bibitem[\protect\citeauthoryear{Veytsman}{Veytsman}{[n.d.]}]%
        {CTANacmart}
\bibfield{author}{\bibinfo{person}{Boris Veytsman}.}
  \bibinfo{year}{[n.d.]}\natexlab{}.
\newblock \bibinfo{booktitle}{\emph{acmart---{C}lass for typesetting
  publications of {ACM}}}.
\newblock
\urldef\tempurl%
\url{http://www.ctan.org/pkg/acmart}
\showURL{%
Retrieved May 27, 2017 from \tempurl}


\end{thebibliography}
